\documentclass[11pt,a4paper]{article}
\usepackage{jheppub,amsmath,  amssymb,slashed,url,bm,textgreek,upgreek}
\usepackage{graphicx}
\usepackage{epstopdf}
\def\t{{ \sf t}} 
\def\Sigmax{{\Sigma^*}}

\def\J{\mathcal J}

\def\sPi{\Phi}

\def\RR{{\mathcal R}}

\def\fD{ \eusm D}

\def\Chi{\chi}

\def\be{\begin{equation}}
\def\ee{\end{equation}}

\def\NS{{\mathrm {NS}}}
\def\Ra{{\mathrm R}}
\def\Ber{{\mathrm{Ber}}}
\def\BBer{{\textit{Ber}}}

\def\Im{{\mathrm{Im}}}
\def\hat{\widehat}

\def\tilde{\widetilde}

\def\frak{\mathfrak}
\def\h{\widehat}

\def\sPhi{\Psi}

\def\D{{\mathcal D}}
\def\S{{\mathcal S}}
\def\SIgma{\Sigma}

\def\V{{\mathcal V}}
\def\O{{\mathcal O}}

\def\red{{\mathrm{red}}}

\def\d{{\mathrm d}}

\def\R{{\mathbb R}}
\def\C{{\mathbb C}}
\def\U{{\mathcal U}}
\def\D{{\mathcal D}}
\def\[{\bigl [}
\def\voli{{\eurm {vol}}^{-1}\cdot}
\def\]{\bigr ]}

\def\CP{{\mathbb{CP}}}

\def\F{{\mathcal F}}

\def\Z{{\mathbb Z}}

\def\L{{\mathcal  L}}
\def\t{\widetilde }
\def\h{\widehat}

\def\V{{\mathcal V}}

\def\M{{\mathcal M}}

\def\MM{{\mathfrak M}}

\def\W{{\mathcal W}}

\def\Ra{{\mathrm{R}}}

\def\st{{\mathrm{st}}}
\def\spin{{\mathrm{spin}}}

\def\tilde{\widetilde}

\def\bar{\overline}

\def\Ber{{\mathrm {Ber}}}

\def\dets{\det{}\negthinspace}
\font\teneurm=eurm10 \font\seveneurm=eurm7  \font\fiveeurm=eurm5
\newfam\eurmfam
\textfont\eurmfam=\teneurm \scriptfont\eurmfam=\seveneurm
\scriptscriptfont\eurmfam=\fiveeurm
\def\eurm#1{{\fam\eurmfam\relax#1}}
\font\teneusm=eusm10 \font\seveneusm=eusm7 \font\fiveeusm=eusm5
\newfam\eusmfam
\textfont\eusmfam=\teneusm \scriptfont\eusmfam=\seveneusm
\scriptscriptfont\eusmfam=\fiveeusm
\def\eusm#1{{\fam\eusmfam\relax#1}}
\font\tencmmib=cmmib10 \skewchar\tencmmib='177
\font\sevencmmib=cmmib7 \skewchar\sevencmmib='177
\font\fivecmmib=cmmib5 \skewchar\fivecmmib='177
\newfam\cmmibfam
\textfont\cmmibfam=\tencmmib \scriptfont\cmmibfam=\sevencmmib
\scriptscriptfont\cmmibfam=\fivecmmib

\title{Notes On Holomorphic String And Superstring Theory Measures Of Low Genus}

 \author{Edward Witten}
\affiliation{School of Natural Sciences, Institute for Advanced Study,\\ 1 Einstein Drive, Princeton, NJ 08540 USA}
\abstract{
It has long been known that in principle, the genus $g$  vacuum amplitude for  bosonic strings or superstrings in  26 or 10  dimensions
can be entirely determined
from conditions of holomorphy.  Moreover, this has been done in practice for bosonic strings of low genus.  Here we describe in a unified
way how to determine the bosonic string and superstring vacuum amplitude in genus 1 and 2 via holomorphy.  The main novelty is the superstring analysis in genus 2, where we use holomorphy to get a new understanding of some of the results
that previously have been obtained by more explicit calculations.}

\begin{document}\maketitle

\section{Introduction}\label{intro} 

The vacuum amplitude of the bosonic string in 26 dimensions -- in other words the measure on the moduli space of Riemann surfaces
that is determined by the worldsheet path integral -- can be entirely determined by considerations of holomorphy \cite{BK},
and moreover, this gives a practical basis for calculation \cite{BMmore,Moore,Mooremore,Morozov,BeM}.  In fact, although the vacuum amplitude
of the bosonic string at genus 1 was computed explicitly in the 1970's, for genus $\geq 2$ most computations have relied 
heavily on holomorphy.

The holomorphic methods in question are based on the Mumford isomorphism \cite{Mumford} between certain
line bundles on $\M_g$, the moduli space of Riemann surfaces of genus $g$.  There is an analogous though less widely known
super analog of the Mumford isomorphism, in this case an isomorphism between line bundles on $\MM_g$, the moduli space
of super Riemann surfaces of genus $g$ (see  \cite{YMa,BMFS,BS}, and especially \cite{RSV}).   The super Mumford isomorphism in some respects is more explicit than the ordinary one, but
it has been less exploited because to do so requires coming to grips with the subtleties of super Riemann surfaces.  Accordingly,
our present knowledge of superstring vacuum amplitudes is based primarily not on arguments of holomorphy but on explicit
computations -- the foundational computations in genus 1 that date back to the 1970's, and much more recently a tour de force in genus 2 that
is reviewed with references in \cite{DPHgold}.   (For an introduction to earlier attempts to understand the genus 2 superstring measure,
the reader may consult \cite{Ortin}.)

Our main goal here is to reconsider the genus 2 superstring measure from the point of view of holomorphy. For orientation, we begin by reviewing from the viewpoint of holomorphy
the bosonic string measure in genus 1 and 2, and in the superstring case we also analyze the genus 1 measure via holomorphy.  
In section \ref{bosonic}, we study the bosonic
string and in section \ref{super}, we study superstring theory.    We use the fact that a Riemann surface of genus 1 or 2 is hyperelliptic.  For a hyperelliptic Riemann
surface, the Mumford isomorphism can be made particularly explicit  \cite{BeS}.  A genus 2 super Riemann surface is not hyperelliptic (in the sense
that it is not a double cover of a genus 0 super Riemann surface), but in the case of an even spin structure, with the aid of the splitting
of $\MM_2$ that was exploited in \cite{DPHgold}, one can use the hyperelliptic nature of a genus 2 ordinary Riemann surface to analyze the genus 2 superstring measure.  For the case of an odd spin structure, this method is not available, though in that case the vacuum amplitude vanishes. The behavior of the 
superstring amplitude at a separating or nonseparating degeneration is the subject of sections \ref{separsuper} and \ref{nonseparating}.

Can similar methods be applied to superstring theory beyond genus 2?  The literature contains a proposal \cite{cacc} for a genus 3 superstring measure.
The main tool used in computing the superstring measure in genus 2  -- a holomorphic projection $\pi$ from the moduli
space of super Riemann surfaces to its reduced space, defined using the super period matrix \cite{DPHgold} -- has an analog in genus 3, with the important difference that
in genus 3, $\pi$ is only meromorphic (with a pole on what is sometimes called the theta-null divisor, defined in section \ref{superwhatfor}).  
Still, the pushforward $\pi_*(\Psi_{3,+})$ (where $\Psi_{3,+}$, as described
later, is the super Mumford form on super moduli space) is holomorphic, since fermion zero-modes more than compensate for the pole in $\pi$.  Whether this pushforward satisfies the assumptions made in \cite{cacc} for
a genus 3 superstring measure can be determined by an extension of the analysis of the super period matrix made in the present paper.  The literature also contains proposals (for example, see \cite{grush,cacctwo,matvol}) for superstring measures for genus $>3$. At the moment,
it is difficult to suggest even an optimistic interpretation of these proposals, since a natural analog of the projection $\pi$ -- even as a meromorphic
projection -- is not known above genus 3.  (Meromorphic projections certainly exist for all genus, but one would one need a nice one to have any hope of
getting the sort of formulas that have been proposed in the literature.)  A holomorphic projection does not exist for genus $\geq 5$ \cite{DonW}.  

One last comment is that understanding what one can say using holomorphy about vacuum amplitudes for superstrings in $\R^{10}$ -- which will be our
goal here -- is rather different from describing a general procedure for superstring perturbation theory.  The latter problem calls for quite different methods;
see for example  \cite{Revisited,more}.

\section{The Bosonic String In Genus 1 And 2}\label{bosonic}

\subsection{The Mumford Isomorphism}\label{mumf}

For $V$ a vector space of dimension $n$, we write $\det V$ for the top exterior power $\wedge^nV$. If $\Sigma$
is a Riemann surface and $\L\to\Sigma$ is a holomorphic line bundle, then the sheaf cohomology of $\Sigma$ with
values in $\L$ consists of the two cohomology groups $H^0(\Sigma,\L)$ and $H^1(\Sigma,\L)$.  The determinant of cohomology
of $\L$, denoted $\det H^*(\Sigma,\L)$ or just $\det H^*(\L)$, is defined to be
\begin{equation}\det H^*(\L)=\det H^0(\SIgma,\L)\otimes (\det H^1(\Sigma,\L))^{-1}. \end{equation}
If $\Sigma$ and $\L$ vary holomorphically with some parameter space $B$, then $\det H^*(\L)$ is a holomorphic line bundle over $B$.\footnote{\label{jumping} In our simple
definition of $\det H^*(\L)$ as a line bundle over $B$, we have assumed that the cohomology groups $H^i(\Sigma,\L)$ vary holomorphically with the parameters in $B$.
This is so if and only if the dimensions of $H^i(\Sigma,\L)$ are constant.  However, a more sophisticated definition of $\det H^*(\L)$ as a holomorphic line bundle over $B$
can be given without this assumption \cite{Quillen}.}

We apply this to the case that $\Sigma$ is a Riemann surface of genus $g$, and $B=\M_g$ is the moduli space of Riemann surfaces of genus $g$.
Moreover, we take $\L$ to be a power of $K=T^*\Sigma$, the canonical bundle of $\Sigma$ (in other words, the relative canonical bundle of the universal
curve over $\M_g$).  The Mumford isomorphism is the statement that
\begin{equation}\label{zolbot} \det H^*(K^2)\cong \left(\det H^*(K)\right)^{13}. \end{equation}
Accordingly,  $\det H^*(K^2)\otimes \det{}^{-13}\, H^*(K)$ is trivial  (we abbreviate the $q^{th}$ tensor power of $\det H^*(K)$ as $\det^qH^*(K)$), and it has a global and everywhere nonzero holomorphic section $\sPi_g$ 
\begin{equation}\label{molbo}\sPi_g \in H^0(\M_g,\det H^*(K^2)\otimes \det{}^{-13}\,H^*(K))\end{equation}
that is uniquely determined\footnote{This statement is oversimplified as $\M_g$ is not compact. {\it A priori}, to determine $\Phi_g$ up
to a constant multiple, one may expect to need some knowledge about its behavior at infinity. In practice, not much such knowledge is needed
and conformal field theory provides more than enough information. } 
up to multiplication by a nonzero complex constant.  For a suitable choice of the constant (which depends on
the string coupling constant $g_\st$), $\sPi_g$ is the holomorphic part of the genus $g$ vacuum amplitude of the bosonic string \cite{BK}
(and it is also, therefore, one factor in the vacuum amplitude of the heterotic string).  We will not explain here why this is true, except to note
the following.  The holomorphic part of the bosonic string vacuum amplitude is the product of the path integral of the $bc$ ghost system and the path
integral of the holomorphic modes of the matter fields.  The $bc$ path integral is a holomorphic section of the appropriate determinant line bundle,
which is $\det H^*(K^2)$, and the holomorphic part of the matter path integral for uncompactified bosonic strings is a section of $\det^{-13} H^*(K)$.
In this last statement, the factor of 13 comes from the fact that $\R^{26}\cong\C^{13}$, and the minus sign reflects the fact that the matter fields
describing motion in $\R^{26}$ are bosonic.  

In making  (\ref{molbo}) more explicit, the two cases of $g\geq 2$ and $g=1$ are slightly different. (From our present point of view, $g=0$ is trivial
as there are no moduli.)  For any $g$, $H^0(\Sigma,K^2)$ is the cotangent bundle to $\M_g$ at the point corresponding to $\Sigma$.  We denote
this cotangent bundle as $T^*\M_g$.  For $g>1$, $H^1(\SIgma,K^2)$ vanishes.  
So in this case, $\det H^*(K^2)\cong \det\,T^*\M_g$.  For any $g$, $H^0(\Sigma,K)$ is the
$g$-dimensional space of holomorphic 1-forms on $\Sigma$.  On the other hand, $H^1(\Sigma,K)$ is always 1-dimensional and canonically
isomorphic to $\C$, with an isomorphism given by the map that takes a $(1,1)$-form $\mu\in H^1(\SIgma,K)$ to its integral $\int_\SIgma \mu$.
Putting these facts together, for $g\geq 2$, the bosonic string measure is a section
\begin{equation}\label{tobbo} \sPi_g\in H^0(\M_g,\det T^*\M_g\otimes \dets^{-13}\,H^0(\SIgma,K)).\end{equation}

The only difference for $g=1$ is that in this case, $H^1(\Sigma,K^2)$ is nonzero.  By Serre duality, it is dual to $H^0(\Sigma,T)$, where $T=T\Sigma$
is the tangent bundle to $\Sigma$.  In turn, for $g=1$, $H^0(\Sigma,T)$ is naturally dual to $H^0(\SIgma,K)$.  Indeed, for a genus 1 curve
$y^2=P(x)$ (where $P(x)$ is a cubic or quartic  polynomial), $H^0(\Sigma,T)$ is generated by the everywhere nonzero holomorphic vector
field $y \partial_x$, and $H^0(\Sigma,K)$ is generated by the inverse of this, the everywhere nonzero holomorphic 1-form $\d x/y$.
Putting these facts together, $H^1(\Sigma,K^2)$ for $g=1$ is naturally isomorphic to $H^0(\Sigma,K),$ which also coincides with $\det H^0(\Sigma,K)$,
since $H^0(\Sigma,K)$ has rank 1.   Hence
$\det H^*(K^2)\cong \det T^*\M_g\otimes \det^{-1}H^0(\Sigma,K)$, and so in genus 1, we have 
\begin{equation}\label{obbo}\sPi_1\in H^0(\M_1,\det T^*\M_1\otimes \dets^{-14}\,H^0(\SIgma,K))=H^0(\M_1, T^*\M_1\otimes H^0(\SIgma,K)^{-14}).\end{equation}
We use the fact that for $g=1$, $T^*\M_1$ and $H^0(\Sigma,K)$ are both 1-dimensional and hence equal to their own determinants.

\subsubsection{What Is $\sPi_g$ Good For?}\label{what for}

What does one do with $\sPi_g$? It is  a holomorphic section  of the line bundle $\det T^*\M_g\otimes \RR$, where $\RR=\dets^{-r}\,H^0(\Sigma,K)$, with $r=13$ or 14, 
depending on $g$.
$\M_g$ is a complex manifold of dimension $3g-3$.  A section of $\det T^*\M_g$ would be a differential form of degree $(3g-3,0)$, so $\sPi_g$ is a $(3g-3,0)$-form on $\M_g$
with values in $\RR$.  Its complex conjugate $\bar\sPi_g$ is accordingly a $(0,3g-3$)-form with values in $\bar\RR$ (the complex conjugate of $\RR$).  The product $|\sPi_{g}|^2
=\bar\sPi_{g}\sPi_{g}$ is thus a $(3g-3,3g-3)$-form with values in $\bar \RR\otimes \RR$.  However, the line bundle $\RR$ has a natural hermitian metric, as we explain shortly.
A hermitian metric can be viewed as a bilinear 
map $\frak H:\bar\RR\otimes \RR\to \O$, where $\O$ is a trivial line bundle.    So $\frak H(|\sPi_g|^2)$ is a $(3g-3,3g-3)$-form on $\M_g$.
Such a form can be integrated, as least locally.  In bosonic string theory, the integral diverges because of infrared instabilities ($\M_g$ is not compact and $\sPi_g$
has a pole at the compactification divisor at infinity, as we discuss later).  In superstring theory, the analogous procedure actually leads to well-defined integrals.

To define a hermitian metric on the line bundle $\RR$, one begins with the fact that there is a natural hermitian metric on $H^0(\Sigma,K)$.  Indeed, if $\omega$ is
a holomorphic 1-form on $\Sigma$, one defines $|\omega|^2=-\frac{i}{2}\int_\Sigma\bar\omega\wedge\omega$.  This hermitian metric on $H^0(\Sigma,K)$ induces
one on $\det H^0(\Sigma,K)$ and hence on any power of $\det H^0(\Sigma,K)$, such as $\RR$.  To make all this a little more explicit, locally in moduli space one can
pick $A$- and $B$-cycles $A^i,B_j$, $i,j=1,\dots,g$ on $\Sigma$, and a basis of holomorphic 1-forms $\omega_k$ normalized so that $\oint_{A^i}\omega_k=\delta^i_k$;
the period matrix $\Omega$ is defined by $\Omega_{jk}=\oint_{B_j}\omega_k$.
Then the expression $\sigma=\omega_1\wedge \omega_2\wedge \dots\wedge \omega_g$ defines a local  holomorphic trivialization of $\det H^0(\Sigma,K)$;
its norm with respect to the hermitian metric on $\det H^0(\Sigma,K)$ is $\det\Im\,\Omega$.
Locally $\sPi_g=
\Lambda_g \cdot \sigma^{-r}$, where $\Lambda_g$ is a holomorphic $(3g-3,0)$-form, and 
\begin{equation}\label{lowf} \frak H(|\sPi_g|^2)=\frac{\bar\Lambda_g\wedge\Lambda_g}{(\det\,\Im\,\Omega)^r}. \end{equation}

In terms of the path integral of bosonic string theory, one can think of the pairing of holomorphic and antiholomorphic 1-forms via
$\frak H$ as coming from the integral over the zero-modes of the bosonic matter fields.
These are the modes that cannot be simply interpreted as part of the holomorphic or antiholomorphic degrees of freedom.

Though in this paper we consider primarily closed string theories, we will briefly indicate the relevance of the holomorphic form $\Phi_g$
 to open and/or unoriented bosonic
strings (the superanalog that we describe in section \ref{supermumford} is similarly applicable to open and/or unoriented superstring theories).  
Let $\Sigma$ be an open and/or unoriented
Riemann surface whose closed oriented double cover $\Sigma'$ has genus $g$.  Then the space $\Gamma$ that parametrizes the moduli of 
 $\Sigma$ is
a component of the fixed point set of a real involution $\tau$ of $\M_g$. $\Sigma'$ has a natural real structure which
 induces a real structure on $H^0(\Sigma',K)|_\Gamma$.
Together with the hermitian metric, this determines a trivialization of $\det H^*(\Sigma',K)|_\Gamma$
(up to sign) and $\Phi_g|_\Gamma$ is a differential form on $\Gamma$ of top degree (more precisely a density) which defines, up to a constant multiple,
 the vacuum amplitude
for this component of the moduli space of open and/or unoriented bosonic strings.  

\subsection{Hyperelliptic Curves}\label{hyper}

For practical purposes, to calculate, we will use the fact that a Riemann surface $\Sigma$ of genus 1 or 2 is hyperelliptic, governed by an equation
\begin{equation}\label{zolbo} y^2 = \prod_{i=1}^s (x-e_i),\end{equation}
where $s=4$ for genus 1, and $s= 6$ for genus 2.  We require the $e_i$ to be distinct, which ensures that $\Sigma$ is smooth.  To make $\Sigma$ compact,
we include two points with $x\to\infty$, $y\sim\pm x^{s/2}$.  The map that forgets $y$ exhibits $\Sigma$ as a double cover of $\CP^1$, with branch points
at $x=e_1,\dots, e_s$.  To describe all covers of $\CP^1$ with $s$ branch points, it is convenient to include the limit that one of the $e_i$ goes to infinity.
Then one takes $y\to \sqrt{- e_i}\, y$, and in the limit that $e_i\to\infty$, we get an equation with the same form as (\ref{zolbo}), but with one less branch point on the finite
$x$-plane.  

Since each $e_i$ is a point in $\CP^1$, the collection of the $e_i$ parametrizes a space that we will call $M$; it is the product of $s$ copies of $\CP^1$, with diagonals
removed as we require the $e_i$ to be distinct.  The group $SL(2,\C)$ acts on $\CP^1$ and therefore on $M$.  To construct the moduli space $\M_1$ or $\M_2$
of Riemann surfaces of genus 1 or 2, we take the quotient $M/SL(2,\C)$ and also divide by the group $\Theta$ of permutations of the $e_i$.

Let us first describe a convenient way to construct an $SL(2,\C)$-invariant differential form on $M$ of top degree.  We begin with an $(s,0)$-form 
$\Lambda = F(e_1,\dots,e_s) \d e^1\dots \d e^s$ on $M$.    Let us assume that $F$ is chosen so that $\Lambda$ is $SL(2,\C)$-invariant and also invariant
under permutation of the $e_i$.  (In our application,
$\Lambda$ will be an $SL(2,\C)$-invariant form valued in a certain line bundle, not an ordinary form.)  This does not mean that $\Lambda$ is a pullback from
$M/SL(2,\C)$; for this it should vanish if contracted with one of the vector fields that generate $SL(2,\C)$.  These vector fields are
\begin{equation}\label{tort} v^a=\sum_{i=1}^s e_i^a\partial_{e_i},~~a=0,1,2.  \end{equation}
For a vector field $v$, let $\iota_v$ be the operation of contraction with respect to $v$.  Since $\iota_v\iota_{v'}=-\iota_{v'}\iota_v$ for any $v,v'$ (and in particular
$\iota_v^2=0$), it follows  that given any form $\Lambda$, the triple contraction $\iota_{v^0}\iota_{v^1}\iota_{v^2}\Lambda$ vanishes when contracted
with any of the $v^a$.  We denote this triple contraction as $\voli \Lambda$ (where the notation is meant to suggest that the triple contraction is a way to remove
the volume form $\eurm{vol}$ of $SL(2,\C)$).  If $\Lambda$ is $SL(2,\C)\times\Theta$-invariant, then $\voli\Lambda$ is a pullback from $M/(SL(2,\C)\times\Theta)$,
which for $s=4$ or $s=6$ is the moduli space $\M_1$ or $\M_2$ of Riemann surfaces of genus 1 or 2.  This is a convenient way to construct forms on these
moduli spaces.

A convenient way to make $\voli\Lambda$ more explicit is as follows.  Let $a<b<c$ be any three elements of the finite set $\{1,2,\dots,s\}$.  Any $SL(2,\C)$
orbit on $M$ has a unique point with specified values of $e_a,e_b$, and $e_c$.    So instead taking the quotient $M/SL(2,\C)$, we could restrict to a subspace
 $M'\subset M$ in which $e_a$, $e_b$, and $e_c$ are fixed.  When restricted to $M'$, $\d e_a=\d e_b=\d e_c=0$.  So when we compute $\voli\Lambda$,
 we need only keep the terms in which the contractions remove $\d e_a$, $\d e_b$, and $\d e_c$.  Since
 \begin{equation}\iota_{v^0}\iota_{v^1}\iota_{v^2}\d e_a\d e_b\d e_c=(e_a-e_b)(e_b-e_c)(e_c-e_a),\end{equation}
 it follows that when restricted to $M'\cong M/SL(2,\C)$,
 \begin{equation}\label{umonk}\voli \d e^1\d e^2\dots\d e^s =(-1)^{a+b+c}(e_a-e_b)(e_b-e_c)(e_c-e_a)\d e_1\dots \hat{\d e_a}\dots \hat {\d e_b}\dots
 \hat{\d e_c}\dots \d e_s. \end{equation}
 This gives convenient formulas in which the moduli space is parametrized by the $e_i$ for $i\not=a,b,c$.  Perhaps the main drawback of such a parametrization is that
 it does not make manifest the group $\Theta$ of permutations.
 
 Now we can easily make the Mumford isomorphism explicit in genus 1 and 2 and determine the bosonic string measure.
 
 \subsubsection{The Mumford Isomorphism In Genus 1}\label{mumone}
 
 For $\Sigma$ of  genus 1, we set $s=4$. The space $H^0(\Sigma,K)$ of holomorphic differentials is 1-dimensional, generated by $\d x/y$.  The bosonic string measure $\sPi_1$
 is supposed to be a holomorphic 1-form on $\M_1$ valued in $H^0(\Sigma,K)^{-14}$, so it has the form
 \begin{equation}\label{tombi}\sPi_1=\voli \frac{F(e_1,\dots,e_4)\d e_1\dots\d e_4}{(\d x/y)^{14}},\end{equation}
 for some function $F$.  The Mumford isomorphism tells us that $\sPi_1$ has neither zeroes nor poles as long as the $e_i$ are distinct, and we will assume\footnote{The
 assumption follows either from a stronger version of the Mumford isomorphism than we have stated, or from some knowledge of the conformal field theory,
 according to which the orders of the poles are determined by the ground state energy of the string.}
 that the singularities as $e_i\to e_j$ are poles (rather than essential singularities).  It follows that $F$ is rational, and when its
 numerator and denominator are factored in 
 irreducible factors, each factor must have the form $e_i-e_j$, for some $i,j$.
 Requiring also invariance under permutation of the $e_i$, we learn that $F=\prod_{1\leq i<j\leq 4}(e_i-e_j)^t$ for some odd integer $t$.  ($t$ must be odd
 since the four-form $\d e_1\dots \d e_4$ is completely antisymmetric.)
 
 We can determine $t$ from $SL(2,\C)$-invariance.
 For any $t$, $\sPi_1$ will be invariant under constant translations of the $e_i$.  Invariance under scalings $e_i\to \lambda e_i$ forces $t=-3$.    Indeed,
 for $s=4$, $e_i\to \lambda e_i$ is a symmetry of the hyperelliptic equation (\ref{zolbo}) if accompanied by $x\to \lambda x$, $y\to \lambda^2 y$.
 Thus, $\d x/y$ scales as $\lambda^{-1}$, and $\d e_1\dots \d e_4/(\d x/y)^{14}$ scales as $\lambda^{18}$.  To compensate for this, we choose $t=-3$,
 so that the bosonic string measure in genus 1 is
 \begin{equation}\label{relme}\sPi_1=\frac{\voli \,\d e_1\dots \d e_4}{\prod_{1\leq i<j\leq 4}(e_i-e_j)^3(\d x/y)^{14}}. \end{equation}
 The numerator and denominator both scale as $\lambda^4$.  To complete the proof of $SL(2,\C)$-invariance of the expression (\ref{relme}),
 it suffices to verify invariance under the inversion 
 \begin{align}\label{thono} e_i & \to \frac{1}{e_i}\end{align}
 accompanied by
 \begin{align}x & \to \frac{1}{x} \cr y & \to \pm \frac{y}{x^2(\prod_i e_i)^{1/2}}.\end{align}
 Under this inversion, one has
 \begin{align}\label{bono} \d e_i & \to -\frac{\d e_i}{e_i^2} \cr e_i-e_j & \to -\frac{e_i-e_j}{e_ie_j}  \cr
 \frac{\d x}{y}& \to \mp \frac{\d x}{y}(\prod_i e_i)^{1/2}  .\end{align}
 Given these formulas, inversion symmetry is equivalent to the statement that $\Phi_1$ has vanishing $e_i$ weight for each $i$,
  where we define the $e_i$ weights to be 2 for  $\d e_i$, 1 for $e_i-e_j$  (for each $j$),
  $-1/2$ for $\d x/y$, and zero for $\d e_j$ and $e_j-e_k$, $j,k\not=i$.  While inversion symmetry is the vanishing of the $e_i$ weights of $\Phi_1$ for each $i$,
  scale invariance  is the vanishing of the sum over $i$ of these $e_i$ weights.  So inversion symmetry is equivalent to scale-invariance plus the statement
 that the $e_i$ weights of $\Phi_1$ are all equal, which is an immediate consequence of the permutation symmetries.  The verification of inversion symmetry is analogous in many  similar formulas considered later in this paper; we will comment
 on this verification only when some novelty is involved.
  
 Now we can determine the behavior of $\sPi_1$ when $\Sigma$ degenerates.  The only possible degeneration in genus 1 is a nonseparating degeneration
 in which $\Sigma$ reduces to a genus 0 curve with two points glued together.  This happens for\footnote{For instance, if $e_1=e_2=e$,
 the equation for $\SIgma$ becomes $y^2=(x-e)^2(x-e_3)(x-e_4)$.  Setting $y =\t y(x-e)$, the equation $\t y^2=(x-e_3)(x-e_4)$ describes a Riemann
 surface $\Sigmax$ of genus 0, and $\Sigma$ is obtained from $\Sigmax$ by gluing together the two points with $x=e$, $\t y=\pm \sqrt{(e-e_3)(e-e_4)}$,
 since on $\SIgma$ those points both have $x=e, \,y=0$. $\Sigmax$ is called the normalization of $\Sigma$.} $e_i\to e_j$.  Let us fix the $SL(2,\C)$ symmetry by keeping fixed $e_1,e_2,e_3$, so that $\M_1$ is parametrized
 by $e_4$ (modulo the finite group $\Theta$).    Then up to a constant multiple, $\sPi_1\sim \d e_4/(e_3-e_4)^3\sim \d q/q^2$, where $q=(e_3-e_4)^2$.
 The reason to express the result in terms of $q$ rather than $e_3-e_4$ is that $q$ is invariant under the permutation $e_3\leftrightarrow e_4$, so $q$ and
 not its square root is a well-defined parameter on $\M_1$.
 
 The result that $\sPi_1\sim \d q/q^2$ is a standard result in bosonic string theory.  
 Conformal field theory predicts that $\sPi_1\sim \d q \,q^{L_0-1}$, where $L_0=-1$ is the ground state energy of the bosonic string.  See for example
 section 6.4.4 of \cite{Revisited}.
 
 The reason that we have been able to completely determine $\sPi_1$ with no assumptions about its behavior for $e_i\to e_j$ (except the absence of an essential
 singularity) is that actually, though this is not manifest in what we have said, $\M_1$ is a copy of $\C$ (parametrized by the usual $j$-invariant of an elliptic curve).
 Though $\C$ is not compact, it has the property that an everywhere nonzero holomorphic function with no essential singularity at infinity is constant.  Here ``infinity''
 is the limit on $\M_1$ with $e_i\to e_j$ for some $i,j$.  In this limit, the $j$-invariant has a pole, $j\sim 1/q=1/(e_i-e_j)^2$. 
  
 \subsubsection{The Mumford Isomorphism In Genus 2}\label{mumtwo}
 
 We can determine the bosonic string measure in genus 2 in the same way.  For this, we set $s=6$, and observe that the scaling is now $e\to \lambda e$,
 $x\to \lambda x$, $y\to \lambda^3y$.  The space $H^0(\Sigma,K)$ is now two-dimensional, generated by $\d x/y$ and $x \d x/y$, so the expression
 $\d x/y\wedge x\d x/y$ represents a section of $\det H^0(\Sigma,K)$.  The analog of  (\ref{tombi}) is now
  \begin{equation}\label{tombix}\sPi_2=\voli \frac{F(e_1,\dots,e_6)\d e_1\dots\d e_6}{\left(\d x/y\wedge x \d x/y\right)^{13}}.\end{equation}
 Once again, because $F$ may have no zeroes or poles except for $e_i\to e_j$ (and assuming it has no essential singularity in that limit), its numerator and
 denominator are products of powers of $e_i-e_j$.  
 Imposing also permutation symmetry, we must have $F=\prod_{i<j}(e_i-e_j)^t$ for some odd integer $t$.  The same scaling argument as before determines
 again that $t=-3$, so that the genus 2 bosonic string measure is
  \begin{equation}\label{relmit}\sPi_2=\frac{\voli \,\d e_1\dots \d e_6}{\prod_{1\leq i<j\leq 6}(e_i-e_j)^3(\d x/y\wedge x\d x/y)^{13}}. \end{equation}
  The numerator and denominator now both scale as $\lambda^6$, and $SL(2,\C)$ symmetry is clear since the $e_i$ weight of $\Phi_2$ is clearly independent of $i$.
  
  A genus 2 Riemann surface has two types of degeneration, separating and nonseparating.  A nonseparating degeneration occurs when $e_i\to e_j$ for some $i,j$.
  Clearly, if we set $q=(e_i-e_j)^2$, $\sPi_2$ has the same $\d q/q^2$ behavior for $q\to 0$ as $\sPi_1$.  This is as expected from string theory and conformal field
  theory; the order of the pole depends only on the ground state energy of the string, not on the genus of the string worldsheet.  Separating degenerations will be studied
  next.

 \subsection{Separating Degenerations}\label{degentwo}

\subsubsection{Behavior Of The String Measure At A Degeneration}\label{condo}

A separating degeneration of a  Riemann surface $\Sigma$ occurs when 
$\Sigma$ splits up into a pair of surfaces $\Sigma_\ell$ and $\Sigma_r$, joined at a point.
Let $\phi_\ell$ be a local parameter on $\Sigma_\ell$ and $\phi_r$ a local parameter on $\Sigma_r$.  
To glue the point $\phi_\ell=a$ in $\Sigma_\ell$ to $\phi_r=b$
in $\Sigma_r$, we would write an equation
\begin{equation}\label{nolf}(\phi_\ell-a)(\phi_r-b)=0,\end{equation}
which describes two branches, one parametrized by $\phi_\ell$ with $\phi_r=b$, and 
one by $\phi_r$ with $\phi_\ell=a$, and meeting at $\phi_\ell=a$, $\phi_r=b$.
To deform this union of two components to a smooth Riemann surface $\Sigma$, we deform the equation to
\begin{equation}\label{tolf}(\phi_\ell-a)(\phi_r-b)=q, \end{equation}
with $q$ a small parameter.  $\Sigma$ is covered by three open sets: one is the complement
of $\phi_\ell=a$ in $\Sigma_\ell$, one is the complement of $\phi_r=b$ in $\Sigma_r$, and
the third is parametrized by $\phi_\ell$ and $\phi_r$ with the relation (\ref{tolf}).

We will suppose that $\Sigma$ has genus $g$, while $\Sigma_\ell$ and $\Sigma_r$ have genera $g_\ell$ and $g_r$, with $g=g_\ell+g_r$.  
To postpone explaining
some details that arise for genus 1, we suppose to begin with that $g_\ell,g_r> 1$ and hence $g\geq 4$.  
(In any event, for the vacuum amplitudes studied in this
paper, we will always assume that $g_\ell,g_r> 0$, since $\M_g$ can be compactified while only allowing stable degenerations.)  We write
$\M_g$, $\M_{g_\ell}$, and $\M_{g_r}$ for the respective moduli spaces, and we observe that $\dim\,\M_g=\dim\,\M_{g_\ell}+\dim\,\M_{g_r}+3$.
Indeed, locally $\Sigma$ can be parametrized by the moduli of $\Sigma_\ell$ and $\Sigma_r$ and three extra parameters, namely the
points $a,b$ at which the gluing occurs and the gluing parameter $q$.

The behavior of the holomorphic string measure $\sPi_g$ for $q\to 0$ is
\begin{equation}\label{mornex}\sPi_g \sim \sPi_{g_\ell}\cdot \d a \frac{\d q}{q^2}\d b\cdot \sPi_{g_r}, \end{equation}
where the symbol $\sim$ means that this is the most singular term for $q\to 0$.  
The point of this formula is that unlike the individual factors, the product $\Omega=\d a\cdot \d q/q^2\cdot \d b$ is well-defined -- 
independent of the choice of local coordinates $\phi_\ell $ and $\phi_r$
(modulo terms less singular for $q\to 0$).  Under the scaling $\phi_\ell\to \lambda \phi_\ell$, $a\to \lambda a$, 
along with $\phi_r\to \lambda' \phi_r,$ $b\to \lambda' b$,
and $q\to \lambda\lambda'q$, clearly $\Omega$ is invariant.  This would not be the case if we replace $\d q\, q^{-2}$ 
with $\d q \,q^r$ with $r\not=-2$.  From the point of
view of conformal field theory, $\d q \,q^{-2}$ is $\d q \,q^{L_0-1}$, where $L_0=-1$ for the ground state of the string.  Under more general
reparametrizations of the local parameters $\phi_\ell$ and $\phi_r$ -- not just scalings -- $\Omega$ is still invariant, modulo less singular
terms.

The left hand side of (\ref{mornex}) is a differential form valued in $\dets^{-13}H^*(\Sigma,K)$ and the right hand side
is a differential form valued  -- in an obvious notation -- 
in $\dets^{-13}H^*(\Sigma_\ell,K_\ell)\otimes \dets^{-13}H^*(\Sigma_r,K_r)$.  For (\ref{mornex}) to make sense,
these line bundles must be naturally isomorphic.  Indeed, when $\Sigma$ undergoes a separating degeneration
to a union of 2 components $\Sigma_\ell$ and $\Sigma_r$, there is a corresponding decomposition of the space of holomorphic
differentials:
\begin{equation}\label{tunx}H^0(\Sigma,K)\cong H^0(\Sigma_\ell,K_\ell)\oplus H^0(\Sigma_r,K_r).\end{equation}
This ensures that $\det H^*(\Sigma,K)\cong \det H^*(\Sigma_\ell,K_\ell)\otimes \det H^*(\SIgma_r,K_r)$, and hence
(after taking the $-13$ power of this isomorphism) that the left and right hand sides of (\ref{mornex}) take values in the same
line bundle when restricted to the divisor (in the compactified moduli space) that parametrizes the separating degeneration.

Eqn. (\ref{mornex}) has a simple analog for nonseparating degenerations.  In this case, we start with a Riemann surface $\Sigmax$ of
genus $g-1$ (initially, we assume $g-1\geq 2$).  By gluing together
 2 points in $ \Sigmax$, we can make a singular Riemann surface $\Sigma$ of (arithmetic) genus $g$.  Picking local coordinates
 $\phi_1$ and $\phi_2$ such that $\phi_1=a$ and $\phi_2=b$ at the two points that are to be glued, and then smoothing by
 deforming to $(\phi_1-a)(\phi_2-b)=q$,  we deform $\Sigma$ to a family of smooth genus $g$ surfaces.  The moduli of $\SIgma$ are
 those of $\Sigmax$ along with $a,q$, and $b$, and in this situation we have the obvious analog of eqn. (\ref{mornex}):
 \begin{equation}\label{ornex} \sPi_g\sim \sPi_{g-1} \cdot \d a\cdot\frac{\d q}{q^2}\cdot \d b. \end{equation}
 That the left and right hand sides of this relation are valued in the same line bundle now depends on the following.  A $g-1$-dimensional
 subspace of the $g$-dimensional space $H^0(\Sigma,K)$ consists of holomorphic differentials that, in the limit that $\Sigma$ degenerates
 to $\Sigmax$ with 2 points $a$ and $b$ glued together, are pullbacks from the $g-1$-dimensional space $H^0(\Sigmax,K^*)$.  The ``last'' differential
 on $\Sigma$ corresponds to a one-form on $\Sigmax$ that has poles with equal and opposite residues at the points $a$ and $b$ (and otherwise
 is holomorphic).  So along the divisor that parametrizes the nonseparating degeneration, there is an exact sequence
 \begin{equation}\label{mutiro} 0\to H^0(\Sigmax,K^*)\to H^0(\SIgma,K)\xrightarrow{\;\mbox{Res}_a\;} \C\to 0,                  \end{equation}
where the last map is the residue at $a$.  Taking determinants, we learn that $\det H^*(\SIgma,K)\cong \det H^*(\Sigmax,K^*)$ along this divisor.

A minus sign in the above formulas actually requires some explanation.  In (\ref{ornex}), the sign of the two-form $\d a\,\d b$ depends on
an ordering of the two points $a$ and $b$.  But likewise the sign of the residue map in (\ref{mutiro}) depends on a choice of one of the points $a$ or $b$, and is
reversed if the two points are exchanged.  So the product of $\d a \,\d b$ times the $13^{th}$ power of the residue map does not depend on the ordering of the points.
A similar remark applies to eqn. (\ref{mornex}); exchanging $\Sigma_\ell$ and $\Sigma_r$ reverses the sign of $\d a\,\d b$ but also reverses the sign of the product
$\Phi_{g_\ell}\Phi_{g_r}$, as these are differential forms of odd degree.

The reader might wonder if the facts stated in this section would be more naturally formulated in terms of a Mumford isomorphism
for Riemann surfaces with punctures.  One can certainly do this, though it is not clear if it is helpful.  See appendix \ref{telmo}.

\subsubsection{Details For Genus 1}\label{detone}

Now let us discuss how the above is modified if $\Sigma_\ell$ and/or $\Sigma_r$ (or $\Sigmax$, in the nonseparating case) has genus 1.

As we have already remarked, from  the point of view of conformal field theory, the $\d q/q^2$ in (\ref{mornex}) represents propagation of
the string ground state, which has $L_0=-1$.  The operator representing this ground state is the ghost field $c$.  For $g_\ell>1$, the position
$a$ at which $c$ is inserted is a modulus, and the usual passage from unintegrated to integrated vertex operators replaces $c$ with the 1-form
$\d a$.

For $g_\ell=1$, $a$ is not a modulus, so we cannot take this last step.  Instead, for $g=1$, the ghost field $c$ has a zero-mode,
valued in  $H^0(\Sigma_\ell,T_\ell)\not=0$ ($T_\ell\cong K_\ell^{-1}$ is the tangent bundle of $\Sigma_\ell$) and the ghost field $c$ should
be used to absorb this zero-mode.

In eqn. (\ref{obbo}), we identified $\sPi_1$ for a genus 1 surface $\Sigma_\ell$
as a trivialization of $T^*\M_1\otimes H^0(\Sigma_\ell,K_\ell)^{-14}$.  But the derivation used the fact that $H^0(\SIgma_\ell,K_\ell)$ is dual
to $H^0(\Sigma_\ell, T_\ell)$.  Hence we could equally well think of $\sPi_1$ as a trivialization of
\begin{equation}\label{medico}  \frac{  T^*\M_1\otimes H^0(\SIgma_\ell, T_\ell)}{H^0(\SIgma_\ell,K_\ell)^{13}}.        \end{equation}
Now imitating what one does in conformal field theory, we can contract a section of $H^0(\Sigma_\ell,T_\ell)$ with the 1-form $\d a$,
eliminating this 1-form  in a situation in which $\d a$ is not a modulus.  (Since $\d\phi_\ell=\d a$ when restricted to $\phi_\ell=a$, we can
view $\d a$ as a 1-form on $\Sigma_\ell$, and it can be contracted with a section of $T_\ell$, evaluated at $a$.)

We write $\sPi_1^{\d a}$ for this contraction of $\sPi_1$ and $\d a$.    
When $\Sigma_\ell$ has genus 1, we should think of the product of $\sPi_{g_\ell}\cdot \d a$ in (\ref{mornex})
 as this contraction $\sPi_1^{\d a}$.
Note that $\sPi_1^{\d a}$ is a section of 
\begin{equation}\label{medox}\frac{T^*\M_1}{\dets^{13} H^0(\Sigma,K)}.\end{equation}

If $\Sigma_r$ has genus 1, we must interpret the product $\d b\,\sPi_{g_r}$  in (\ref{mornex}) in the same way.  
Thus, for a genus 2 surface $\Sigma$ splitting to two genus 1 components,
we should replace (\ref{mornex}) with
\begin{equation}\label{todzo}\sPi_2\sim \sPi_1^{\d a} \cdot\frac{\d q}{q^2}\cdot\sPi_1^{\d b}. \end{equation}
This formula is consistent with the fact that $\sPi_2$ is supposed to be a 3-form on $\M_2$; indeed,  $\sPi_1^a$ and $\sPi_1^b$ are 1-forms on the 
divisor $\D\subset
\M_2$ that parametrizes separating degenerations, and $\d q/q^2$ is a 1-form in the normal direction.  (All these 1-forms are valued in suitable line bundles and
the product on the right hand side of (\ref{todzo}) is well-defined though the individual factors depend on the choices of gluing parameters $\phi_\ell,\,\phi_r$.)

For a nonseparating degeneration from genus 2 to genus 1, eqn. (\ref{ornex}) must be interpreted similarly.  

\subsubsection{Separating Degeneration From A Hyperelliptic Point Of View}\label{separhyp}

\begin{figure}
 \begin{center}
   \includegraphics[width=3in]{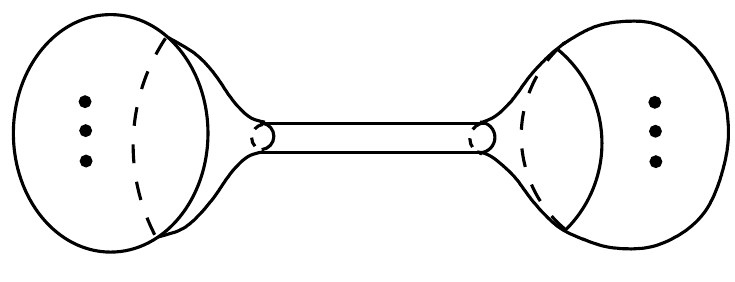}
 \end{center}
\caption{\small A separating degeneration of a hyperelliptic Riemann surface $\Sigma$ of genus 2 occurs when the base $B$ of the hyperelliptic
covering splits into 2 components, each of which contains 3 of the 6 branch points, as shown here. In the limit, $B$ degenerates to a union of 2 genus 0
components $B_\ell$
and $B_r$ meeting at a point $p$; $\Sigma$ degenerates to a union of two components $\Sigma_\ell$ and $\Sigma_r$, which respectively are
hyperelliptic coverings of $B_\ell$ or $B_r$ with 4 branch points, namely the 3 points visible in the figure and $p$. (In the notation of the text, $p$
corresponds to $x_\ell=0$ or $x_r=\infty$.)}
 \label{dumbbell}
\end{figure}
In section \ref{hyper}, we studied a genus 2 Riemann surface $\Sigma$ as a hyperelliptic curve $y^2=\prod_{i=1}^6(x-e_i)$.  $\Sigma$ is a double cover
of a genus 0 Riemann surface $B$ (parametrized by $x$), which contains 6 marked points corresponding to the branch points of the map
$\Sigma\to B$.  For $\Sigma$ to undergo a separating degeneration, we want $B$ to degenerate to 2 components each containing 3 of the 6 branched
points, connected by a long tube (fig. \ref{dumbbell}).  To reach this situation, up to
an $SL(2,\C)$ transformation, we can take 3 of the $e_i$ to be of order $q$ and 3 of order $q^{-1}$,
with $q\to 0$, or we can take 3 of them to be fixed and 3 to be of order $q^2$. 

We will follow the second route, so we keep 3 branch points fixed at $e_1,e_2,e_3$, and place the others at $e_{3+j}=q^2f_j$, $j=1,2,3$, where we
keep the $e_i$ and $f_j$ fixed for $q\to 0$. (This scaling
will turn out to match properly with the standard gluing relation (\ref{tolf}).)
Thus the hyperelliptic equation is
\begin{equation}\label{elbo}y^2=\prod_{i=1}^3(x-e_i)\prod_{j=1}^3(x-q^2f_j). \end{equation}
It will be convenient to take a slice of the $SL(2,\C)$ action in which $e_1,e_2$ and $f_1,f_2$ are kept fixed, and $\M_2$ is parametrized by $e_3, f_3$, 
and $q$.  By explicitly evaluating the $\eurm{vol}^{-1}$ operation in a way similar to what is explained in eqn. (\ref{umonk}), one can show that in this
parametrization, $\voli \d e_1\dots \d e_6$ becomes   $2(e_1-e_2)e_1e_2 \d e_3 q^3\d q (f_1-f_2)\d f_3+\dots$ where higher order terms in $q$
have been dropped (for instance $e_1-q^2 f_1$ has been replaced by $e_1$).  
Accordingly eqn. (\ref{relmit}) for the genus 2 holomorphic string measure becomes
\begin{equation}\label{tinedo}\sPi_2\sim 2 \frac{e_1e_2(e_1-e_2) \d e_3 \cdot q^3 \d q \cdot (f_1-f_2)\d f_3}
{\prod_{i<j}(e_i-e_j)^3\cdot \prod_{k=1}^3 e_k^9 \cdot q^{18}\prod_{i<j}(f_i-f_j)^3 (\d x/y \wedge x\d x/y)^{13}}\end{equation}
where  again terms of higher order in $q$ have been dropped.  

As $q\to 0$, there are 2 different ways to look at the equation (\ref{elbo}).  First, we can keep $x$ and $y$ fixed while $q\to 0$.
The limit of the equation is $y^2=x^3\prod_{i=1}^3(x-e_i)$.  It is convenient to set $y_\ell=y/x$, $x_\ell=x$.  The equation becomes
\begin{equation}\label{sigel} y_\ell^2=x_\ell \prod_{i=1}^3(x_\ell-e_i). \end{equation}
This describes a genus 1 Riemann surface $\Sigma_\ell$, a branched cover of the $x_\ell$ plane, with branch points at $0,e_1,e_2,e_3$.
For a local parameter near the branch point at $x_\ell=y_\ell=0$, it is convenient to take
\begin{equation}\label{mizo} \phi_\ell=\Delta^{-1/2} y_\ell,\end{equation}
 with 
\begin{equation}\label{ryy} \Delta=-\prod_{i=1}^3e_i. \end{equation}
This implies that
\begin{equation}\label{pizo}\phi_\ell^2\sim  x_\ell,\end{equation}
near $x_\ell=0$.

Alternatively, we can set $x=q^2x_r$, $y=\Delta^{1/2}q^3y_r$.
The equation becomes
\begin{equation}\label{defo} y_r^2=\prod_{i=1}^3(x_r-f_r).  \end{equation}
This describes a genus 1 Riemann surface $\Sigma_r$, a branched cover of the $x_r$ plane, with branch points at $f_1,f_2,f_3,\infty$.
For a local parameter near the branch point at $x_r=\infty$, we can take
\begin{equation}\label{nizo}\phi_r=\frac{x_r}{y_r}. \end{equation}
This implies that
\begin{equation}\label{tizo}\phi_r^2\sim \frac{1}{x_r},~~x_r\to \infty.\end{equation}

The definition of $x_\ell$ and $x_r$ was such that $x_\ell=q^2 x_r$.  According to (\ref{pizo}) and (\ref{tizo}), this means that near
$x_\ell=0$, $x_r=\infty$ (for example, in the region with $x_\ell\sim q$, $x_r\sim q^{-1}$), we have $\phi_\ell^2\phi_r^2=q^2$ or
\begin{equation}\label{zembo}\phi_\ell \phi_r=q. \end{equation}  In other words, $\Sigma$ can be built by gluing together $\Sigma_\ell$
and $\Sigma_r$, using the local parameters $\phi_\ell$ and $\phi_r$ near $x_\ell=0$, $x_r=\infty$, and the usual gluing relation $\phi_\ell\phi_r=q$.

Going back to formula (\ref{tinedo}) for the holomorphic measure $\sPi_2$, we also need to study the behavior of the holomorphic
1-forms $\d x/y$ and $x \d x/y$ for $q\to 0$.  On $\Sigma_\ell$, $x\d x/y=\d x_\ell/y_\ell$, while on $\Sigma_r$, it is of order $q^2$.
Conversely, on $\Sigma_\ell$, $\d x/y$ is of order 1 while on $\Sigma_r$, it is of order $q^{-1}$:
\begin{equation}\label{mex}\frac{\d x}{y}\sim \Delta^{-1/2}q^{-1}\frac{\d x_r}{y_r}.\end{equation}
So for $q\to 0$, we keep the dominant terms in $\d x /y \wedge x\d x/y$ and  replace $\d x/y\wedge x\d x/y$ with $\Delta^{-1/2}q^{-1} \d x_\ell/y_\ell\wedge \d x_r/y_r$.   Then (\ref{tinedo})
becomes
\begin{equation}\label{inedo}\sPi_2\sim 2\frac{e_1e_2(e_1-e_2) \d e_3}{\prod_{i<j}(e_i-e_j)^3\prod_k e_k^3(\d x_\ell/y_\ell)^{13}}
\cdot \Delta^{1/2}\frac{\d q}{q^2}\cdot \frac{(f_1-f_2)\d f_3}{\prod_{i<j}(f_i-f_j)^3 (\d x_r/y_r)^{13}}.\end{equation}  

We have exhibited the expected $\d q/q^2$ pole.  However, to compare to the more precise prediction (\ref{todzo}) takes a little more
work.  

\subsubsection{Comparison}\label{compar}

In this derivation,
$\Sigma_\ell$ is described by a standard hyperelliptic equation $y^2=\prod_{i=1}^4(x-e_i)$,  in the special case $e_4=0$.  The corresponding
holomorphic measure $\sPi_1$ is given in eqn. (\ref{relme}).  If we parametrize $\M_1$ by $e_3$, keeping $e_1,e_2$, and $e_4=0$ fixed, then
we can use (\ref{umonk}) to find
\begin{equation}\label{toldo}\sPi_1=\frac{e_1e_2(e_1-e_2)\d e_3}{\prod_{1\leq i<j\leq 3}(e_i-e_j)^3\prod_{k=1}^3 e_k^3\,(\d x_\ell/y_\ell)^{14}}.\end{equation}
However, as explained in section \ref{detone}, for the present calculation, it is more illuminating to replace one factor of the holomorphic
differential $\d x_\ell/y_\ell$ in the denominator with a factor of its inverse, the holomorphic vector field $y_\ell\partial_{x_\ell}$, in the numerator:
\begin{equation}\label{tolodo}\sPi_1=\frac{e_1e_2(e_1-e_2)\d e_3\cdot (y_\ell\partial_{x_\ell})}{\prod_{i<j}(e_i-e_j)^3\prod_k e_k^3\,(\d x_\ell/y_\ell)^{13}}.\end{equation}  Then we are supposed to evaluate what in eqn. (\ref{todzo}) is called $\sPi_1^{\d a}$ by replacing the holomorphic vector field $y_\ell\partial_{x_\ell}$ in the numerator of (\ref{tolodo}) with $y_\ell\partial_{x_\ell}\phi_\ell|_{\phi_\ell=0}$,
which in the present context turns out to equal $\Delta^{1/2}/2$.  So
\begin{equation}\label{whox}\sPi_1^{\d a}=\frac{\Delta^{1/2}}{2}\frac{e_1e_2(e_1-e_2)\d e_3}{\prod_{i<j}(e_i-e_j)^3\prod_k e_k^3\,(\d x_\ell/y_\ell)^{13}}.\end{equation}

To make a similar analysis for $\Sigma_r$, we first have to slightly generalize our formulas to cover the case of an elliptic curve with a branch point at infinity.
For the familiar elliptic curve $y_0^2=\prod_{i=1}^4(x_r-f_i)$, with $\M_1$ parametrized by $f_3$,  $\sPi_1$ is given as in (\ref{relme}),
with the obvious subsitutions.  Now setting $y_0=(-f_4)^{1/2}y_r$ and taking $f_4\to\infty$, the equation becomes
\begin{equation}\label{oldox}y_r^2=\prod_{j=1}^3(x_r-f_j),\end{equation} and if we parametrize $\M_1$ by $f_3$,  the formula for $\sPi_1$ becomes
\begin{equation}\label{lodox}\sPi_1=\frac{(f_1-f_2)\d f_3}{\prod_{i<j}(f_i-f_j)^3(\d x_r/y_r)^{14}}.\end{equation}
Once again we replace one factor of $\d x_r/y_r$ in the denominator with the inverse vector field $y_r\partial_{x_r}$ in the numerator:
\begin{equation}\label{loddox}\sPi_1=\frac{(f_1-f_2)\d f_3\cdot (y_r\partial_{x_r})}{\prod_{i<j}(f_i-f_j)^3(\d x_r/y_r)^{}}.\end{equation}
And to compute $\sPi_1^{\d b}$, we replace $y_r\partial_{x_r}$ with $y_r\partial_{x_r}\phi_r|_{\phi_r=0}$, which turns out to equal $ -1/2 $:
\begin{equation}\label{lodidox}\sPi_1^{\d b}=-\frac{1}{2}\frac{(f_1-f_2)\d f_3}{\prod_{i<j}(f_i-f_j)^3(\d x_r/y_r)^{}}.\end{equation}

With the aid of these expressions for $\sPi_1^{\d a}$ and $\sPi_1^{\d b}$, we find that the $q\to 0 $ asymptotics of $\sPi_2$, as found in (\ref{inedo}),
does indeed agree with the prediction (\ref{todzo}), modulo an overall constant that depends on the normalizations.

\section{Superstrings In Genus 1 And 2}\label{super}

\subsection{The Superanalog Of The Mumford Isomorphism}\label{supermumford}

We began our discussion of the Mumford isomorphism in bosonic string theory by introducing, for any ordinary vector space $V$ of dimension $n$, the top
exterior power $\det V=\wedge^nV$.

Suppose instead that $V$ is a $\Z_2$-graded vector space of dimension $n|m$.  The analog of the top exterior power is a 1-dimensional vector space
(of statistics $(-1)^m$) called the Berezinian (or the Berezinian line), $\Ber \,V$.  The
definition of $\Ber\,V$ is a little subtle (see for example \cite{DM} or section 3.1 of \cite{wittennotes}).  For our purposes, 
it will suffice to know that if we are given a decomposition 
$V=A\oplus B$ of $V$ as the direct sum of an even subspace $A$ and an odd subspace $B$, then there is a 
natural isomorphism $\Ber\,V\cong\det A\otimes \dets^{-1}B$.

Now consider the case that $V$ is not a $\Z_2$ graded vector space, but a $\Z_2$-graded vector bundle over a 
supermanifold $B$.   Then one defines the Berezinian
line bundle of $V$, denoted $\BBer (V)$, by taking the Berezinian of each fiber of $V\to B$.  In other words, if $V_b$ is the fiber of $V$ above $b\in B$, then 
$\Ber(V_b)$ is the fiber of $\BBer(V)$ at $b$.  An important application of this is to define the analog of the canonical bundle of a super Riemann surface.
A super Riemann surface $\Sigma$ has a cotangent bundle $T^*\Sigma$ that is of rank $1|1$.  Its Berezinian $\BBer(T^*\Sigma)$, or
simply $\BBer(\Sigma)$, is the analog of the canonical bundle of an ordinary Riemann surface.  

Suppose now that $\Sigma$ is a super Riemann surface, and let $\L\to \Sigma$ be a holomorphic line bundle.  
Then one defines cohomology groups $H^i(\Sigma,\L),
\,i=0,1$, just as for ordinary Riemann surfaces, with the difference that the cohomology groups are now $\Z_2$-graded vector spaces.  The analog of the determinant
of cohomology for an ordinary Riemann surface is now the Berezinian of the cohomology, which we define as
\begin{equation}\label{wonko} \Ber H^*(\L)=\Ber H^0(\Sigma,\L)\otimes \Ber^{-1} H^1(\Sigma,\L). \end{equation}

Combining these constructions, for any integer $k$, we have the line bundle $\BBer^k(\Sigma)=(\BBer(\Sigma))^{\otimes k}$ 
over $\Sigma$, with cohomology groups $H^i(\Sigma,\BBer^k(\SIgma))$,
and a Berezinian line $\Ber_k(\Sigma)= \Ber(H^*(\BBer^k(\Sigma))$.  If $\Sigma$ varies\footnote{A comment is necessary here that mirrors what we said in footnote \ref{jumping} for
bosonic string theory.  The simple definition of the Berezinian line bundle $\BBer_k(\Sigma)$ given in the text assumes that the cohomology groups $H^i(\Sigma,\BBer^k(\SIgma))$ vary holomorphically, with
no jumping in their dimensions.  A more sophisticated definition  can be given without this assumption.  One approach is to use the relations between $\BBer H^*(\L)$ for different $\L$ described
in \cite{RSV} to reduce to a locally free situation in which no jumping of cohomology occurs.} in a family parametrized by a super manifold $B$, then $\Ber_k(\SIgma)$
is the fiber of a holomorphic line bundle $\BBer_k\to B$. In particular, if  $\MM_g$ is the moduli space of super Riemann surfaces of genus $g$ (of even or odd spin structure),
we can let $\Sigma$ be the fiber of the universal super Riemann surface over $\MM_g$ and then define a line bundle $\BBer_k\to\M_g$ whose fiber at the point
corresponding to a given super Riemann surface $\Sigma$ is $\Ber_k(\Sigma)$.  

The super Mumford isomorphism is the statement that $\BBer_3\cong \BBer_1^5$, or 
equivalently that $\BBer_3\otimes \BBer_1^{-5}$ is trivial.\footnote{This assertion is
eqn. (27) in \cite{RSV}, where what we call $\BBer^k(\SIgma)$ is denoted as $\Sigma_k$, and our 
$\BBer(H^*(\L))$ is denoted $m_\C(\L)$.}  The holomorphic measure
of superstring theory in $\R^{10}$ is a holomorphic trivialization $\sPhi_g$ of $\BBer_3\otimes \BBer_1^{-5}$, sometimes
called the super Mumford form.
The qualitative interpretation of this statement is the same
as it is in bosonic string theory: the path integral of the $BC$ ghost system is a holomorphic section of $\BBer_3$, and the holomorphic part of the
matter path integral is a holomorphic section of $\BBer_1^{-5}$.  
(The exponent $5$ reflects the fact that $\R^{10}\cong \C^5$, and the minus sign reflects the statistics
of the matter fields.)  A holomorphic trivialization $\sPhi_g$ of $\BBer_3\otimes \BBer_1^{-5}$ is 
uniquely determined, up to a constant multiple, if one has some knowledge
of how $\sPhi_g$ should behave at infinity.  From our point of view, this knowledge will come from our knowledge of superconformal field theory and string theory. 

In principle, the proof of the super Mumford isomorphism in \cite{RSV} implies 
much more than we have claimed in the last paragraph, or will exploit in this  paper.
The proof is not just an existence proof, but comes with a procedure to construct $\sPhi_g$,  so there is no 
undetermined constant multiple and the behavior at infinity is predicted,
with no need for external input from superconformal field theory or any other source.  Moreover, the procedure to 
compute $\sPhi_g$ is local on $\MM_g$ and requires no global
knowledge of $\MM_g$.  Hopefully these  facts (which have partial analogs for the bosonic Mumford isomorphism \cite{BeM}) will be exploited
in future work. 

The physical application of a section $\sPhi_g$ of $\BBer_3\otimes \BBer_1^{-5}$, at least in the case of an even spin structure,
 is similar to what we explained for bosonic string theory in section \ref{what for}.
Analogous to $\det H^*(K^2)$ in the bosonic case, for $g>1$, $\BBer_3$ is the Berezinian of $T^*\MM_g$, so a section of $\BBer_3$ is a volume form on $\MM_g$
in the holomorphic sense.  Thus $\sPhi_g$ is a holomorphic (and everywhere nonzero) volume form on $\MM_g$, valued in the line bundle $\BBer_1^{-5}$:
\begin{equation}\label{meltdo}\sPhi_g\in H^0(\MM_g,\BBer \,T^*\MM_g\otimes \BBer_1^{-5}). \end{equation}
Using facts analogous to those that were explained in section \ref{what for} in the bosonic case, a holomorphic measure with values
in a line bundle can be combined with an analogous antiholomorphic object to make an ordinary measure that can be integrated over the appropriate
integration cycle to compute the superstring vacuum amplitude.
 Further details of this are described in section \ref{superwhatfor}.

To get some more insight about the super Mumford isomorphism, let us ask what it says when restricted to the reduced space of $\MM_g$.  This reduced space is
$\M_{g,\spin}$, which parametrizes a split super Riemann surface $\Sigma$, or equivalently an ordinary Riemann surface 
$\Sigma_0$ of genus $g$ with a choice of spin structure.
We write $K$ for the canonical bundle of $\Sigma_0$ and $K^{1/2}$ for the square root of $K$ that is determined by the choice of spin structure.
On a super Riemann surface $\Sigma$ in local superconformal coordinates $z|\theta$, a function $f(z|\theta)$ can be expanded $f(z|\theta)=a(z)+\theta b(z)$,
where locally $a(z)$ is a function on $\Sigma_0$ and $b(z)$ is a section of $K^{1/2}$.   In the case of a split super Riemann surface, this decomposition is
valid globally, and therefore if we write $\O_\Sigma$ or $\O_{\Sigma_0}$ for a trivial line bundle over $\Sigma$ or $\Sigma_0$,
and  identify a line bundle with its sheaf\footnote{An open set on $\Sigma$ is defined to be the same thing as an open set on $\Sigma_0$, so a sheaf on $\Sigma$
can be understood as a sheaf on $\Sigma_0$.  Eqn. (\ref{zork}) is a relation between sheaves on $\Sigma_0$.} of sections, we have
\begin{equation}\label{zork}\O_\Sigma=\O_{\SIgma_0}\oplus K^{1/2},\end{equation}
where the first summand is even and the second one is odd.  The analogous decomposition for $\BBer^q(\Sigma)$ is
\begin{equation}\label{mork}\BBer^q(\Sigma)=K^{q/2}\oplus K^{(q+1)/2},\end{equation}
where the two summands have statistics $(-1)^q$ and $(-1)^{q+1}$, respectively.  

The decomposition (\ref{mork}) leads to a formula for the restriction to $\M_{g,\spin}$ of the line bundle $\BBer_k\to \MM_g$:
\begin{equation}\BBer_k|_{\M_g,\spin}\cong \dets^{(-1)^k} H^*(K^{k/2})\otimes \dets^{(-1)^{k+1}}H^*(K^{(k+1)/2}).\end{equation}
  (Here $\det H^*(\L)$, as in section \ref{mumf},
is the determinant of cohomology of a line bundle $\L$ over an ordinary Riemann surface $\Sigma_0$.)  So \begin{equation}\label{zelf}
\BBer_3|_{\M_{g,\spin}}\cong
\det H^*(K^2)\otimes \dets^{-1}H^*(K^{3/2}),\end{equation} where we associate the two factors respectively with the $bc$ and $\beta\gamma$
path integrals.   Similarly, 
 \begin{equation}\label{uzelf}
\BBer_1^{-5}|_{\M_{g,\spin}}\cong \dets^{-5}H^*(K)\otimes \dets^{5}H^*(K^{1/2}),\end{equation}
where the two factors  respectively come from integration over the bosonic and fermionic matter fields.
So for $g>1$, we can identify the restriction of the holomorphic string path integral $\sPhi_g$ to the reduced space $\M_{g,\spin}$ as
a section of
\begin{equation}\label{udono}\frac{\det T^*\M_{g,\spin}\otimes \det^5 H^*(K^{1/2})}{\det H^*(K^{3/2})\otimes \det^5H^*(K)} \end{equation}
where we use the fact that $\det H^*(K^2)$ can be identified with the canonical bundle $\det T^*M_{g,\spin}$, and we write a ratio of line
bundles suggestively as a fraction.  For $g=1$, there is a correction to this, just as in the case of the bosonic string.  We discuss the details later.

This restriction of the super Mumford isomorphism to the reduced space of $\MM_g$ has a simple interpretation;
the path integrals of the $bc$ ghosts, the $\beta\gamma$ ghosts, the bosonic matter fields, and the fermionic matter fields take values in the four
factors in the numerator and the denominator of (\ref{udono}).  The point of the super Mumford isomorphism, however, is that it is valid on $\MM_g$, without restriction
to the reduced space.  From a supermanifold point of view, there are only two factors, $\BBer_3$ and $\BBer_1^{-5}$, which correspond respectively to the
$BC$ and matter path integrals.

Using the fact that the restriction  of $\sPhi_g$ to $\M_{g,\spin}$ is a holomorphic trivialization of the line bundle indicated in (\ref{udono}) -- with
a behavior at infinity that is predicted from conformal field theory -- this restriction can be computed by methods similar to those that one uses
to compute the holomorphic part of the bosonic string path integral.  We will carry this out explicitly in sections \ref{evenone} and \ref{eventwo} for genus 1 and 2.
However, though useful in string theory (as input in computing scattering amplitudes), this restriction is not what is usually called the superstring vacuum amplitude. Let $m_1,\dots,m_{3g-3}$ be local holomorphic coordinates
on $\M_{g,\spin}$, and, picking a local holomorphic projection $\pi:\MM_g\to \M_{g,\spin}$, pull back $m_1,\dots,m_{3g-3}$ to functions on $\MM_g$
and complete them by adding odd functions $\eta_1,\dots,\eta_{2g-2}$ 
to make a local coordinate system $m_1,\dots,m_{3g-3}|\eta_1,\dots,\eta_{2g-2}$ on $\MM_g$. (Apart from the case of $g=1$ with an odd spin structure,
the odd dimension of $\MM_g$ is always $2g-2$.)   In this coordinate system, we write $\sPhi_g=[\d m_1,\dots,\d m_{3g-3}|\d\eta_1,\dots,\d\eta_{2g-2}]\Upsilon$,
where according to eqn. (\ref{meltdo}), $\Upsilon$ is a holomorphic section of $\BBer_1^{-5}$.  After locally trivializing this line bundle (or at least identifying
it with a pullback from $\M_{g,\spin}$), we  can expand $\Upsilon$ in powers of the odd
variables:
\begin{align}\label{cobmo} \Upsilon=&\Upsilon^{(0)}(m_1,\dots,m_{3g-3})+
\sum_{1\leq i<j\leq 2g-2}\eta_i\eta_j \Upsilon^{(1)}_{ij}(m_1,\dots,m_{3g-3}) +\dots\cr &+\eta_1\eta_2\dots\eta_{2g-2}\Upsilon^{(g-1)}(m_1,\dots,m_{3g-3})
 .\end{align} Only even powers appear since $\Upsilon$ is even.
Naively speaking, the superstring vacuum amplitude is associated to the top term in this expansion, since this is the term that survives
in the Berezin integral over the $\eta$'s, but that is too naive if the ingredients that were used to make the expansion --- the projection $\pi$ and
the identification of $\BBer_1^{-5}$ as a pullback -- are only locally-defined. 
The bottom term $\Upsilon^{(0)}$ in the expansion of $\sPhi_g$ has an invariant meaning, since it controls the restriction of $\sPhi_g$ to $\M_{g,\spin}$,
but the higher terms depend on the choices of $\pi$ and of the identification of $\BBer_1^{-5}$ as a pullback.
In general, as far as is known, in the absence of a global holomorphic projection $\pi:\MM_g\to \M_{g,\spin}$ such that $\BBer_1^{-5}$ is a pullback, one needs to know the full $\sPhi_g$
to compute a superstring vacuum amplitude. The requisite procedure involves all the complexities of supermanifold integration.

 A global holomorphic projection does not exist for $g\geq 5$ \cite{DonW}, but there is such a projection in the
important case of $g=2$ with an even spin structure \cite{DPHgold}, and more trivially also for $g=1$.  So let us ask what happens
if a global holomorphic projection $\pi:\MM_g\to \M_{g,\spin}$ does exist.  Suppose further that the line bundle $\BBer_1\to \MM_g$ is the pullback
of a line bundle $\S\to \M_{g,\spin}$; this is so  in the $g=2$ situation studied in \cite{DPHgold} (and again more trivially for $g=1$), 
as we explain at the end of section \ref{superwhatfor}. 
Note that if such an $\S$ does exist, it is simply the restriction of $\BBer_1$ to $\M_{g,\spin}$, and so is given in eqn. (\ref{uzelf}). 
With these hypotheses, $\sPhi_g$ is a section of $\BBer(\MM_g)\otimes\pi^* \S^{-5}$, and there is a natural map 
\begin{equation}\label{telmex}\pi_*:H^0(\MM_g,\BBer(\MM_g)\otimes \pi^*\S^{-5})\to H^0(\M_{g,\spin},\det\,T^*\M_{g,\spin}\otimes \S^{-5}) \end{equation}
given by integration over the fibers of $\pi$, in other words integration over the odd variables.
Under these circumstances, what one would call the superstring vacuum amplitude is $\pi_*(\sPhi_g)$, which corresponds to the ``top'' term
$\Upsilon^{(g-1)}$ in eqn. (\ref{cobmo}).  When a global holomorphic projection $\pi$ exists, $\pi_*(\sPhi_g)$ is a global holomorphic section of
\begin{equation}\label{udonox}\det T^*\M_{g,\spin}\otimes \dets^5 H^*(K^{1/2})\otimes \dets^{-5}H^*(K).\end{equation}
This differs from (\ref{udono}) only in that the factor $\dets^{-1} H^*(K^{3/2})$, which reflects the measure for integration over the odd variables,
is absent, since in arriving at $\pi_*(\sPhi_g)$, we have already integrated over the odd variables.

There is, however, a crucial difference between the statement that the restriction $\sPhi_g|_{\M_{g,\spin}}$ is a holomorphic
trivialization of one line bundle, given in eqn. (\ref{udono}), and the statement that  (under certain hypotheses) $\pi_*(\sPhi_g)$ is a holomorphic
section of another line bundle, given in eqn. (\ref{udonox}).  The super Mumford isomorphism asserts that $\sPhi_g$ is everywhere holomorphic
and nonzero, where ``nonzero'' in the context of supermanifolds means nonzero modulo the odd variables, or in other words nonzero after
restriction to the reduced space.  So $\sPhi_g|_{\M_{g,\spin}}$ is an everywhere holomorphic {\it and nonzero} section of the
indicated line bundle, while $\pi_*(\sPhi_g)$ (under the hypotheses leading to (\ref{udonox})) is merely a holomorphic section, but possibly with zeroes.  Indeed, 
$\pi_*(\sPhi_2)$ certainly turns out to have zeroes.

\subsubsection{What Is $\Psi_g$  Good For?}\label{superwhatfor}

Here we will explain
the relation of $\Psi_g$ for even spin structure to superstring vacuum amplitudes.
See \cite{RSV} for an explanation of this in the context of type 0 string theory, in which holomorphic and antiholomorphic odd moduli are
complex conjugates; we will adapt the reasoning given there for superstring theory, in which they are independent. 
The contribution of an odd spin structure to the vacuum amplitude vanishes, because of fermion zero-modes.  $\Psi_g$ with an odd spin structure
is therefore an input to more complicated string theory computations of scattering amplitudes, but not to the vacuum amplitude.   We will write $\MM_{g,+}$
and $\MM_{g,-}$ for the components of $\MM_g$ with even or odd spin structure, $\M_{g,\spin\pm}$  for the  reduced space of $\MM_{g,\pm}$,
and $\Psi_{g,\pm}$ for the restriction of $\Psi_g$ to $\MM_{g,\pm}$.

For the case of an even spin structure, $H^0(\SIgma,\BBer(\Sigma))$ is generically of dimension $g|0$, and naturally isomorphic to the space
$\V$ of closed holomorphic 1-forms on $\Sigma$.  (For these observations, see \cite{RSVtwo} and also \cite{wittennotes}, section 8 and appendix D.)
On the other hand, $H^1(\Sigma,\BBer(\SIgma))$ is generically of dimension $1|0$ and canonically trivial (the trivialization is given by integration:
a 1-form valued in $\BBer(\Sigma)$ can be naturally integrated, analogous to integrating a $(1,1)$-form on an ordinary Riemann surface).  
These statements fail on a divisor $\eusm S\subset \MM_{g,+}$.  Our considerations will be valid on the complement of this divisor.

On the complement of $\eusm S$, we identify $\Ber H^*(\BBer(\Sigma))$ with $\det \V$, so the super Mumford form $\Psi$ is a holomorphic trivialization of
\begin{equation}\label{zyreno} \frac{\BBer\, T^*\MM_g}{\dets^5\V}.\end{equation}
We can identify the reduced space $\eusm S_\red $ of $\eusm S$ as follows.  The reduced space of $\MM_{g,+}$ parametrizes split super Riemann
surfaces.  For a split super Riemann surface $\Sigma$, 
we have the decomposition of eqn (\ref{mork}): $\BBer(\SIgma)\cong K^{1/2}\oplus K$, where $K$ is the canonical
bundle of the reduced space $\Sigma_\red$. From this, it follows that for a reduced surface $\Sigma$, the condition that $H^0(\Sigma,\BBer(\Sigma))$
is not of dimension $g|0$ is that
\begin{equation}\label{noz}H^0(\Sigma_\red,K^{1/2})\not=0. \end{equation}
This condition characterizes a divisor in $\M_{g,\spin+}$ that is sometimes called the theta-null divisor, and this divisor is the reduced space $\eusm S_\red$
of the divisor $\eusm S\subset \MM_{g,+}$ on which the isomorphism $\BBer_1\cong \det \V$ breaks down.
We will use the super Mumford form to construct a measure on the complement of $\eusm S$, but actually the measure we construct extends over $\eusm S$
and in fact vanishes along $\eusm S$ (at least in low genus) because of the fermion zero-modes that appear along $\eusm S$.

To proceed farther, we need to pick a particular superstring theory.  For convenience, we will begin with Type II superstring theory.
We  use the characterization of a Type II superstring worldsheet that is described very briefly in \cite{DM} and in more detail in \cite{wittennotes}, section 5.
The complexification of a Type II superstring worldsheet $\Sigma$ is simply a product $\Sigma_L\times \Sigma_R$ of two super Riemann surfaces,
such that the reduced space of $\Sigma_L$ is the complex conjugate\footnote{The complex conjugate of a complex manifold is the same manifold
with opposite complex structure.  It suffices here if the reduced space of $\Sigma_L$ is sufficiently close to the complex conjugate of $\Sigma_R$.}
 of the reduced space of $\Sigma_R$.  In particular, $\Sigma_L$ and $\Sigma_R$ have the same genus $g$.
$\Sigma$ itself is characterized, up to homology, by saying that its reduced space $\Sigma_\red$
is the diagonal in $(\SIgma_L\times \SIgma_R)_\red$, while the odd dimension of $\Sigma$ is the sum of the odd dimensions of $\Sigma_L$ and $\Sigma_R$.

If we allow $\Sigma_L$ and $\Sigma_R$ to vary independently, then the product $\Sigma_L\times \SIgma_R$ is parametrized by a copy of $\MM_L\times \MM_R$,
that is, two copies of the moduli space of super Riemann surfaces (of the appropriate genus).  
The reduced space $(\MM_L\times \MM_R)_\red$ parametrizes a pair of ordinary Riemann surfaces $\Sigma_{0,L}$, $\Sigma_{0,R}$, each endowed with a
spin structure.
The integration cycle $\varGamma$ for Type II superstring theory
is a cycle $\varGamma\subset \M_L\times\M_R$ chacterized up to homology by the following conditions:  {\it (1)} the reduced space $\varGamma_\red$ of 
$\varGamma$ is the subspace of $(\MM_L\times \MM_R)_\red$ that parametrizes pairs $\Sigma_{0,L},\,\Sigma_{0,R}$, such that $\Sigma_{0,L}$ is
the complex conjugate of $\Sigma_{0,R}$ (but with no relation between the two spin structures); {\it (2)} the odd dimension of $\varGamma$ is the same as that
of $\MM_L\times \MM_R$.  (Since $\MM_L$ and $\MM_R$ are not compact, one
 also requires a condition on the behavior of $\varGamma$ at infinity, but this need not concern us here.)

  Let $\V_L$  and $\V_R$ be the spaces of closed holomorphic
1-forms on $\Sigma_L$ and $\Sigma_R$, respectively.  Let $\Psi_L$ and $\Psi_R$ be the super Mumford forms of $\MM_L$ and $\MM_R$.
On the complement of the divisors $\eusm S_L\subset \MM_L$ and $\eusm S_R\subset \MM_R$  on which the relevant cohomology is non-generic,
$\Psi_L$ trivializes $\BBer \,T^*\MM_L\otimes \dets^{-5}\V_L$, and $\Psi_R$ trivializes $\BBer\, T^*\MM_R\otimes \dets^{-5}\V_R$.
So the product $\Psi_L\otimes \Psi_R$ trivializes $\BBer\,T^*(\MM_L\times \MM_R)\otimes \dets^{-5}\V_L\otimes \dets^{-5}\V_R$.
When restricted to $\varGamma\subset \MM_L\times \MM_R$, we can identify $\BBer\,T^*(\MM_L\times \MM_R)$ with $\BBer\, T^*\varGamma$,
whose sections are complex-valued measures on $\varGamma$.  Thus to turn $\Psi_L\otimes \Psi_R$ into a measure on $\varGamma$,
we need a trivialization of $(\det \V_L\otimes \det \V_R)^{-5}$.

Rather as in the bosonic case discussed in section \ref{what for}, integration gives a natural nondegenerate pairing $\V_L\otimes \V_R\to \C$, as follows.
Unless $\Sigma_L$ and $\Sigma_R$ are split, there is no natural embedding of the reduced space $\Sigma_\red$ in the Type II superstring worldsheet
$\Sigma$.
However, we can always pick such an embedding, in a way that is unique up to homology.  Given $\omega_L\in \V_L$ and $\omega_R\in\V_R$,
the product $\omega_L\wedge\omega_R$ is a closed holomorphic 2-form on $\Sigma_L\times \SIgma_R$. Because this form is closed, the integral
\begin{equation}\label{telp}\langle\omega_L,\omega_R\rangle =\int_{\SIgma_\red}\omega_L\wedge\omega_R \end{equation}
does not depend on the precise embedding of $\Sigma$ in $\Sigma_L\times\Sigma_R$ or of $\SIgma_\red$ in $\Sigma$.  The pairing $\langle~,~\rangle$ is nondegenerate,
since this is true if $\Sigma_L$ and $\Sigma_R$ are split, so its determinant gives a natural isomorphism $\det \V_L\otimes \det\V_R\cong \O$
(where $\O$ is a trivial line bundle). The $-5$ power of this is an isomorphism $\frak H:\dets^{-5}\V_L\otimes \dets^{-5}\V_R\cong \O$.
Finally, $\frak H(\Psi_L\otimes\Psi_R)$ is a measure on $\varGamma$, or at least on the complement of the divisors $\eusm S_L$ and $\eusm S_R$.
This is the measure that one uses to compute the vacuum amplitude.

The derivation actually shows that the measure  $\frak H(\Psi_L\otimes \Psi_R)$ is defined and everywhere nonzero on 
the complement of $\eusm S_L$ and $\eusm S_R$. ($\eusm S_L$ and $\eusm S_R$
 intersect $\varGamma$ in distinct loci, as we have placed no relation between the spin
structures of $\Sigma_L$ and $\Sigma_R$.)   At least in low genus, along $\eusm S_L$ and $\eusm S_R$,  $\frak H(\Psi_L\otimes \Psi_R)$ actually develops not a pole but a zero of rather high
order,  because of fermion zero-modes.\footnote{In high genus, there may not be such a zero, since in conventional language, there are for $g\geq 11$ sufficient
picture-changing operators to absorb the fermion zero-modes.}

For the heterotic string, only a few minor changes are needed.
$\Sigma_L$ becomes an ordinary Riemann surface and $\MM_L$ is
replaced by the the moduli space  $\M_L$ that parametrizes $\Sigma_L$.
 In constructing the heterotic string, 16 of the 26 dimensions of the
bosonic string are compactified using the root lattice of $E_8\times E_8$ or $\mathrm{Spin}(32)/\Z_2$. To make the left-moving part of the heterotic
string vacuum amplitude, the Mumford form $\Phi_g$ for the bosonic string
is multiplied by a certain theta function appropriate to the lattice. The product is a section (not  a trivialization) of $\det T^*\M_L\otimes \det^{-5} \V_L$.  
The rest of what we have said, including the use of integration to define a nondegenerate pairing $\V_L\otimes\V_R\to
\C$, is applicable to the heterotic string.

There is one more important observation about this situation.  In genus 2, let $\pi:\MM_{2,+}\to \M_{2,\spin+}$ be the holomorphic projection
that maps a super Riemann surface $\Sigma$ to the ordinary Riemann surface $\Sigma_0$ that has the same period matrix.  If $\omega$
is a closed holomorphic 1-form on $\Sigma$, there is a corresponding holomorphic 1-form $\omega_0$ on $\Sigma_0$ with the same periods.
This map exhibits the vector bundle $\V\to \MM_{2,+}$, and therefore also its determinant $\det\V$, as a pullback from $\M_{2,\spin+}$.
Since $\BBer_1\to\MM_{2,+}$ is isomorphic to $\det\V$, it follows that $\BBer_1$ is such a pullback, as claimed in
 the explanation of eqn. (\ref{telmex}).

\subsection{Even Spin Structure In Genus 1}\label{evenone}

The reduced space of a genus 1 super Riemann surface $\Sigma$ is an ordinary Riemann surface $\Sigma_0$
described by a familiar hypelliptic equation:
\begin{equation}\label{flunky} y^2=\prod_{i=1}^4(x-e_i).\end{equation}
$\Sigma_0$ is also endowed with a spin structure, which is a line bundle $K^{1/2}\to \Sigma_0$
with an isomorphism $\varphi:K^{1/2}\otimes K^{1/2}\cong K$.  We can characterize a spin structure by saying which
meromorphic sections of $K$ can be written as $\varphi(s\otimes s)$ for some meromorphic section of $K^{1/2}$.  For
example, if we assume that $K^{1/2}$ has a holomorphic section $s$ obeying $\varphi(s\otimes s)=\d x/y$,
we get what is called an odd spin structure on $\Sigma$.  Indeed, as $\omega=\d x/y$ has neither zeroes
nor poles, $s$ likewise has no zeroes or poles, and hence is a global trivialization of $K^{1/2}$.  As $K^{1/2}$
is trivial, it has a 1-dimensional space of holomorphic sections, generated by $s$; since the dimension of $H^0(\SIgma_0,K^{1/2})$
is odd, $K^{1/2}$ is said to define an odd spin structure.  In what follows, we usually write an equation such as $\varphi(s\otimes s)=\d x/y$ more informally
as $s^2=\d x/y$ or  $s=(\d x/y)^{1/2}$.

The odd spin structure on $\Sigma_0$ that we have just described is unique up to isomorphism.  The choice of an even spin structure on $\Sigma_0$
depends on a division of the 4 branch points $e_1,\dots,e_4$ into 2 sets of 2, say $u_1,u_2$ and $v_1,v_2$.  Thus we write the hyperelliptic
equation as
\begin{equation}\label{lunky}y^2=\prod_{i=1,2}(x-u_i)\prod_{j=1,2}(x-v_j). \end{equation}
Having made this division, we define an even spin structure $K^{1/2}$ by saying that it has a meromorphic
section $s$ with $\varphi(s\otimes s)=(\d x/y)(x-v_1)/(x-v_2)$ or in other words
\begin{equation}\label{morz}s^2=\frac{\d x}{y}\frac{x-v_1}{x-v_2}.\end{equation}
It immediately follows that $K^{1/2}$ also has a rational section $s'=s(x-v_2)/(x-v_1)$ with $(s')^2=(\d x/y)(x-v_2)/(x-v_1)$,
and similarly a rational section $s''=s y/(x-u_2)(x-v_1)$ with $(s'')^2=(\d x/y)(x-u_1)/(x-u_2)$, so actually the choice
of $K^{1/2}$ is invariant under exchange of the 2 $u$'s, or of the 2 $v$'s, or exchange of the $u$'s with the $v$'s.   All 3 even spin structures on
$\Sigma_0$ are associated to such a division of the 4 branch points into 2 sets of 2.
$\M_{1,\spin+}$ is parametrized by the choice of the 4 branch points divided into 2 groups of 2, modulo $SL(2,\C)$ and the permutations
of branch points that preserve the pairwise separation.

A genus 1 curve $\Sigma_0$ with no additional structure has only one stable degeneration, namely the degeneration to a genus 0 curve with 
2 points glued together.  We discussed 
this degeneration in the context of the bosonic string in section \ref{mumone}; it occurs when two branch points $e_i$ and $e_j$ colllide.  
When $\Sigma_0$ is endowed with a spin structure, we have to distinguish two possible degenerations, according to
whether the string state propagating through the singularity is in the Ramond or Neveu-Schwarz (NS) sector; we refer to these as degenerations
of Ramond or NS type.  (They have been discussed from the standpoint of the super Mumford form in \cite{BS}.)

An odd spin structure corresponds to the case that $K^{1/2}$ is trivial, so fermions propagating in any channel are untwisted and in the Ramond
sector; hence only a Ramond degeneration will occur.  We can verify this by examining the behavior of the holomorphic differential $\omega=\d x/y$
as two branch points coincide, for instance $e_1,e_2\to e$.  For $e_1=e_2=e$, we set $y=(x-e)\t y$, whence equation (\ref{flunky}) becomes
\begin{equation}\label{zurtz} \t y^2=(x-e_3)(x-e_4), \end{equation}
which describes a smooth curve $\Sigmax$ of genus 0 (called the normalization of $\Sigma_0$).  $\Sigma_0$ is obtained from $\Sigmax$ by gluing together the two points $x=e$,
$\t y=\pm \sqrt{(e-e_3)(e-e_4)}$, since both of these points correspond on $\Sigmax$ to $x=e,$ $y=0$.  Let us call these points $p$ and $p'$.
The holomorphic differential $\omega =\d x/y$
on $\Sigma_0$ becomes $\d x/\t y(x-e)$ on $\Sigmax$, and has simple poles of equal and opposite residue at $p$ and $p'$.  So to define
$s=\omega^{1/2}$ on $\Sigmax$ amounts to taking the square root of a simple pole at $p$ and at $p'$, introducing what in superconformal field theory are usually
called  square root branch points.  Such a branch point is associated to a Ramond vertex operator, so this is a Ramond degeneration.

We can  similarly understand the degenerations of a genus 1 curve $\Sigma_0$ endowed with an even spin structure.  For $u_1\to u_2$,
$s=\sqrt{(\d x/y)(x-v_1)/(x-v_2)}$ behaves exactly as found in the last paragraph, so this is a Ramond degeneration.   Since $K^{1/2}$ is invariant under the exchange $\{u_1,u_2\}\leftrightarrow\{v_1,v_2\}$, it follows that $v_1\to v_2$ is
similarly a Ramond degeneration.  The opposite
type of degeneration in which one of the $u$'s approaches one of the $v$'s is an NS degeneration.  For example, if $u_1,v_1\to e$, we set again
$y=\t y(x-e)$, whereupon $\omega'=(\d x/y)(x-v_1)/(x-v_2)$ becomes $\d x/(\t y\,(x-v_2))$, which is regular at $x=e$, that is at $p$ and $p'$.  So $s=\sqrt{\omega'}$ is also
regular at $p$ and $p'$, corresponding to an NS degeneration.  

We concentrate here on the case of an even spin structure.  Since a genus 1 super Riemann surface with even spin structure has no odd moduli,
its moduli space $\MM_{1,+}$ is equal to the corresponding reduced space $\M_{1,\spin+}$, so the holomorphic superstring
amplitude $\sPhi_{1,+}$ coincides with its restriction to the reduced space.  For an even spin structure in genus 1,  $H^i(\Sigma_0,K^{1/2})=H^i(\Sigma_0,K^{3/2})=0$, $i=0,1$,
and we can drop the factors $\dets^{5} H^*(K^{1/2})$ and $\dets^{-1}H^*(K^{3/2})$ in eqn. (\ref{udono}).  However, rather as in the case of the bosonic
string, there is a correction in going from (\ref{zelf}) and (\ref{uzelf}) to (\ref{udono}) that comes from the fact that for $\Sigma_0$ of genus 1, $H^1(\Sigma_0,K^2)\not=0$.
By the same reasoning as for the bosonic string, the effect of this correction is to increase the power of $H^0(K)$ in the denominator by 1, and accordingly
$\sPhi_{1,+}$ is a holomorphic trivialization of
\begin{equation}\label{zondox}T^*\M_{1,\spin+}\otimes (H^0(K))^{-6}.  \end{equation}

Such a trivialization can be described concretely using the same ideas as in section \ref{mumone}.  As before, $H^0(K)$ is trivialized by the section $\d x/y$,
and so by analogy with eqn. (\ref{tombi}),  $\sPhi_{1,+}$ can be written
\begin{equation}\label{tombiz}\sPhi_{1,+}=	\voli \frac{F(u_1,u_2,v_1,v_2) \d u_1\d u_2\d v_1 \d v_2}{(\d x/y)^6}, \end{equation}
where the function $F$ is regular and nonzero as long as the $u_i$ and $v_j$ are distinct, and moreover must be odd under the exchange $u_1\leftrightarrow u_2$
or $v_1\leftrightarrow v_2$, and even under exchange of the $u$'s with the $v$'s.   Moreover, from conformal field
theory, the only singularities of $F$ are poles (as opposed to essential singularities).  The most general function with these properties is $F=(u_1-u_2)^a(v_1-v_2)^a\prod_{i,j=1}^2 (u_i-v_j)^b$,
with $a$ and $b$ integers and $a$ odd.  Finally, to ensure invariance under the scaling $u_i\to\lambda u_i$, $v_i\to \lambda v_i$, $x\to \lambda x$, $y\to
\lambda^2 y$ , we require $2a+4b=-10$.  As in our discussion of bosonic string theory, scale-invariance is necessary and sufficient for $SL(2,\C)$ invariance of $\Psi_{1,+}$
(for the sufficiency, one needs the fact that under inversion, the right hand side of (\ref{tombiz}) transforms with the same weight in each of the $u_i$ and $v_j$; this follows from the obviouss permutation symmetries).

In contrast to our study of the bosonic string, this is not quite enough to determine $\sPhi_{1,+}$ up to a constant multiple; we also need to know either $a$ or $b$.
The reason that this has happened is that although not manifest in our description of it, $\M_{1,\spin+}$ is a copy of $\C^*$, isomorphic to the complex
$z$-plane with the origin omitted.  Given a meromorphic function $f(z)$ with no poles or zeroes except possibly at $0$ and $\infty$, to determine $f$
up to a constant multiple we need one integer, which is the order of growth of $f(z)$ at either 0 or $\infty$ (in other words, such a function is $f(z)=c z^n$ with 
a constant $c$ and some integer $n$).  This corresponds to the fact that in the last
paragraph, to get a unique answer we need to know $a$ or $b$.

As we explain momentarily, superconformal field theory determines that $a=-1$, $b=-2$,  so
\begin{equation}\label{rombiz} F=(u_1-u_2)^{-1}(v_1-v_2)^{-1}\prod_{i,j=1}^2(u_i-v_j)^{-2}=\frac{(u_1-u_2)(v_1-v_2)}{\prod_{1\leq i<j\leq 4}(e_i-e_j)^2},\end{equation}
where in the last formula the $e_i$ are all four branch points $u_i$ and $v_j$.  So
\begin{equation}\label{yombix}\sPhi_{1,+}= \voli \frac{\d u_1\d u_2\d v_1 \d v_2}{(u_1-u_2)(v_1-v_2)\prod_{i,j=1}^2(u_i-v_j)^2(\d x/y)^6}.\end{equation}
As a check on this, one can verify the GSO cancellation, which says that $\sPhi_{1,+}$ vanishes if summed over the three even spin structures, keeping the $e_i$
fixed.  This amounts to saying that $F$ vanishes if summed over cyclic permutations of $u_2, v_1$, and $v_2$.  Using the second formula in (\ref{rombiz}),
the requisite identity is
\begin{equation}\label{wombiz}(u_1-u_2)(v_1-v_2)+(u_1-v_1)(v_2-u_2)+(u_1-v_2)(u_2-v_1)=0.\end{equation}

Finally, let us explain the predictions of superconformal field theory for $a$ and $b$.  Consider first the Ramond degeneration for $u_1\to u_2$.
Allowing for permutation symmetry between $u_1$ and $u_2$, the natural parameter describing this degeneration is $q_\Ra=(u_1-u_2)^2$.
Superconformal field theory says that the holomorphic superstring path integral behaves for $q_\Ra\to 0$ as $\d q_\Ra\,q_\Ra^{L_0-1}$, where $L_0$ is the ground
state energy in the Ramond sector. (This formula is analogous to the bosonic string formula that we used at the end of section \ref{mumone}.)
 Since $L_0=0$ for the Ramond ground state of uncompactified superstrings, we expect $\sPhi_{1,+}$
to be proportional to $\d q_\Ra/q_\Ra$, which is equivalent, if we hold $u_2$ fixed and let $u_1$ vary, to $\d u_1/(u_1-u_2)$.  This is the behavior
found in (\ref{tombiz}) if $\M_{1,\spin+}$ is parametrized by $u_1$, holding fixed the other branch points.   Similarly, let us consider the NS degeneration
as $u_1\to v_1$.  There is no exchange symmetry between $u_1$ and $v_1$ (while keeping $u_2$ and $v_2$ fixed), so the natural parameter is
simply $\varepsilon=u_1-v_1$. The expected behavior (see for instance section 6.4.4 of \cite{Revisited}) is $\sPhi_{1,+}\sim\d\varepsilon\, \varepsilon^{2L_0-1}$,
where now $L_0$ is the ground state energy in the NS sector.  As this ground state energy is $-1/2$, we expect $\sPhi_{1,+}\sim \d\varepsilon/\varepsilon^2$,
or equivalently (if we again parametrize $\M_{1,\spin+}$ by $u_1$ with the other branch points held fixed) $\sPhi_{1,+}\sim \d u_1/(u_1-v_1)^2$.
This is the behavior seen in eqn. (\ref{yombix}).  In superconformal field theory, one usually uses the variable $q_\NS=\varepsilon^2$ instead
of $\varepsilon$, and then $\sPhi_{1,+}\sim \d q_\NS/q_\NS^{3/2}$ for $q_\NS\to 0$.

\subsection{Even Spin Structure In Genus 2}\label{eventwo}

A genus 2 Riemann surface $\Sigma_0$ is a hyperelliptic curve with 6 branch points.  To endow $\Sigma_0$ with an even spin structure,
we divide the branch points into 2 groups of 3, say $u_1,u_2,u_3$ and $v_1,v_2,v_3$, and write the hyperelliptic equation as
\begin{equation}\label{dono}y^2=\prod_{i=1}^3(x-u_i)\prod_{j=1}^3(x-v_j).    \end{equation}
A spin structure $K^{1/2}$ associated to this division of the branch points is described by saying that it has a meromorphic section $s$
with, for example, 
\begin{equation}\label{melbop}s^2=\frac{\d x}{y}\frac{(x-u_1)(x-u_2)}{(x-u_3)}.\end{equation}  One can permute the $u_i$ by 
replacing $s$ with, for example, $s'=s(x-u_3)/(x-u_2)$,
and one can exchange the $u$'s and $v$'s by replacing $s$ with, for example, $s''=s y/(x-u_1)(x-u_2)(x-v_3)$, whose square is
$(\d x/y)(x-v_1)(x-v_2)/(x-v_3)$.  All 10 even spin structures on $\Sigma_0$ are of this form, for some splitting of the 6 branch points into 2 sets of 3.

Since the moduli space $\MM_2$ of genus 2 super Riemann surfaces has odd dimension 2, there are only 2 terms in the expansion (\ref{cobmo})
of the holomorphic string amplitude.
Luckily, these are the 2  terms that we can most easily compute -- the ``bottom'' term is the restriction of $\sPhi_{2,+}$ to $\M_{2,\spin+}$, and the
``top'' term is the projection $\pi_*(\sPhi_{2,+})$.

We can analyze the restriction to the reduced space using (\ref{udono}).  A genus 2 Riemann surface $\Sigma_0$ with even spin structure
always has $H^0(\Sigma_0,K^{1/2})=H^1(\Sigma_0,K^{1/2})=0$, so $\det H^*(K^{1/2})$ is trivial.  $H^0(\Sigma_0,K)$ is generated by
$\d x/y$ and $x\,\d x/y$ (while $H^1(\Sigma_0,K)$ is canonically isomorphic to $\C$, as explained in section \ref{mumf}), 
so $\det H^*(K)$ is trivialized by the section $\d x/y\wedge x\,\d x/y$.   Finally, if $s$ is a meromorphic section of $K^{1/2}$ obeying (\ref{melbop}), so $s$
has simple zeroes at $u_1$ and $u_2$ and a simple pole at $u_3$,
then $H^0(\Sigma_0,K^{3/2})$ is generated by $s(x-u_3)$ and $sy/(x-u_1)(x-u_2)$, which we can write more informally as $\Chi_1=((\d x/y)^3\prod_{i=1}^3(x-u_i))^{1/2}$
and $\Chi_2=((\d x/y)^3\prod_{j=1}^3(x-v_j))^{1/2}$.  On the other hand, $H^1(\Sigma_0,K^{3/2})=0$.  So $\det H^*(K^{3/2})$ is trivialized by the section 
$\Chi_1\wedge \Chi_2$.

We can now write down a formula for $\sPhi_{2,+}|_{\M_{2,\spin+}}$ in terms of an unknown function $F$:
\begin{equation}\label{teffor}\sPhi_{2,+}|_{\M_{2,\spin+}}=\voli \frac{F(u_i,v_j) \,\d u_1\,\d u_2\,\d u_3\,\d v_1\,\d v_2\,\d v_3 }{\chi_1\wedge \chi_2\cdot(\d x /y \wedge x\d x/y)^5}. \end{equation}
By now, all of the ingredients needed to determine $F$ (up to a constant multiple) are familiar.  The super Mumford isomorphism says that $F$ has
no zeroes or poles as long as the branch points are all distinct.  On the other hand, from the superconformal analysis explained in section \ref{evenone}, 
$F$ should have, for all $i,j$, a simple pole for $u_i-u_j\to 0$ or $v_i-v_j\to 0$, and a double pole for $u_i-v_j\to 0$.   Thus, we must have
\begin{equation}\label{teffora}\sPhi_{2,+}|_{\M_{2,\spin+}}=\voli \frac{\d u_1\,\d u_2\,\d u_3\,\d v_1\,\d v_2\,\d v_3 }{\prod_{i<j}((u_i-u_j)(v_i-v_j))\prod_{k,l}(u_k-v_l)^2\cdot
\chi_1\wedge \chi_2\cdot(\d x /y \wedge x\d x/y)^5}. \end{equation}
Happily, this expression possesses all the requisite symmetries.  Symmetry under permutation of the $u_i$ or of the $v_j$ is obvious.  The exchange of the $u_i$
with the $v_j$, say $u_i\leftrightarrow v_i$ for $i=1,2,3$, changes the sign of the differential form in the 
numerator but also exchanges $\chi_1$ and $\chi_2$ and hence reverses the sign of $\chi_1\wedge \chi_2$.
  It remains to check $SL(2,\C)$ invariance.
The non-trivial point  is invariance under the familiar scaling $u_i\to \lambda u_i$, $v_i\to \lambda v_i$, $x\to\lambda x$, $y\to\lambda^3y$.  A short
check shows that numerator and denominator both scale as $\lambda^6$.  (The weights in each of the $u_i$ and $v_j$ are manifestly the same, so scale-invariance implies
$SL(2,\C)$ symmetry.)

One can go on and learn what the constraints of holomorphy  say about the ``top component'' $\pi_*(\sPhi_{2,+})$.   Comparing
(\ref{udonox}) to (\ref{udono}), the only changes are that to describe $\pi_*(\sPhi_{2,+})$, we should omit the factor $\chi_1\wedge\chi_2$ from the denominator;
also, we should allow additional zeroes, but no additional poles.  Since $1/\chi_1\wedge\chi_2$ scales as $\lambda^3$ under the scaling considered
in the last paragraph, to maintain scale-invariance, we must replace $1/\chi_1\wedge\chi_2$ by a homogeneous cubic polynomial $Q(u_i,v_j)$.
Thus,
\begin{equation}\label{beffor}\pi_*(\sPhi_{2,+})=\voli \frac{Q(u_i,v_j)\cdot \d u_1\,\d u_2\,\d u_3\,\d v_1\,\d v_2\,\d v_3 }{\prod_{i<j}((u_i-u_j)(v_i-v_j))\prod_{k,l}(u_k-v_l)^2\cdot
(\d x /y \wedge x\d x/y)^5}. \end{equation}
Here we are tacitly assuming that integration over the fibers of the projection $\pi:\MM_2\to\M_{2,\spin}$ does
not change the nature of the singularities when a pair of branch points collide, so that $\pi_*(\Psi_{2,+})$ has the same behavior as $\Psi_{2,+}|_{\M_{2,\spin+}}$ 
for $u_i\to u_j$ or for $u_i\to v_j$.  This is true but far from trivial; it is explained in section \ref{nonseparating}.

$Q(u,v)$ is partly constrained by the usual permutation symmetries. It must be invariant under permutations of the $u_i$ or of the $v_j$.  It  must be odd under
the exchange of all $u_i$ with $v_i$ to provide the minus sign that previously came from  $\chi_1\wedge \chi_2$.  
As part of $SL(2,\C)$ symmetry, it must also be invariant under translations:
\begin{equation}\label{guelf}\sum_{i=1}^3\left(\partial_{u_i}+\partial_{v_i}\right)Q(u_i,v_j)=0. \end{equation}  
Three polynomials with the symmetries that $Q$ is supposed to have are \begin{align}\label{donkeys}
                                                  P_1&=(c_1(u)-c_1(v))^3\cr
                                                   P_2&=  c_1(u)c_2(u)-(c_2(u)+(2/3)c_1(u)^2)c_1(v)\cr &~~~~+c_1(u)(c_2(v)+(2/3)c_1(v)^2)-c_1(v)c_2(v) \cr
                                                    P_3& =3c_3(u)-c_2(u)c_1(v)+c_1(u)c_2(v)-3c_3(v) , \end{align}
 where  the $c_i$ are  the elementary symmetric functions $c_1(p_1,p_2,p_3)=p_1+p_2+p_3$, $c_2(p_1,p_2,p_3)=p_1p_2+p_2p_3+p_3p_1$, and $c_3(p_1,p_2,p_3)=p_1p_2p_3$,
  and we have picked a basis of polynomials
 that reduce at $v=0$ to $c_1(u)^3$, $c_1(u)c_2(u)$, and $3c_3(u)$.  Since this is a basis of cubic homogeneous symmetric polynomials in the $u$'s,
 we have shown that every such polynomial can be extended to a polynomial $P(u,v)$ that possesses the desired symmetries.  Conversely, a polynomial
 $P(u,v)$ possessing the desired symmetries and vanishing at $v=0$ is identically 0 (the permutation symmetries would force
 $P(u,v)=(a c_2(u)+bc_1(u)^2)c_1(v)-c_1(u)(ac_2(v)+bc_1(v)^2)$ with constants $a,b$, but this does not satisfy (\ref{guelf}) unless $a=b=0$).  So $Q$ must be a linear combination
 of the $P_i$.  
 
 To learn more, we simply use $SL(2,\C)$ symmetry and more specifically inversion symmetry, which is not as straightforward as in the previous examples.
  Invariance under the inversion $u_i\to 1/u_i$, $v_i\to 1/v_i$ (accompanied as usual by $x\to 1/x$ and $y\to\pm  y/x^3\prod_i(u_iv_i)^{1/2}$)
 is equivalent to
 \begin{equation}\label{xexo} Q(1/u_1,\dots, 1/v_3)=-\frac{Q(u_1,\dots,v_3)}{u_1u_2u_3v_1v_2v_3}\end{equation}
 (where the minus sign compensates for a sign in the transformation of $\d x/y \wedge x \d x/y$).
 This immediately implies that $Q(u_1,\dots, v_3)$ grows only linearly as $u_1\to\infty$ keeping the other variables fixed.  A linear combination of the $P_i$ that has this property must
 be a multiple of $P_3$, and so  (up to a constant multiple)
\begin{equation}\label{morzono}Q(u,v)=3c_3(u)-c_2(u)c_1(v)+c_1(u)c_2(v)-3c_3(v).\end{equation}
 In appendix \ref{comp}, we show that (\ref{beffor}) together with (\ref{morzono}) is equivalent to the formula for $\pi_*(\Psi_{2,+})$
originally obtained by D'Hoker and Phong \cite{DPHgold}.

\subsubsection{Sum Over Spin Structures}\label{zum}

In superstring theory, before integrating over odd moduli, it does not make sense to sum over spin structures.  That is because the definition
of the odd moduli depends on the spin structure and there is no notion of changing the spin structure on a super Riemann surface
while otherwise leaving it unchanged.  Accordingly, a general proof of the vanishing of the cosmological constant in genus $g$ is not based
on a direct imitation of the GSO cancellation in genus 1.  See \cite{Bel} or section 8 of \cite{Revisited}.  

However, if one does have a preferred method to integrate over odd moduli, then after doing so it makes sense to sum over spin structures.
It has indeed been shown \cite{DPHgold} that in genus 2, $\pi_*(\Psi_{2,+})$ vanishes upon summation over spin structures.  Let us
verify this in the context of the formula (\ref{beffor}).  First we  rewrite this formula as follows:
\begin{equation}\label{neffor}\pi_*(\sPhi_{2,+})=\voli \frac{Q(u_i,v_j)\prod_{k<l}((u_k-u_l)(v_k-v_l))\cdot \d u_1\,\d u_2\,\d u_3\,\d v_1\,\d v_2\,\d v_3 }{\prod_{1\leq s<t\leq 6}(e_s-e_t)^2\cdot
(\d x /y \wedge x\d x/y)^5}. \end{equation}
In the denominator, the $e$'s are all 6 branch points $u_i$ and $v_j$.

Summing over spin structures means symmetrizing this expression under permutations of the 6 branch points. Since the denominator
in (\ref{neffor}) has the full permutation symmetry, we must symmetrize the numerator.   Since the expression $\voli \,\d u_1 \dots \d v_3$ is completely
antisymmetric, we must antisymmetrize the polynomial $\mathcal Q=Q\prod_{k<l}((u_k-u_l)(v_k-v_l))$ with respect to all its arguments.

The polynomial $\mathcal Q$  vanishes if antisymmetrized with respect to all 6 variables, since it has degree 9, but a completely
antisymmetric polynomial in 6 variables $e_1,\dots,e_6$ is divisible by $\prod_{i<j}(e_i-e_j)$ and has degree at least 15.  The same argument
shows that $\mathcal Q$ vanishes if antisymmetrized with respect to any 5 of the 6 variables.  

It is not true in general that $\mathcal Q$ vanishes if antisymmetrized with respect to 4 variables.  But there is a weaker property that is still
interesting.  A nonseparating NS degeneration corresponds to, say, $u_3\to v_3$.  The limiting behavior at such a degeneration can be extracted
by simply setting $u_3=v_3$ in $\mathcal Q$, in which case by translation symmetry we may as well take $u_3=v_3=0$.  $\mathcal Q$
then vanishes if antisymmetrized in the other 4 variables.  This is a GSO-like cancellation for a nonseparating NS degeneration.  A nonseparating Ramond degeneration corresponds
to, say, $v_2\to v_3$.  If we simply set $v_2=v_3$, $\mathcal Q$ vanishes; to extract the leading behavior at a degeneration of this
type, we should instead factor $\mathcal Q=(v_2-v_3)\mathcal Q'$ and set $v_2=v_3$ in $\mathcal Q'$.  Again it turns out that $\mathcal Q'$
vanishes if antisymmetrized over the remaining 4 variables; this is a GSO-like cancellation for a nonseparating Ramond degeneration.

\subsection{Odd Spin Structures}\label{oddone}

The object $\sPhi_g$ described by the super Mumford isomorphism has a less immediate physical interpretation when the spin structure is odd,
as we have seen in section \ref{superwhatfor}.  A super Riemann surface with an odd spin structure does not contribute to the vacuum amplitude;
it contributes to certain parity-violating scattering amplitudes.  Nevertheless, the super Mumford isomorphism is valid for an odd spin structure and the object
that it describes does have applications in string theory.  So let us see what we can say.  We will start in genus 2, to postpone grappling with the exceptional
behavior that occurs for genus 1.

If a  Riemann surface $\Sigma_0$ of genus 2, constructed as usual as a hyperelliptic curve,
 is endowed with an odd spin structure, then $H^0(\Sigma_0,K^{1/2})$ is always of dimension 1, generated
by a section that vanishes at one of the 6 branch points.  So one of the branch points will play a distinguished role, and we write the hyperelliptic
equation in the form
\begin{equation}\label{zommo}y^2=(x-u)\prod_{j=1}^5(x-v_j). \end{equation}
We can describe the line bundle $K^{1/2}$ by saying that it has a global holomorphic section $s$ such that $s^2=(\d x/y)(x-u)$; informally
we write $s=((\d x/y)(x-u))^{1/2}$.  The 6 odd spin structures on $\Sigma$ are all constructed in this way, with one of the 6 branch points playing a distinguished
role.  $\M_{2,\spin-}$ is parametrized by $u$ and the $v_i$ (required to be all distinct), modulo $SL(2,\C)$ and the permutations of the $v_i$.

We do not know a holomorphic projection $\pi:\MM_{2,-}\to \M_{2,\spin-}$, so we do not have a convenient way to describe the ``top'' component
of $\sPhi_{2,-}$.  But we can certainly use holomorphy to analyze the ``bottom'' component, the restriction $\sPhi_{2,-}|_{\M_{2,\spin-}}$.  
Perhaps the main novelty is that in (\ref{udono}), we must
now include the factor $\dets^5 H^*(K^{1/2})$, since $H^*(\Sigma_0,K^{1/2})$ is nonzero.  The definition of the determinant of cohomology\footnote{For a genus 2
surface $\Sigma_0$ with odd spin structure, the dimensions of $H^i(\Sigma_0,K^{1/2})$ are constant as the moduli of $\Sigma_0$ vary, 
so one can use the naive definition of the determinant 
of cohomology.} is 
that $\det H^*(K^{1/2})=\det H^0(\Sigma_0,K^{1/2})\otimes \det H^1(\Sigma_0,K^{1/2})^{-1}$.  But in the particular case of $K^{1/2}$, Serre
duality says that $H^1(\Sigma_0,K^{1/2})$ is dual to $H^0(\Sigma_0,K^{1/2})$, so $\det H^*(K^{1/2})=\dets^2 H^0(\Sigma_0,K^{1/2})$.
For $\Sigma_0$ of genus 2, $H^0(\Sigma_0,K^{1/2})$ is 1-dimensional, so $\det H^0(\Sigma_0,K^{1/2})=H^0(\Sigma_0,K^{1/2})$.
The latter is generated by $s$, so finally $\det H^*(K^{1/2})$ is generated by $s^2$, and $\dets^5 H^*(K^{1/2})$ is generated by $s^{10}$.
To evaluate (\ref{udono}), we also need to know that $H^0(\Sigma_0,K^{3/2})$ with an odd spin structure is generated by $\chi_1=s\, \d x/y$ and $\chi_2=s x\,\d x/y$,
while $H^1(\SIgma_0,K^{3/2})=0$. So $\det H^*(\Sigma_0,K^{3/2})$ is generated by $\chi_1\wedge \chi_2$.  And as usual, $\dets^5 H^*(K)$ is
generated by $(\d x/y \wedge x\d x/y)^5$.  Finally, $\sPhi_{2,-}|_{\M_{2,\spin-}}$ must have the usual double poles for $u\to v_i$ and simple poles for 
$v_i\to v_j$.  Putting all this together, we must have
\begin{equation}\label{noch} \sPhi_{2,-}|_{\M_{2,\spin-}}=\voli \frac{\d u\,\d v_1\d v_2\dots \d v_5\cdot s^{10}}{\prod_{i=1}^5(u-v_i)^2\prod_{1\leq k<l\leq5}(v_k-v_l)\cdot
(\chi_1\wedge\chi_2)\cdot (\d x/y\wedge x\d x/y)^5}.    \end{equation}
As usual, for $SL(2,\C)$ invariance,  the right hand side must be invariant under the scaling in which $u,v_i, x$ have weight 1
and $y$ has weight 3.  It is straightforward to verify that numerator and denominator both have weight 1, given that
 $s$ has weight $-1/2$, $\chi_1$ and $\chi_2$ have weights $-5/2$ and $-3/2$, and $\d x/y$
and $x\d x/y$ have weights $-2$ and $-1$.  (The proof of inversion symmetry requires a little care; one uses the transformation $s\to s\sqrt{\prod_i v_i/u}$ under the inversion
$u\to 1/u$, $v_i\to 1/v_i$, $x\to 1/x$, $y\to\pm y/x^3(u\prod_iv_i)^{1/2}$.  A similar remark applies to eqn. (\ref{merot}) below.)

An odd spin structure on a Riemann surface $\Sigma_0$ of genus 1 has already been described in section \ref{evenone}.  Describing $\Sigma_0$
by the usual hyperelliptic equation
\begin{equation}\label{olm}y^2=\prod_{i=1}^4(x-e_i),\end{equation}
$H^0(\Sigma_0,K^2)$ is generated by a section that we can informally denote as $s=(\d x/y)^{1/2}$.  It is still true, by the same reasoning as in the genus 2
case, that $\dets^5H^*(K^{1/2})$ is generated by $s^{10}=((\d x/y)^{1/2})^{10}$.  Likewise, $H^0(\SIgma_0,K^{3/2})$ is 1-dimensional, generated by a section that is
naturally understood as $(\d x/y)^{3/2}$.  However, for genus 1, $H^1(\Sigma_0,K^{3/2})$ is nonzero.  It is Serre dual to $H^0(\SIgma_0,K^{-1/2})$,
which in turn is dual to $H^0(\SIgma_0,K^{1/2})$  -- indeed, $H^0(\SIgma_0,K^{1/2})$ and $H^0(\SIgma_0,K^{-1/2})$ are generated by global sections
that are dual under the natural duality between $K^{1/2}$ and $K^{-1/2}$.  So $H^1(\Sigma_0,K^{3/2})$ is naturally isomorphic to $H^0(\SIgma_0,K^{1/2})$,
and hence $\det H^*(K^{3/2})\cong H^0(\SIgma_0,K^{3/2})\otimes (H^0(\SIgma_0,K^{1/2}))^{-1}$ is trivialized by a section that we can write as
$(\d x/y)^{3/2}\otimes (\d x/y)^{-1/2}$.  Finally, as usual in genus 1, $\det H^*(K)$ is generated by $\d x/y$.  Just as in (\ref{zondox}), we must
remember in genus 1 to include $\det H^*(K)$ with the power $-6$, not $-5$.  Putting this together and allowing 
for the usual Ramond sector simple poles as $e_i\to e_j$, we have
\begin{equation}\label{merot}\sPhi_{1,-}|_{\M_{1,\spin-}}=
\voli \frac{\d e_1\d e_2\d e_3\d e_4 \,s^{10}}{\prod_{i<j}(e_i-e_j) ((\d x/y)^{3/2}\otimes (\d x/y)^{-1/2}) (\d x/y)^6}.
\end{equation}  Under the usual scaling in which $e_i$ and $x$ have degree 1 and $y$ has degree 2, the numerator and denominator both have degree $-1$,
ensuring $SL(2,\C)$ invariance of this formula.

The supermoduli space $\MM_{1,-}$ has dimension $1|1$, with just 1 odd modulus.  Accordingly, there is a unique holomorphic
projection $\pi:\MM_{1,-}\to\M_{1,\spin-}$.  With only 1 odd modulus, there is no way to make a nontrivial expansion like the one in eqn. (\ref{cobmo}).
Thus trivially $\sPhi_{1,-}$ is the pullback via $\pi$ of its restriction to $\M_{1,\spin-}$, written in eqn. (\ref{merot}).  

One usually says that the superstring vacuum amplitude vanishes with an odd spin structure because each of the 10 RNS fermions has a zero mode.
This is reflected in the factors of $s^{10}$ that appear in the numerator of eqns. (\ref{merot}) and (\ref{noch}).  There is no spontaneous way to replace
the factors of $s^{10}$ by numbers, but if one includes external vertex operators to compute a scattering amplitude, 10 fermions in the vertex operators
would be used to absorb the zero-modes; in an algebrogeometric description, the factors of $s^{10}$ in the vacuum amplitude would be part of the description
of this process. For an algebrogeometric description of bosonic string scattering amplitudes, see \cite{Voronov}.  It would be interesting to compare this
procedure with the sort of holomorphic decomposition of scattering amplitudes described in \cite{DPHtwo,DPHgold}.

\section{Superstring Amplitude At A Separating Degeneration}\label{separsuper}

\subsection{Factorization Of The Super Mumford Form}\label{factor}

Local superconformal coordinates $\phi|\theta$ on a super Riemann surface $\Sigma$ are coordinates in which the subbundle $\D\subset T\Sigma$
that defines the superconformal structure is generated by $D_\theta=\partial_\theta+\theta\partial_\phi$.  

A separating degeneration of a super Riemann surface $\Sigma$ occurs when $\Sigma$ splits up into a pair of super Riemann surfaces $\Sigma_\ell$
and $\Sigma_r$, joined at a point.   In the absence of Ramond punctures associated to external vertex operators (we do not consider these in the present
	paper except in appendix \ref{superram}),  $\Sigma_\ell$ and $\Sigma_r$ will meet at a smooth
point of their superconformal structures (as opposed to a Ramond puncture). In other words, the string state propagating
between $\Sigma_\ell$ and $\Sigma_r$ is in the NS sector. 

Pick local superconformal coordinates $\phi_\ell|\theta_\ell$ on $\Sigma_\ell$, and $\phi_r|\theta_r$
on $\Sigma_r$.  The gluing of the point $\phi_\ell|\theta_\ell=a|\alpha$ in $\Sigma_\ell$ to the point
$\phi_r|\theta_r=b|\beta$ in $\Sigma_r$, along with smoothing by a small parameter $\varepsilon$
to a smooth surface $\Sigma$, is described by
\begin{align}\label{toddo}
  (\phi_\ell-a+\alpha\theta_\ell)(\phi_r-b+\beta\theta_r)& = -\varepsilon^2 \cr
   (\phi_r-b+\beta\theta_r)(\theta_\ell-\alpha)& =\varepsilon (\theta_r-\beta)\cr
   (\phi_\ell-a+\alpha\theta_\ell)(\theta_r-\beta)&=-\varepsilon(\theta_\ell-\alpha)\cr
   (\theta_\ell-\alpha)(\theta_r-\beta)& = 0.
\end{align}
All that we really need to know about these formulas for our present purposes is the scaling behavior.
The gluing formulas are invariant under  rescaling of the local superconformal coordinates $\phi_\ell|\theta_\ell\to \lambda_\ell\phi_\ell|\lambda_\ell^{1/2}\theta_\ell$,
$\phi_r|\psi_r\to \lambda_r\phi_r|\lambda_r^{1/2}\psi_r$,  together with $a|\alpha\to \lambda_\ell|\lambda_\ell^{1/2}\alpha$, $b|\beta\to \lambda_rb|\lambda_r^{1/2}
\beta$, and $\varepsilon\to \lambda_\ell^{1/2}\lambda_r^{1/2}\varepsilon$.  

The oddness or evenness of a spin structure is additive in a separating degeneration (basically because the number of fermion zero modes is additive).
So a Riemann surface $\Sigma$ with even spin structure can degenerate to a pair of components that both have even spin structures or both have odd ones;
we call these $++$ and $- - $ degenerations, respectively.  If the spin structure of $\Sigma$ is odd, then at a separating degeneration, one component
has an even spin structure and one has an odd one; we call this a $+-$ degeneration.

In discussing the factorization of the super Mumford form $\Psi_g$, let us first assume that the spin structures of 
$\Sigma$, $\Sigma_\ell$, and $\Sigma_r$
are all even.  
We suppose that $\Sigma_\ell$ and $\Sigma_r$ are of respective genus $g_\ell$ and $g_r$.  We also assume to begin with that $g_\ell,g_r>1$.
For the moduli of $\Sigma$, we can take the moduli of $\Sigma_\ell$ and $\Sigma_r$ together with the gluing data $a|\alpha$, $b|\beta$, and $\varepsilon$.
The asymptotic behavior of $\Psi_{g,+}$ for $\varepsilon\to 0$ at a $++$ degeneration is
\begin{equation}\label{cosmo}\Psi_{g,+}\sim \Psi_{g_\ell,+}[\d a|\d\alpha]\cdot\frac{\d \varepsilon}{\varepsilon^2}\cdot[\d b|\d\beta]\Psi_{g_r,+}.    \end{equation}
As in the bosonic case, the expression $\Omega=[\d a|\d\alpha]\cdot\d \varepsilon/\varepsilon^2\cdot[\d b|\d\beta]$ is uniquely determined by the
fact that it is invariant under
rescaling of the local superconformal coordinates, and moreover modulo terms that are less singular for $\varepsilon\to 0$, it is independent of more general
changes of those local coordinates.  From the point of view of superconformal field theory, $\d\varepsilon/\varepsilon^2$ is $\d\varepsilon\, \varepsilon^{2L_0-1}$,
where the ground state of the string in the NS sector has $L_0=-1/2$.  (This ground state is represented by the operator $\delta^{1|1}(C)=\delta(c)\delta(\gamma)
=c\delta(\gamma)$; the usual operation of passing from unintegrated to integrated vertex operators converts $\delta^{1|1}(C)$ to $[\d a|\d\alpha]$.)

Now suppose that $\Sigma$ has even spin structure but $\Sigma_\ell$ and $\Sigma_r$ have odd spin structures.
   In this case, each of the 10 worldsheet matter fields $X^I(\phi|\theta)=x^I(\phi)+\theta\psi^I(\phi),\,I=1,\dots,10$ that describe
motion of the string in $\R^{10}$ has a fermionic zero-mode on $\Sigma_\ell$ and one on $\Sigma_r$.
There is no contribution to $\Psi_g$ from the string ground state propagating between $\Sigma_\ell$ and $\Sigma_r$; such a contribution vanishes
because of the fermion zero-modes on the two sides.  The lowest dimension operator that can absorb the zero modes is $\U=D_\theta X^1 D_\theta X^2
\dots D_\theta X^{10}$; one can characterize this operator as the superconformal primary of lowest dimension -- namely dimension 5 --  
that is invariant under orientation-preserving symmetries of $\R^{10}$ but not under orientation-reversing ones.  Changing
the operator propagating between the two
branches of $\Sigma$ from $\delta^{1|1}(C)$ to $\delta^{1|1}(C)\,\U$ increases $L_0$ from $-1/2$ to $-1/2+5$, and shifts 
$\d\varepsilon \,\varepsilon^{2L_0-1}$ from $\d\varepsilon/\varepsilon^2$
to $\d\varepsilon\,\varepsilon^{10-2}=\d\varepsilon\,\varepsilon^8$.   The asymptotic behavior of $\Psi_{g,+}$ at a $--$ degeneration is accordingly
\begin{equation}\label{mordo}  \Psi_{g,+}\sim \Psi_{g_\ell,-}(\U) \,[\d a|\d\alpha] \cdot \d\varepsilon \,\varepsilon^8 \cdot 
[\d b|\d\beta]\,\Psi_{g_r,-}(\U),      \end{equation}
where $\Psi_{g_\ell,-}(\U)$ is computed by inserting the operator $\U$ on $\Sigma_\ell$ at $\phi|\theta=a|\alpha$, and similarly for $\Psi_{g_r,-}(\U)$.

Finally, if $\Sigma$ has an odd spin structure, we can assume that $\Sigma_\ell$ has an even spin structure and $\Sigma_r$ has an odd one.
Generically, no new fermion zero modes appear at a degeneration of this type, so again the dominant contribution comes from the string ground state and
\begin{equation}\label{mdox}\Psi_{g,-}\sim \Psi_{g_\ell,+}[\d a|\d\alpha]\cdot \frac{\d \varepsilon}{\varepsilon^2}\cdot [\d b|\d\beta]\Psi_{g_r,-}. \end{equation}

\subsection{Details For Genus 1}\label{superdetone}

As in section \ref{detone}, these formulas need some slight changes if $\Sigma_\ell$ and/or $\Sigma_r$ has genus 1.  Suppose that $\Sigma_\ell$
has genus 1 and even spin structure.  Then $\Sigma_\ell$ has a continuous bosonic symmetry group that can be used to shift $a$, but no corresponding
fermionic symmetries.  So in gluing of $\Sigma_\ell$, $\alpha$ is a modulus but $a$ is not.  To deal with this situation, we should view $\Psi_{1,+}$ as
a trivialization of
\begin{equation}\label{ydex}\frac{T^*\M_{1,\spin+}\otimes H^0(\Sigma_\ell,T)}{H^0(\Sigma_\ell,K)^5} .\end{equation}
Using the pairing between $H^0(\SIgma_\ell,T)$ and 1-forms, we contract $\Psi_{1,+}$ with $\d a$ to get what we call $\Psi_{1,+}^{\d a}$,
a section of 
\begin{equation}\label{ydexy}\frac{T^*\M_{1,\spin+}}{H^0(\Sigma_\ell,K)^5} .\end{equation}
If $\Sigma_r$ has genus 1, we likewise replace $\d b \cdot \Psi_{g_r,+}$ by the contraction $\Psi_{1,+}^{\d b}$.  In particular, for  a $++$ degeneration of a genus 2
super Riemann surface
splitting to 2 components each of genus 1, we should replace (\ref{cosmo}) by
\begin{equation}\label{rydex}\Psi_{2,+}\sim \Psi_{1,+}^{\d a}\cdot[\d \alpha]\cdot \frac{\d\varepsilon} {\varepsilon^2}\cdot[\d\beta]\cdot \Psi_{1,+}^{\d b}. \end{equation}
This is the analog of eqn. (\ref{todzo}) for the bosonic string.  

If $\Sigma_\ell$ has genus 1 with an odd spin structure, then neither $a$ nor $\alpha$ should be treated as a modulus.  That is because a split
super Riemann surface of genus 1 with odd spin structure has an automorphism group $F$ of dimension $1|1$ that can be used to transform away
$a$ and $\alpha$.  We can think of $\Psi_{1,-}$ as a trivialization of
\begin{equation}\label{uuu}\frac{ T^*\M_{1,-} \otimes H^0(\Sigma_\ell,K^{1/2})^{10} \otimes H^0(\Sigma_\ell,T)}{H^0(\SIgma_\ell,K)^5\otimes
H^0(\Sigma_\ell,T^{1/2})}=\frac{ T^*\M_{1,-} \otimes H^0(\Sigma_\ell,K^{1/2})^{10} }{H^0(\SIgma_\ell,K)^5}\otimes \Ber(\frak f) ,\end{equation}
where $\frak f$ is the Lie algebra of $F$, and $\Ber(\frak f)\cong H^0(\Sigma_\ell,T)\otimes H^0(\Sigma_\ell,T^{1/2})^{-1}$.  There is a natural
pairing of $\Ber(\frak f)$ with $[\d a|\d\alpha]$, which we consider to be valued in the fiber at $\phi_\ell|\theta_\ell=a|\alpha$ of $\Ber \,T^*\Sigma$.
We write $\Psi_{1,-}^{[\d a|\d \alpha]}$ for the output of this pairing.  It is a section of 
\begin{equation}\label{vv}\frac{ T^*\M_{1,-} \otimes H^0(\Sigma_\ell,K^{1/2})^{10} }{H^0(\SIgma_\ell,K)^5}.\end{equation}
The behavior of $\Psi_{2,+}$ at a $--$ degeneration is
\begin{equation} \label{pyrdex}\Psi_{2,+}\sim \Psi_{1,-}^{[\d a|\d\alpha]}(\U)\cdot \d\varepsilon\,\varepsilon^8\cdot \Psi_{1,-}^{[\d b|\beta]}(\U),
\end{equation}
where as in eqn. (\ref{mordo}), the operator $\U$ is inserted on each side.  Concretely, this insertion means that the factor of $s^{10}$ in
eqn. (\ref{merot}) is replaced with the expectation value of $\U$.  After this replacement, (\ref{pyrdex}) gives the leading behavior as $\varepsilon\to 0$
of a differential form that is valued -- as usual in superstring theory -- in $\det H^0(K)^{-5}$.

\subsection{Analog For Nonseparating Degenerations}\label{nonsepan}

As in the case of the bosonic string (compare eqns. (\ref{mornex})  and (\ref{ornex})), there is also an analog of the above formulas for a nonseparating degeneration.
Here we must distinguish two cases according to whether the string state propagating through the singularity is in the NS or Ramond sector.
The NS case is much more straightforward, and we defer the Ramond case to section \ref{ramondnon}.  At a nonseparating degeneration of NS type,
we suppose that a super Riemann surface $\Sigma$ of genus $g$ is built by gluing together the points $a|\alpha$ and $b|\beta$ in a super Riemann
surface $\Sigmax$ of genus $g-1$, and then smoothing the singularity by the usual procedure of eqn. (\ref{toddo}).   ($\Sigmax$ is called the normalization
of $\Sigma$.)
For $g-1>1$ and thus $g>2$, the analog of the bosonic string formula (\ref{ornex}) is the obvious
close cousin of eqn.  (\ref{cosmo}):
\begin{equation}\label{supernon} \Psi_g\sim \Psi_{g-1}[\d a|\d\alpha] \frac{\d \varepsilon}{\varepsilon^2}[\d b|\d\beta]. \end{equation}
For $g=2$ and so $g-1=1$, this formula needs a slight correction; either $a$
or $b$ should not be treated as a modulus, and $\d a$ or $\d b$ should be combined with one factor of\footnote{For $g-1=1$, $\Sigma^*$ is
split, with reduced space $\Sigma^*_\red$.} $H^0(\Sigma^*_\red,K)^{-1}$ in $\Psi_1$ in the manner
that was described in section \ref{superdetone}.  If we write $\Psi_1^{\d a}$ for the contraction of $\Psi_1$ with $\d a$, and treat $b$  as a modulus,
then the analog of (\ref{supernon}) for $g=2$ is 
\begin{equation}\label{upernon}\Psi_2\sim \Psi_1^{\d a} [\d\alpha]\frac{\d\varepsilon}{\varepsilon^2}[\d b|\d\beta]. \end{equation}
Note that in these formulas for the behavior of $\Psi_g$ at a nonseparating degeneration, it is not necessary to specify whether the spin structure of $\Sigma$
is even or odd.  

\subsection{Eliminating The Odd Variables In Genus 2}\label{elim}

We want to use eqn. (\ref{rydex}) to improve our understanding of $\pi_*(\Psi_{2,+})$, which was computed in eqn. (\ref{beffor}).  At first sight, we face a quandary.  The $\pi_*$ operation represents integration over
$\alpha$ and $\beta$.  But the right hand side of eqn. (\ref{rydex}) appears to be independent of $\alpha$ and $\beta$, so will it not be annihilated
by integration over $\alpha$ and $\beta$?

Here we have to ask what is held fixed when we integrate over $\alpha$ and $\beta$.  If we integrate over $\alpha$ and $\beta$ holding $\varepsilon$ (and the moduli
of $\Sigma_\ell$ and $\Sigma_r$) 
fixed, this will certainly annihilate the right hand side of (\ref{rydex}).  However, the $\pi_*$ operation was defined using the procedure of
\cite{DPHgold}, in which the super period matrix is kept fixed while integrating over the odd variables.  It was shown in \cite{more}, section 3.3,
that in the case of a $++$ degeneration in genus 2, 
it is not $\varepsilon$ but $\varepsilon+\alpha\beta$ that can be expressed in terms of the super period matrix.\footnote{\label{depe} This calculation
was performed in that reference using specific choices of the local parameters in the gluing relation (\ref{toddo}), and with those choices, the precise relation of $\varepsilon+\alpha\beta$
 to
the super period matrix was determined. Because of scale-invariance, the assertion that the combination of $\varepsilon$, $\alpha$, and $\beta$
that can be expressed in terms of the super period matrix is $\varepsilon+\alpha\beta$ does not depend on the choices of local parameters.}  
Thus, to implement the
$\pi_*$ operation, we should integrate over $\alpha$ and $\beta$ keeping fixed $\varepsilon+\alpha\beta$ or equivalently keeping
fixed 
\begin{equation}\label{cute}q=-(\varepsilon+\alpha\beta)^2.\end{equation} 
 It is convenient to express our results in terms of $q$ rather than its square root, because $q$ is a matrix
element of the super period matrix (with the normalizations used in \cite{more}, the off-diagonal matrix element of the super period matrix is
$\Omega_{\ell r}=2\pi q$), and because $q$ rather than its square root is the variable most similar to  the gluing parameter of an ordinary Riemann
surface; the last statement is visible in the first line of (\ref{toddo}), where $-\varepsilon^2$ takes the place of the usual bosonic
parameter $q$.

To integrate over $\alpha$ and $\beta$ with fixed $q$, we eliminate $\varepsilon$ in favor of $q$, giving
\begin{equation}\label{zed}[\d\alpha]\cdot \frac{\d \varepsilon}{\varepsilon^2}\cdot [\d\beta]=[\d\alpha]\cdot \left(\frac{\d q}{q^2}\alpha\beta+\dots\right)\cdot
[\d\beta],\end{equation}
where a term independent of $\alpha$ and $\beta$ has been dropped.  Integration over $\alpha$ and $\beta$ with fixed $q$ maps this
to $\d q/q^2$, and accordingly (\ref{rydex}) implies that
\begin{equation}\label{fixly}\pi_*(\Psi_{2,+})\sim \Psi_{1,+}^{\d a}\cdot \frac{\d q}{q^2}\cdot \Psi_{1,+}^{\d b}. \end{equation}
The double pole for $q\to 0$ was originally described in eqn. (10.4) of \cite{DPHgold}.

What is the analog of this for a $--$ degeneration?
To go from (\ref{pyrdex}) to the asymptotic behavior of $\pi_*(\Psi_{2,+})$ at a $--$ degeneration, we again need to know the
asymptotic behavior of the super period matrix, so as to determine what to hold fixed for $\varepsilon\to 0$.  We will show
in section  \ref{zenot} that the  off-diagonal part of the super period matrix at a $--$ degeneration in genus 2 is proportional to 
\begin{equation}\label{whiff}q\sim-(\varepsilon+C\alpha\beta/\varepsilon^2)^2=-\varepsilon^2-\frac{2C\alpha\beta}{\varepsilon},
\end{equation} where here the constant $C$ does depend on the local parameters, in contrast
to the remark in footnote \ref{depe}.  
This leads to
\begin{equation}\label{thiff}[\d\alpha]\cdot \d\varepsilon\,\varepsilon^8\cdot [\d\beta]\sim [\d\alpha]\cdot \left(4C\alpha\beta\,\d q\,q^2+\dots\right)\cdot
[\d\beta] \end{equation}
where the omitted term is independent of $\alpha$ and $\beta$. 
So (\ref{pyrdex}) implies that at a $--$ degeneration
\begin{equation}\label{ghiff}\pi_*(\Psi_{2,+})\sim \Psi_{1,-}^{[\d a|\d\alpha]}(\U)\cdot 4C \d q\cdot q^2 \cdot \Psi_{1,-}^{[\d b|\d\beta]}(\U).\end{equation}
The $\d q\cdot q^2$ behavior was again originally found in eqn. (10.4) of \cite{DPHgold}.  

\subsection{Super Period Matrix At A Separating Degeneration Of Type $--$}\label{zenot}

We claimed in eqn. (\ref{whiff}) that near a separating degeneration of type $--$ of a genus 2 super Riemann surface, 
the off-diagonal matrix element of the super 
period matrix has a contribution proportional to $\alpha\beta/\varepsilon$.  We will deduce this claim from 
the general formula  for the dependence of the super period matrix of a super Riemann surface on odd moduli. 
(This formula was obtained in \cite{DPHtwo}; see 
section 8.3 of \cite{wittentwo} for a recent explanation.)
We will explain this formula for a case such as the present of varying only 2 odd moduli. 
  Let $\omega_i$, $i=1,\dots,g$
be holomorphic differentials on a genus $g$ Riemann surface $\SIgma_0$, which is the reduced space of a split super Riemann
surface $\Sigma$.  We want to deform $\Sigma$ by turning on odd moduli.   We take the gravitino field to be
\begin{equation}\label{comedy}\chi_{\t z}^\theta=\sum_{s=1}^2 \eta_s f^\theta_{s\,\t z}, \end{equation}
where $\eta_s$, $s=1,2 $ are the moduli and $f^\theta_{s\,\t z}$ are $c$-number gravitino wavefunctions.  
The difference between the ordinary period matrix $\Omega_{ij}$ of $\Sigma_0$ and the super period matrix $\h\Omega_{ij}$ of $\Sigma$ is then
\begin{equation}\label{menilo}\h\Omega_{ij}-\Omega_{ij}=-\frac{1}{2\pi}\sum_{s,t=1}^2 \eta_s\eta_t \int_{\Sigma_0\times \Sigma_0'} \omega_i(z)
f_{s\,\t z}^\theta(\t z;z)\d \t z\,S(z,z') f_{t\,\t z'}^\theta(\t z';z')\d\t z' \omega_j(z'),\end{equation}
where the integral runs over a product $\Sigma_0\times \Sigma_0'$ of two copies of $\Sigma_0$, and $S(z,z')$ is the Dirac propagator,
normalized to have a simple pole of residue 1 on the diagonal.  This formula has been used in \cite{more}, section 3.3.2, to analyze the behavior
of the super period matrix in a separating degeneration of $++$ type, and here we will determine what happens in the $--$ case.  

The Dirac propagator $S(z,z')$ only exists when the Dirac operator has no zero-modes, that is when $H^0(\Sigma_0,K^{1/2})=0$.
When this fails, $S(z,z')$ acquires a pole (as a function of the moduli parametrizing $\Sigma_0$) and eqn. (\ref{menilo}) then shows
that the super period matrix $\hat\Omega_{ij}$ 
likewise acquires a pole.  In genus $g>2$, there is a divisor $\eusm S$  in the moduli space $\M_{g,\spin+}$ on which
$H^0(\Sigma_0,K^{1/2})\not=0$ and the super period matrix has a pole.  In genus 2, a smooth Riemann surface $\Sigma_0$ with an 
even spin structure always has $H^0(\Sigma_0,K^{1/2})=0$,  but when $\Sigma_0$ degenerates to a pair of components $\Sigma_{0,\ell}$
and $\Sigma_{0,r}$, each with an odd spin structure, then there is a Dirac zero-mode on each component and hence two such modes on $\Sigma_0$.
Thus, the divisor in the Deligne-Mumford compactification of $\M_{g,\spin+}$  that parametrizes separating degenerations of type $--$ can be viewed as a component of $\eusm S$
at infinity.  The singular behavior that we are about to find in the super period matrix reflects this fact.  

Suppose that $\Sigma_0$ is a singular surface obtained by gluing together two components $\Sigma_{0,\ell}$ and $\Sigma_{0,r}$ at a point.  
Then its classical period
matrix is block-diagonal; if $\omega_\ell$ is a holomorphic differential supported on $\Sigma_{0,\ell}$ and $\omega_r$ is a holomorphic
differential supported on $\Sigma_{0,r}$, then $\Omega_{\ell r}=0$.  If we deform away from this singular situation, so that $\Sigma_0$
is described in local coordinates $\phi_\ell$, $\phi_r$ by the bosonic gluing relation $(\phi_\ell-a)(\phi_r-b)=-\varepsilon^2$ (which is the reduced
version of the super Riemann surface gluing relation (\ref{toddo}) with odd variables $\alpha,\beta$ set to 0),  then $\Omega_{\ell r}$
becomes nonzero and proportional to $\varepsilon^2$.  Extending this to the case of a super Riemann surface $\Sigma$ with reduced space $\Sigma_0$
and including $\alpha$ and $\beta$ 
 in the gluing, the behavior of  $\h\Omega_{\ell r}-\Omega_{\ell r}$ was computed for small $\varepsilon$ in the case of a $++$
degeneration in section 3.3.2 of \cite{more}.  Here we will extend this analysis to the $--$ case.  

We assume that one of the gravitino wavefunctions $f_{s\,\t z}^\theta$ is supported on $\Sigma_{0,\ell}$ and one on $\Sigma_{0,r}$;
let us call them $f_{\ell }$ and $f_r$, respectively, and write $\alpha$ and $\beta$ for the corresponding odd moduli.\footnote{For $g_\ell, g_r$ greater
than 1
(or equal to 1 in the case of an even spin structure),
we can take  $\alpha,\beta$ to be the usual odd moduli appearing in  eqn. (\ref{toddo}). For $g_\ell$ or $g_r$ equal to 1 with an odd spin  
structure, this parameter can be transformed away by an automorphism of the split super Riemann surface $\Sigma_\ell$ or $\SIgma_r$
whose reduced space is $\Sigma_{0,\ell}$ or $\Sigma_{0,r}$, but $\Sigma_{\ell}$ or $\SIgma_{r}$ can still be deformed by an odd modulus that we call
$\alpha$ or $\beta$.}
  Then eqn. (\ref{menilo}) becomes
\begin{equation}\label{enilo}\h\Omega_{\ell r}-\Omega_{\ell r}=-\frac{\alpha\beta}{\pi}\int_{\Sigma_{0,\ell}\times \SIgma_{0,r}}
\omega_\ell(z) f_{\ell \t z}^\theta(\t z,z) \d \t zS(z,z') f_{r\t z'}^\theta (\t z',z')\d\t z'\omega_r(z'). \end{equation}

To determine the small $\varepsilon$ behavior of (\ref{enilo}), we simply need to determine the small $\varepsilon$ behavior of $S(z,z')$ with
$z\in\Sigma_{0,\ell}$, $z'\in\Sigma_{0,r}$.  In \cite{more}, it is shown that for spin structures of type $++$, one has $S(z,z')\sim \varepsilon$.
This leads in (\ref{enilo}) to $\hat\Omega_{\ell r}-\Omega_{\ell r}\sim \alpha\beta\varepsilon$, which is an ingredient in showing that $\pi_*(\Psi_{2,+})\sim
\d q/q^2$ for $q\to 0$ at a $++$ degeneration.    By contrast, for $--$ spin structures, one has $S(z,z')\sim \varepsilon^{-1}$, which leads to (\ref{whiff})
and is a step in showing that $\pi_*(\Psi_{2,+})\sim \d q\, q^2$ at a $--$ degeneration.  

In the theory of a free holomorphic fermion field $\psi$ on a Riemann surface, let $\langle \W\rangle$ denote the path integral with insertion of
an operator $\W$:
\begin{equation}\label{tinox} \langle\W\rangle=\int \D \psi \,\exp(-I)\cdot \W. \end{equation}
We call this an unnormalized path integral.  The expectation value of $\W$, which we denote $\langle \W\rangle_N$, is given by a normalized
path integral, or in other words a ratio of two path integrals
\begin{equation}\label{inox}\langle \W\rangle_N=\frac{\langle\W\rangle}{\langle 1\rangle}. \end{equation}
The Dirac propagator $S(z,z')$ is the expectation value $S(z,z')=\langle \psi(z)\psi(z')\rangle_N$ or in other words
\begin{equation}\label{pnox} S(z,z')=\frac{\langle\psi(z)\psi(z')\rangle}{\langle 1\rangle}.\end{equation}
Consider
an  unnormalized path integral on $\Sigma_0$, near a degeneration at which $\Sigma_0$ is built by gluing a point $a\in \Sigma_{0,\ell}$
to a point $b\in \SIgma_{0,r}$.
The small $\varepsilon$ behavior of such a path integral
 is given by a sum over contributions of states propagating between the two components.  The contribution of a state of
given $L_0$ is obtained by inserting a vertex operator $\O(a)$ on $\Sigma_{0,\ell}$ and  a conjugate
operator $\h\O(b)$ on $\SIgma_{0,r}$, and multiplying by $\varepsilon^{2L_0}$.  The small $\varepsilon$ behavior of an unnormalized
path integral is thus determined, in the absence of cancellations, by the operator of smallest $L_0$ whose contribution is nonvanishing. (There will
be no cancellations in our problem as the pertinent operators of lowest dimension will be unique.)

To implement this in our context, we just need to know that on a Riemann surface with even spin structure, the unnormalized
path integrals $\langle 1\rangle$ and $\langle \psi(z_1)\psi(z_2)\rangle$ are generically nonzero, but $\langle \psi\rangle=0$, and that conversely on a Riemann surface with odd spin structure, $\langle 1\rangle=\langle\psi(z_1)\psi(z_2)\rangle=0$ but $\langle\psi\rangle\not=0$.
Given this, the operator of lowest dimension contributing to the denominator in (\ref{pnox}) is the identity, of $L_0=0$,
 in the case of a $++$ degeneration,
but is $\psi$, of $L_0=1/2$, in the case of a $--$ degeneration.  Accordingly, 
\begin{equation}\label{tocc}\langle 1\rangle_{\SIgma_0} \sim \begin{cases} \langle 1\rangle_{\SIgma_{0,\ell}}\;\langle 1\rangle_{\SIgma_{0,r}}
& \mbox{$++$ degeneration}\\ \varepsilon\langle \psi(a)\rangle_{\SIgma_{0,\ell}}\;\langle \psi(b)\rangle_{\Sigma_{0,r}}&\mbox {$--$ degeneration}.\end{cases}\end{equation}
The operators of lowest dimension contributing to the numerator in the same formula are reversed:
\begin{equation}\label{tocco}\langle \psi(z)\psi(z')\rangle_{\SIgma_0} \sim \begin{cases} \varepsilon\langle\psi(z)\psi(a) \rangle_{\SIgma_{0,\ell}}\;\langle \psi(b)\psi(z')\rangle_{\SIgma_{0,r}}
& \mbox{$++$ degeneration}\\ \langle \psi(z)\rangle_{\SIgma_{0,\ell}}\;\langle \psi(z')\rangle_{\Sigma_{0,r}}&\mbox {$--$ degeneration}.\end{cases}\end{equation}
From these statements, it follows that $S(z,z')\sim \varepsilon $ in the $++$ case, but $S(z,z')\sim \varepsilon^{-1}$ in the $--$ case, as promised.
In fact, more specifically,
\begin{equation}\label{occo}S(z,z')\sim \begin{cases} \varepsilon S(z,a)|_{\Sigma_{0,\ell}}S(b,z')|_{\Sigma_{0,r}} & \mbox{$++$ degeneration}\\
\varepsilon^{-1}\Huge{\frac{\langle\psi(z)\rangle_{\SIgma_{0,\ell}}}{\langle\psi(a)\rangle_{\SIgma_{0,\ell}}}
\frac{\langle\psi(z')\rangle_{\SIgma_{0,r}}}{\langle\psi(b)\rangle_{\SIgma_{0,r}}}}& \mbox{$--$ degeneration.} \end{cases}\end{equation}

\subsection{The Superstring Vacuum Amplitude In Genus 2}\label{evaluation}

Finally we want to study 
the  genus 2 superstring measure $\pi_*(\Psi_{2,+})$ from the standpoint of a separating degeneration.  
In eqn. (\ref{beffor}), we expressed this quantity in the form
\begin{equation}\label{neffori}\pi_*(\Psi_{2,+})=\voli \frac{Q(u_i,v_j)\cdot \d u_1\,\d u_2\,\d u_3\,\d v_1\,\d v_2\,\d v_3 }{\prod_{i<j}((u_i-u_j)(v_i-v_j))\prod_{k,l}(u_k-v_l)^2\cdot
(\d x /y \wedge x\d x/y)^5}, \end{equation} where we used $SL(2,\C)$ symmetry to show that
\begin{equation}\label{mizzo} Q=3 c_3(u)-c_2(u)c_1(v)+c_1(u)c_2(v)-3 c_3(v). \end{equation}
We want to show that these formulas agree with the expectation of eqn. (\ref{fixly}).

In (\ref{neffori}), the $u_i$ and $v_j$ are branch points of a genus 2 hyperelliptic Riemann surface $\Sigma_0$ (which we regard as the reduced
space of a super Riemann surface $\Sigma$):
\begin{equation}\label{humfly}y^2=\prod_{i=1}^3(x-u_i)\prod_{j=1}^3(x-v_j).  \end{equation}
The division of the branch points into $u$'s and $v$'s encodes an even spin structure on $\Sigma_0$.
For a separating degeneration, as in eqn. (\ref{elbo}), we keep 3 branch points fixed and let the others be of order $q^2$, with $q\to 0$.
For a $++$ degeneration, the 3 branch points that are kept fixed must be 2 $u$'s and 1 $v$, or vice-versa.  
So we will keep $u_1,u_2,v_1$ fixed, and take $(v_2,v_3,u_3)=(q^2v_2',q^2v_3',q^2u_3')$, with $v_2'$, $v_3'$, $u_3'$ fixed and $q\to 0$.  Thus the hyperelliptic equation will read
\begin{equation}\label{belm}y^2=(x-u_1)(x-u_2)(x-v_1)(x-q^2 v_2')(x-q^2v_3')(x-q^2u_3'). \end{equation} 
Parametrizing the moduli space by $v_1$, $u_3'$, and $q$, the  analog of (\ref{tinedo}) is
\begin{equation}\label{onel} \pi_*(\Psi_{2,+})\sim 2Q_*\frac{u_1u_2(u_1-u_2)\d v_1\cdot q^3\d q\cdot (v_2'-v_3')\d u_3'}{(u_1-u_2)(u_1-v_1)^2(u_2-v_1)^2
u_1^5u_2^5v_1^4  \cdot q^{10}(u_3'-v_2')^2(u_3'-v_3')^2(v_2'-v_3')(\d x/y\wedge x\d x/y)^5}.\end{equation}
Here $Q_*$ is a homogeneous cubic polynomial in $u_1,u_2,v_1$ obtained by restricting $Q$ to $u_3=v_2=v_3=0$.

As in section \ref{separhyp}, we can view the limit $q\to 0$ of the equation (\ref{belm}) in two ways.  Setting $y_\ell=y/x$, $x_\ell=x$, the equation becomes
\begin{equation}\label{trifo} y_\ell^2=x_\ell (x_\ell-u_1)(x_\ell-u_2)(x_\ell-v_1), \end{equation}
which describes a hyperelliptic curve $\Sigma_{0,\ell}$ of genus 1 with branch points $u_1,u_2,v_1$, and $v_2''=0$.  A local parameter on $\Sigma_{0,\ell}$
near the branch
point at 0 is
\begin{equation}\label{xn}\phi_\ell=\Delta^{-1/2} y_\ell,\end{equation}
where it is convenient to set
\begin{equation}\label{mn} \Delta=-u_1u_2v_1.\end{equation}
So
\begin{equation}\label{nurt}\phi_\ell^2\sim x_\ell,  ~~x_\ell\to 0.\end{equation}

Alternatively, setting $x=q^2x_r$, $y=\Delta^{1/2}q^3 y_r$, the equation becomes
\begin{equation}\label{murt}y_r^2=(x_r-u_3')(x_r-v_2')(x_r-v_3'), \end{equation}
which describes a hyperelliptic curve $\Sigma_{0,r}$ with branch points $u_3',v_2',v_3'$, and $u'_2=\infty$.  
The spin structures of both $\Sigma_{0,\ell}$ and $\SIgma_{0,r}$ are defined by the division of the branch points
into $u$'s and $v$'s.
For a local parameter on $\Sigma_{0,r}$ near $x_r=\infty$, we can take 
\begin{equation}\label{plir}\phi_r=x_r/y_r,\end{equation}
so that
\begin{equation}\label{lir}\phi_r^2\sim \frac{1}{x_r^2},~~x_r\to\infty.  \end{equation}
Just as in eqn. (\ref{zembo}), these definitions ensure that
\begin{equation}\label{plirm}\phi_\ell\phi_r=q,\end{equation}
so that $\Sigma_0$ is made by gluing $\Sigma_{0,\ell}$ and $\Sigma_{0,r}$ with this gluing law.
As in our previous analysis, for $q\to 0$, we can replace $\d x/y\wedge x\d x/y$ with $\Delta^{-1/2}q^{-1} \d x_\ell/y_\ell\wedge \d x_r/y_r$.
Substituting these formulas in eqn. (\ref{onel}), we arrive at the analog of eqn. (\ref{inedo}):
\begin{align}\label{plinedo}\pi_*(\Psi_{2,+})\sim 2&\frac{Q_*}{\Delta}\frac{u_1u_2(u_1-u_2)\d v_1} 
{(u_1-u_2)(u_1-v_1)^2(u_2-v_1)^2u_1^2u_2^2v_1 (\d x_\ell/y_\ell)^5}\cr\cdot &\Delta^{1/2} \frac{\d q}{q^2} \cdot \frac{(v_2'-v_3')\d u_3'}{(u_3'-v_2')^2(u_3'-v_3')^2(v_2'-v_3')(\d x_r/y_r)^5}. \end{align}

This formula is extremely similar to eqn. (\ref{inedo}) except for the factor of $Q_*/\Delta$ in front, and at this stage perhaps it is not surprising that consistency with
eqn. (\ref{fixly}) will tell us that $Q_*/\Delta$ must be constant.   We compute $\Psi_{1,+}^{\d a}$ and $\Psi_{1,+}^{\d b}$ by the same steps
that we used to compute $\Phi_{1}^{\d a}$ and $\Phi_1^{\d b}$ in section \ref{compar}.  
Moreover, the results are fairly obvious analogs of eqns. (\ref{whox}) and (\ref{lodidox}). 
From (\ref{yombix}), applied to the curve $\Sigma_{0,\ell}$, with the moduli space parametrized by $v_1$ and with one factor of $\d x_\ell/y_\ell$ inverted and
moved to the numerator, we have
\begin{equation}\label{phex}\Psi_{1,+} = \frac{u_1u_2(u_1-u_2)\d v_1\cdot (y_\ell\partial x_{\ell})}{(u_1-u_2)(u_1-v_1)^2(u_2-v_1)^2u_1^2u_2^2v_1
(\d x_\ell/y_\ell)^5}.\end{equation}
Given this, the same computation that led to (\ref{whox}) leads to 
\begin{equation}\label{phexo}\Psi_{1,+}^{\d a} = \frac{\Delta^{1/2}}{2}\frac{u_1u_2(u_1-u_2)\d v_1}{(u_1-u_2)(u_1-v_1)^2(u_2-v_1)^2u_1^2u_2^2v_1
(\d x_\ell/y_\ell)^5}.\end{equation}
To apply (\ref{yombix}) to $\Sigma_{0,r}$, we first have to slightly generalize (\ref{yombix}) to allow for the case that one of the branch points is at infinity.
This is done as in the derivation of (\ref{loddox}), and the analog of that formula, with the moduli space parametrized by $u_3'$, is
\begin{equation}\label{flexo} \Psi_{1,+}=\frac{(v_2'-v_3')\d u_3' \cdot (y_r\partial x_r)}{(v_2'-u_3')^2(v_3'-u_3')^2(v_2'-v_3')\cdot (\d x_r/y_r)^5}. \end{equation}
The analog of (\ref{lodidox}) is then
\begin{equation}\label{gecko}\Psi_{1,+}^{\d b}=-\frac{1}{2}\frac{(v_2'-v_3')\d u_3'}{   (v_2'-u_3')^2(v_3'-u_3')^2(v_2'-v_3')    \cdot (\d x_r/y_r)^5}. \end{equation}

With the help of these formulas, 
the comparison of (\ref{plinedo}) to (\ref{fixly}) does  tell us that $Q_*/\Delta$ is constant, which indeed follows from the expression (\ref{mizzo}) for $Q$.                                    

\section{Superstring Amplitude At A Nonseparating Degeneration}\label{nonseparating}

\subsection{Overview}\label{overview}
Here we will return\footnote{The results that we will be explaining were first found
 in eqn. (10.5) of \cite{DPHgold}. } to a point raised in section \ref{eventwo}.   In going from eqn. (\ref{teffora}) to eqn. (\ref{beffor}), we assumed that the
$\pi_*$ operation does not affect the singular behavior at a nonseparating degeneration, when 2 branch points collide.  In more detail, a nonseparating
degeneration of NS type (the Ramond case is discussed in section \ref{ramondnon}) corresponds to $u_i\to v_j$ for some $i,j$, and  the singular behavior
of (\ref{beffor}), if we let $u_i$ vary and keep $v_j$ fixed, is $\pi_*(\Psi_{2,+})\sim \d u_i/(u_i-v_j)^2$.  With $\varepsilon\sim u_i-v_j$, this
 agrees with the general behavior
$\Psi_{g} \sim \d \varepsilon/\varepsilon^2$ at a nonseparating degeneration of NS type, as described in section \ref{nonsepan},
and means -- assuming that (\ref{beffor}) is correct -- that integration over the fibers of $\pi:\MM_{2,+}\to \M_{2,\spin+}$ has not changed the singular behavior for $\varepsilon\to 0$. 

By contrast, at a separating degeneration, integration over the fibers of $\pi$ does change
the singular behavior, as we explained in section \ref{elim}.  The reason for the difference is that the projection $\pi$ is better behaved at 
a nonseparating degeneration than at a separating one.

\def\fD{{\mathfrak D}}
Given the holomorphic projection $\pi:\MM_{2,+}\to \M_{2,\spin+}$, we may ask whether $\pi$ extends to a projection  $\hat\pi:\hat\MM_{2,+}\to \hat\M_{2,\spin+}$,
where $\hat \MM_2$ and $\hat\M_{2,\spin}$ are the corresponding Deligne-Mumford compactifications.  This question should be refined in several ways.
First of all, the compactification of $\hat\M_{2,\spin+}$ is achieved by adding several divisors $\D_\sigma$, which are the reduced spaces of divisors $\fD_\sigma\subset\hat\MM_{2,+}$.  The question about whether $\pi$ extends should be asked separately for each of the $\fD_\sigma$.  Moreover, for each $\sigma$, there are really two
versions of this question:

(1)  Does the projection $\pi:\MM_{2,+}\to \M_{2,\spin+}$ extend over a partial compactification of $\MM_{2,+}$ in which a given divisor at infinity $\fD_\sigma$
is included?

(2) If so, does the extended map restrict to a projection $\pi_\sigma:\fD_\sigma\to \D_\sigma$ from $\fD_\sigma$ to its reduced space $\D_\sigma$?

A ``yes'' answer to the first question means that if $f$ is a local holomorphic function on a suitable partial compactification of $\M_{2,\spin+}$,
then $\pi^*(f)$ is a local holomorphic function on the corresponding partial compactification of $\MM_{2,+}$.  A ``yes'' answer to the second question
means that in addition, if $f$ is 0 when restricted to $\D_\sigma$, then $\pi^*f $ is 0 when restricted to $\fD_\sigma$.  
Here we can take $f$ to be any local holomorphic function on $\M_{2,\spin+}$. Since the genus 2 projection that we 
are studying is defined using the period matrix and super period matrix, it is convenient
to take $f$ to be a function of the matrix elements of the period matrix;  then $\pi^*f$ is the corresponding function of the matrix elements of the super
period matrix.  The important case turns out to be the case that $f$ is a function with a simple zero along $\D_\sigma$.

\def\sep{{\mathrm{sep}}}
The bosonic gluing parameter $q$ at a separating degeneration of a genus 2 surface $\Sigma$ splitting into genus 1
components $\Sigma_\ell$ and $\Sigma_r$ is, up to a constant factor, the off-diagonal matrix element $\Omega_{\ell r}$ of the period matrix.
(For a detailed explanation, see section 3.3 of \cite{more}.)
However, if $\Sigma$ is endowed with a spin structure, then the parameter with a simple zero along $\D_\sigma$ is not $q$ but its square root.
The square root enters because one needs to pick a square root of $q$ in order to define a gluing law for the spin structures.  (Thus the parameter
$q$ in the bosonic gluing law $\phi_\ell\phi_r=q$ is replaced by $-\varepsilon^2$ in the super extension (\ref{toddo}) of this gluing law.)  Accordingly,
we rewrite eqns. (\ref{cute}) and (\ref{whiff}) as formulas for the pullback under $\pi$ of $(-q)^{1/2}$, which has a simple zero along $\D_\sigma$.
For a separating degeneration of type $++$, we have 
\begin{equation}\label{zcute} \varepsilon+\alpha\beta =\pi^*((-q)^{1/2}),\end{equation}
and for a separating degeneration of type $--$, we have
\begin{equation}\label{zwhiff} \varepsilon+\frac{C\alpha\beta}{\varepsilon^2}=\pi^*((-q)^{1/2}).\end{equation}
In particular, in the $--$ case,  $\pi^*((-q)^{1/2})$ has a pole at $\varepsilon=0$, so the projection $\pi$ does not extend over the divisor $\fD_{\sep,--}$
that parametrizes degenerations of this type.  Thus for a separating degeneration of type $--$, 
the answer to question (1) is ``no.''   For the $++$ case, we see that $\pi^*((-q)^{1/2})$ is
holomorphic at $\varepsilon=0$ but is not equal to 0 when restricted to $\varepsilon=0$.  
So the answer to question (1) is ``yes,'' but the answer to question
(2) is ``no'': the projection $\pi$ extends over the divisor $\fD_{\sep,++}$ that parametrizes degenerations of this type, but this extension 
does not restrict to a projection of
$\fD_{\sep,++}$ to its reduced space.

For nonseparating degenerations, the answer to both questions is ``yes,'' as we will discuss in 
sections \ref{nsnon} and \ref{ramondnon}.  We explain at the end of
section \ref{nsnon} how this is related to the statement that $\pi_*(\Psi_{2,+})$ has the same behavior as $\Psi_{2.+}$ near a nonseparating degeneration.  
For nonseparating degenerations, question (2) has refined versions that will be explained in due course.

 \subsection{Some Preliminaries About Nonseparating Degenerations}\label{generalities}
 
We describe the reduced space $\Sigma_0$ of a genus 2 super Riemann surface by the familiar hyperelliptic equation:
 \begin{equation}\label{poco} y^2=\prod_{i=1}^3(x-u_i)\prod_{j=1}^3(x-v_j).  \end{equation}
$\Sigma_0$ undergoes a  nonseparating degeneration of NS type  if, for example, $u_3,v_3\to e$.  In the limiting case that $u_3,v_3=e$,
we set $y=(x-e)\t y$, and the equation becomes
\begin{equation}\label{noco}\t y^2=\prod_{i=1}^2(x-u_i)\prod_{j=1}^2(x-v_j).  \end{equation}
This defines a genus 1 Riemann surface $\Sigmax$ which is called the normalization of $\Sigma_0$; $\Sigma_0$ is built by gluing together the two points  $p,p'\in\Sigmax$  
that lie above the point $x=e$ in the $x$-plane. $ \Sigma_0$ can be smoothed by taking $u_3$ not quite equal to $v_3$; in this case, we set 
\begin{equation}\label{zobby}q=(u_3-v_3)^2,~~q^{1/2}=u_3-v_3.\end{equation}
The degeneration is naturally parametrized by $q$ if one forgets the spin structures or by $q^{1/2}$ if one takes the spin structures into account.

A basis of holomorphic differentials on $\Sigma_0$ is given by $\omega_1=\d x(x-e)/y$ and $\omega_2=\d x/y$.  $\omega_1$ is the pullback
from $\Sigmax$ of the holomorphic differential $\omega_1'=\d x/\t y$, but $\omega_2$ is the pullback from $\Sigmax$ of the differential
$\omega_2'=\d x/\t y(x-e)$, which has simple poles, with equal and opposite residues, at the two points  $p,p'\in\Sigmax$ lying above $x=e$.

\begin{figure}
 \begin{center}
   \includegraphics[width=1.5in]{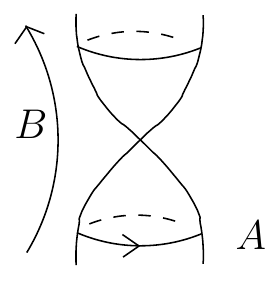}
 \end{center}
\caption{\small For a surface with a nonseparating degeneration, we take one $A$-cycle to wrap once around the singularity on one side,
and one $B$-cycle to pass through the singularity.}
 \label{zorki}
\end{figure}
To define the period matrix of $\Sigma_0$, we first introduce $A$- and $B$-cycles.
Since $\Sigma_0$ has genus 2, we need two $A$-cycles $A^1,A^2$ and two $B$-cycles $B_1,B_2$.
We take $A^1$ and $B_1$ to be the pullbacks of an $A$- and a $B$-cycle on $\Sigmax$.
We take $A^2$ to wrap once around the singular point $x=e$ on one of the two branches, while $B_2$ passes from one branch to the other
near the singular point (fig. \ref{zorki}).    We replace $\omega_1$ and $\omega_2$ by linear combinations $\omega_1^*$, $\omega_2^*$ such that
$\oint_{A^i}\omega_j^*=\delta^i_j$.  Then we define the period matrix by $\Omega_{ij}=\oint_{B_i}\omega^*_j$.

At $q=0$, it is convenient to regard the $\omega_j^*$ as forms on $\Sigmax$, possibly with poles at $p,p'$.
The condition that $\oint_{A^i}\omega_j^*=\delta^i_j$ implies that, for $q=0$, $\omega^*_2$ has  simple poles at $p,p'$, with residues
$1/2\pi i$ and $-1/2\pi i$ (the residues are equal and opposite since the sum of the residues vanishes), 
while $\omega_1^*$ has no such pole.  So (for $q=0$) $\omega^*_2\sim \pm (1/2\pi i)\d x/(x-e)$ near $x=e$, with opposite signs near $p$ and $p'$.
The integral defining $\Omega_{22}$ is divergent: $\Omega_{22}\sim 2\cdot (1/2\pi i)\int_e^\Lambda \d x/(x-e)$ (where a factor of 2 comes
because the two branches contribute equally;  the upper limit $\Lambda$ of the integral does not affect the divergence).  For $q\not=0$, this logarithmically
divergent integral is cut off at $x-e\sim q^{1/2}$ and thus
\begin{equation}\label{zelb}\Omega_{22}\sim \frac{\log q}{2\pi i}, \end{equation}
or equivalently,
\begin{equation}\label{nelb}q^{1/2}\sim \exp(\pi i\Omega_{22}).  \end{equation}
One may define $q$ so that this relationship is an equality.

\subsection{Nonseparating Degenerations Of NS Type}\label{nsnon}

Now let us regard $\Sigma_0$ as the reduced space of a super Riemann surface $\Sigma$ and deform by including odd moduli.  If $\Sigma$ has genus 2,
the odd moduli at a nonseparating degeneration are simply the odd parameters $\alpha,\beta$ in the gluing law (\ref{toddo}).  However, what
we are about to say applies equally in higher genus, in which case there are additional odd parameters.  The general formula for the dependence of the super period matrix $\hat\Omega$ on odd moduli was already given in
eqn. (\ref{menilo}) (for the case of 2 odd moduli\footnote{In genus greater than 2, there are more odd moduli so one
must consider higher order terms in the expansion \cite{DPHtwo} of the super period matrix.  They have the same nonsingular behavior that
we are about to describe, simply because the Dirac propagator has a limit at a nonseparating degeneration.}):  
\begin{equation}\label{zenilo}\h\Omega_{ij}-\Omega_{ij}=-\frac{1}{2\pi}\sum_{s,t=1}^2 \eta_s\eta_t \int_{\Sigma_0\times \Sigma_0'} \omega^*_i(z)
f_{s\,\t z}^\theta(\t z;z)\d \t z\,S(z,z') f_{t\,\t z'}^\theta(\t z';z')\d \t z' \omega^*_j(z'),\end{equation}  where $\Sigma_0\times \Sigma_0'$ is the product of two copies of $\Sigma_0$.
All we really need to know for our present purposes is that when an ordinary Riemann surface $\Sigma_0$  approaches a nonseparating degeneration of
NS type, the Dirac propagator $S(z,z')$ that appears in this formula approaches a limit\footnote{At a separating degeneration, rather than approaching a nonzero limit,
the Dirac propagator has matrix elements proportional to $\varepsilon$ or $\varepsilon^{-1}$, as in eqn. (\ref{occo}).} -- it approaches the Dirac propagator on $\Sigmax$
(the normalization of $\Sigma_0$).  So the difference between the super period matrix $\h\Omega$ and the ordinary period
matrix $\Omega$ has a limit for $q\to 0$, given by an integral on $\Sigmax\times \Sigmax'$ (the product of two copies of $\Sigmax$):
\begin{equation}\label{zenilod}\h\Omega_{ij}-\Omega_{ij}=-\frac{1}{2\pi}\sum_{s,t=1}^2 \eta_s\eta_t \int_{\Sigmax\times \Sigmax'} \omega^*_i(z)
f_{s\,\t z}^\theta(\t z;z)\d \t z\,S(z,z') f_{t\,\t z'}^\theta(\t z';z')\d\t z' \omega^*_j(z').\end{equation}
  It follows in  particular that
\begin{equation}\label{enilop}\exp(i\pi \h\Omega_{22})=\exp(i\pi \Omega_{22}) \cdot e^w, \end{equation}
where the function $w$ is holomorphic at $q=0$.  It will be important momentarily that $w$, in addition to being holomorphic at $q=0$, is actually
nonzero and has a nontrivial dependence on the odd gluing parameters.  All this follows from eqn. (\ref{zenilod}).  Indeed, for the case that the odd moduli
are the odd gluing parameters $\alpha,\beta$,  (\ref{zenilod}) can be explicitly evaluated and shown to be nonzero  by a calculation similar to that in section 3.3.2 of \cite{more}.
For this, take $f_{s\, \t z}^\theta=\partial_{\t z} f^\theta$, $f_{t\,\t z}^\theta=\partial_{\t z}f^{'\theta}$, where $f^\theta$ is nonzero at $p$ but vanishes at $p'$ and reciprocally $f^{'\theta}$ vanishes at $p$ but not at $p'$.
In other words, the $f$'s can be gauged away, but not by  gauge transformations that vanish at $p$ and $p'$; they represent the odd moduli 
associated to the choices of $p$ and $p'$.    Then setting $i=j=2$ in  (\ref{zenilod}), integrating by parts, and using the poles of $\omega_2^*$ at $p$ and $p'$, we find
\begin{equation}\label{zubz} \hat\Omega_{22}-\Omega_{22}=-\frac{\alpha\beta}{2\pi}\int_{\Sigmax\times \Sigmax'}
\omega_2^*(z)\partial_{\t z} f^\theta\d\t z \,S(z,z')\, \partial_{\t z'}f^{'\theta}\d \t z' \omega^*_2(z')=2\pi\alpha\beta  f^\theta(p)f^{'\theta}(p')S(p,p').\end{equation}

\def\nonsep{{\mathrm{nonsep}}}
When we interpret $\Sigma_0$ as the reduced space of a super Riemann surface $\Sigma$, we interpret $q=\exp(2\pi i \Omega_{22})$ as
$-\varepsilon^2$, where $\varepsilon$ is the gluing parameter whose vanishing defines the  divisor $\D_{\nonsep,\NS}\subset\M_{2,\spin+}$ that parametrizes
nonseparating degenerations of NS type.
So \begin{equation}\label{zorn}\varepsilon=\sqrt{-1}\exp(i\pi\Omega_{22}).\end{equation}  On the other hand, the projection $\pi:\MM_{2,+}\to\M_{2,\spin+}$ is defined to map a super Riemann
surface to an ordinary Riemann surface with the same period matrix, so in particular $\pi^*(\Omega_{22})=\h\Omega_{22}$ and
hence $\pi^*(\exp(i\pi\Omega_{22}))=\exp(i\pi\h\Omega_{22})$.  In view of (\ref{nelb}), the last statement is equivalent to  $\pi^*(q^{1/2})=\exp(i\pi\h\Omega_{22})$.  
Combining this with (\ref{enilop}) and (\ref{zorn}), we find \begin{equation}\label{fedo}\pi^*(q^{1/2})=\varepsilon\cdot  e^w/\sqrt{-1}. \end{equation}
Thus, $\pi^*(q^{1/2})$ is equal to $\varepsilon$ times an invertible holomorphic function.  In other words, the local parameter $q^{1/2}$ that has a simple
zero on  $\D_{\nonsep,\NS}$  pulls back to
a function -- namely $\varepsilon$ times the  invertible function  $e^w/\sqrt{-1}$ -- that has a simple zero on the corresponding divisor $\fD_{\nonsep,\NS}$ on the super moduli space $\MM_{2,+}$.  This assertion corresponds to ``yes'' answers to questions (1) and (2)
of section \ref{overview}.  

On the other hand, we can think of the nontrivial dependence of $\hat\Omega_{22}$ on $\alpha$ and $\beta$
as representing a ``no'' answer to a refined version of question (2).  To explain this refined version, observe first that for $\Sigma$ of genus 2,
$\Sigma^*$ has genus 1; since it is endowed with an even spin structure, it has no odd moduli and is automatically split.  Since $\Sigma^*$ is split,
there is a  natural projection $\pi_0:\fD_{\nonsep,\NS}\to \D_{\nonsep,\NS}$ that forgets the odd coordinates $\alpha,\beta$ of the punctures.  The
refined version of question (2) is this: (2$'$) Does the projection $\pi:\fD_{\nonsep,\NS}\to \D_{\nonsep,\NS}$ coincide with $\pi_0$? The answer to this refined question is ``no,'' because $\pi$ is defined to keep fixed the super period matrix, while $\pi_0$ keeps fixed the bosonic moduli of $\Sigma^*$;
these operations differ since the super period matrix depends non-trivially on $\alpha$ and $\beta$, as we saw explicitly in eqn. (\ref{zubz}).

 Now we can explain why integration over the fibers of $\pi$ does not affect the order of the singularity of the super Mumford form
along $\fD_{\nonsep,\NS}$.  Let us return to eqn. (\ref{upernon}), which expresses the genus 2 super Mumford form $\Psi_2$ near $\fD_{\nonsep,\NS}$ in
terms of the genus 1 super Mumford form $\Psi_1^{\d a}$ and the gluing parameters:
\begin{equation}\label{zupernon}\Psi_2\sim \Psi_1^{\d a} [\d\alpha]\frac{\d\varepsilon}{\varepsilon^2}[\d b|\d\beta]. \end{equation}
$\Psi_1^{\d a}$ depends only on the moduli of a genus 1 super Riemann surface $\Sigma^*$ that is the normalization of $\Sigma_0$, not on the gluing parameters
$\varepsilon,\alpha,\beta,b$.
If we integrate over the odd moduli $\alpha,\beta$, keeping fixed $\varepsilon$ (and $b$ and the moduli  of $\Sigma^*$), we get zero. But to evaluate $\pi_*(\Psi_{2,+})$,
we are supposed to integrate over $\alpha,\beta$ keeping fixed not $\varepsilon$ but $\varepsilon e^w$ (and $b$ and the moduli of $\Sigma^*$), where $w$ is holomorphic
and nonzero at $\varepsilon=0$ and proportional to $\alpha\beta$.  These conditions are enough to ensure that $\pi_*(\Psi_{2,+})\sim \d\varepsilon/\varepsilon^2$.  

In this derivation, a ``no'' answer to question (1) or (2) would have caused $\pi_*(\Psi_{2,+})$ to be more singular than
$\d\varepsilon/\varepsilon^2$ (similarly to what we found in section \ref{elim} for separating degenerations), and a ``yes'' answer to question
(2$'$) would have caused $\pi_*(\Psi_{2,+})$ to be less singular than $\d\varepsilon/\varepsilon^2$.  

 \subsection{Nonseparating Degenerations of Ramond Type}\label{ramondnon}
 
 The case of a nonseparating degeneration of Ramond type is qualitatively similar, but with many differences of detail that reflect the special nature
 of Ramond punctures.   Here we will give only an outline of the main points, beginning with a review of the relevant facts about Ramond
 punctures. (Background on Ramond punctures and the associated moduli
 spaces can be found in \cite{BS} and also in \cite{wittentwo}, especially sections 4 and 6.)
 
 A Ramond puncture on a super Riemann surface $\Sigma$ is really a Ramond divisor along which the superconformal structure of $\Sigma$
 is singular.  In the absence of a Ramond puncture, the superconformal structure of $\Sigma$ is defined  locally by
  an odd vector field $D$ (given up to multiplication
 by a scalar function)  with the property that $D$ and $D^2$
 are everywhere linearly independent, and thus furnish a basis of the tangent bundle $T\Sigma$.    A local model of a Ramond divisor
is given by local coordinates $x|\theta$ in which the superconformal structure is defined by
 \begin{equation}\label{mondo}D_\theta^*=\partial_\theta+\theta x\partial_x. \end{equation}
 Since $D_\theta^*{}^2=x\partial_x$, we see that the condition that $D_\theta^*$ and $D_\theta^*{}^2$ should be everywhere linearly independent
 fails precisely along the divisor $\F$ defined by $x=0$.  This is the Ramond divisor.\footnote{For understanding local properties,
 it is often  useful  to introduce a new double-valued coordinate
 $\h\theta=x^{1/2}\theta$  and so to put the superconformal structure in a standard form away from $x=0$, a 
 generator being $D_{\h\theta}=\partial_{\h\theta}+\h\theta\partial_x$, at the cost of introducing a square root branch point
 at $x=0$.  This is less useful for global questions, so we will not follow that route.}  
 
 An important detail is that because a Ramond puncture is a singularity in the superconformal structure of $\Sigma$, there is no notion of changing
 the position of a Ramond puncture without changing the other moduli of $\Sigma$.   So there is no natural notion of an integrated Ramond
 vertex operator and the basic formulas are best expressed in terms of 
 unintegrated ones.  A Ramond vertex operator of picture number $-1/2$
 (usually the most convenient case) is associated to a whole Ramond divisor $\F$, not to a point on $\F$.
 
 A nonseparating Ramond degeneration involves the gluing of two Ramond divisors.  
To give a local model of a Ramond degeneration, we start with two copies
 of $\C^{1|1}$, parametrized respectively by $x|\theta$ and $y|\psi$, and with the superconformal structures defined by $D_\theta^*=\partial_\theta+\theta x\partial_x$ and similarly by $D_\psi^*=\partial_\psi+\psi y\partial_y$.  We let $\F$ and $\F'$ be the Ramond divisors at $x=0$ and
 at $y=0$, respectively. The gluing of the two branches  is described by the equations
 \begin{align}\label{gluing}  xy&=q_\Ra\cr
                                           \theta&=\pm \sqrt{-1}\psi  ,   \end{align}
 which are the Ramond analogs of the NS sector gluing equations (\ref{toddo}).                                          
 To be more precise, if $q_\Ra=0$, these equations describe a simple gluing of the two copies of $\C^{1|1}$, by gluing $\F$ to $\F'$.  For $q_\Ra\not=0$,
 they describe a deformation of the singular glued surface 
 to a smooth, irreducible super Riemann surface (without Ramond divisors).  The factor $\pm \sqrt{-1}$ in the second equation ensures
 that the gluing preserves the superconformal structure (because $D_\psi^*$ is a multiple of $D_\theta^*$); the sum over the sign in this formula leads to the GSO projection on the string state
 propagating through the singularity. 
 
 Suppose that $\Sigma$ is a genus $g$ super Riemann surface (without Ramond punctures)
 that undergoes such a Ramond degeneration.  The normalization 
 $\Sigma^*$ of $\Sigma$
 is then a genus $g-1$ super Riemann surface with 2 Ramond punctures that could be glued together to make $\Sigma$.   A genus $g$ super Riemann surface with $2k$ Ramond punctures
 has $2g-2+k$ odd moduli, so $\Sigma$ has $2g-2$ odd moduli and $\Sigma^*$ has only $2g-3$ odd moduli.  Where is the missing odd modulus?
 The answer is that it is contained in the way the two Ramond divisors in $\Sigma^*$ are glued to make $\Sigma$.  At $q_\Ra=0$, we can generalize
 the second of eqn. (\ref{gluing}) to 
 \begin{equation}\label{thog}\theta=\zeta\pm \sqrt{-1}\psi,\end{equation} where we call $\zeta$ the fermionic gluing parameter.   In the local model
 (\ref{gluing}), the fermionic gluing parameter can be transformed away by redefining the coordinates, but globally it is a modulus of $\Sigma$.
 This modulus  is GSO-odd; indeed, the GSO projection comes from a sum over the sign of $\theta$ relative to $\psi$, and a reversal
 of this sign also changes the sign of $\zeta$.
 
 The fermionic gluing parameter $\zeta$ plays an important qualitative role in string theory (see for example section 6 of \cite{Revisited}).  
 For brevity we state the following in terms of open strings
 or a chiral sector of closed strings.  The usual propagator of a bosonic string or of a superstring in the NS sector is $1/L_0$.  This comes from integration over the gluing parameter $q$ of bosonic string theory, or its superstring analogs  $-\varepsilon^2$ in the NS sector  or
 $q_\Ra$ in the Ramond sector.  The field theory limit of $1/L_0$ is a conventional boson propagator $1/(p^2+M^2)$.  In the Ramond sector, 
 integration over the fermionic gluing parameter acts as $G_0$ on the propagating string
 state (here $G_0$ is the global supersymmetry generator of the string, whose field theory limit is the Dirac operator), 
 and converts the propagator from $1/L_0$ to $G_0/L_0=1/G_0$, whose
 field theory limit is a conventional fermion propagator. 
 
 The divisor $\fD_{\nonsep,\Ra}$ that parametrizes a nonseparating Ramond degeneration of a super Riemann surface is defined by $q_\Ra=0$
 in (\ref{gluing}), and in particular the fermionic gluing parameter $\zeta$ is one of the odd moduli of $\fD_{\nonsep,\Ra}$.  On the other hand,
 $\zeta$ can be changed without changing the normalization $\Sigma^*$ of $\Sigma$; it only enters when the Ramond divisors in $\Sigma^*$ are
 glued back to make $\Sigma$.  So there is a fibration
 \begin{equation}\label{zonely}\begin{matrix} \C^{0|1} & \to & \fD_{\nonsep,\Ra}\cr &&\downarrow \cr
 && \MM^*,\end{matrix}\end{equation}
 where the base space $\MM^*$ parametrizes the moduli of $\Sigma^*$, and 
  the fibers are copies of $\C^{0|1}$ parametrized by $\zeta$.

 Now let us discuss the behavior of the superstring measure $\Psi_g$ at a nonseparating
 Ramond degeneration.  Clearly, there will be no contribution from the
 identity operator flowing through the singularity; contributions can only come from Ramond-sector vertex operators.  The dominant contributions
 will come from  Ramond-sector superconformal primaries of lowest conformal
 dimension, and with zero momentum in spacetime.\footnote{The momentum of
 a string is associated to the motion of its center of mass, which is precisely the degree of freedom that cannot be simply expressed
 as a sum of holomorphic and antiholomorphic degrees of freedom.  In an approach based on the super Mumford form, the factors that conventionally arise from integration over the momentum of the string are derived from the sesquilinear form $\frak H$ of section \ref{superwhatfor}.}
 For strings in $\R^{10}$, the relevant operators are the fundamental spin fields  $\varXi_\alpha$, $\alpha=1,\dots,32$, which transform
 in the spin representation of $SO(10)$ \cite{FMS,K}.  (For reasons that will become evident, 
 we do not impose the GSO projection at this stage, and instead
 include spin fields of both positive and negative chirality.) 
 By $\varXi_\alpha$, we mean what in conventional language is the product of the spin field of the matter system (this operator carries the
 spinor index $\alpha$)
 and the spin field of the $\beta\gamma$
 ghost system, multiplied by the $c$ ghost since $\varXi_\alpha$ is supposed to be an unintegrated vertex operator.
$\varXi_\alpha$ is a superconformal primary of 
 dimension 0, with the $bc$ ghosts, $\beta\gamma$ ghosts, and matter fields contributing respectively $-1$, $3/8$, and $5/8$.
 
 The behavior of $\Psi_g$ at a Ramond degeneration is 
 \begin{equation}\label{unicorn}\Psi_g\sim \sum_{\alpha=1}^{32} \Psi_{g-1;2}(\varXi_\alpha,\varXi^\alpha) [\d\zeta]\frac{\d q_\Ra}{q_\Ra}.
   \end{equation}
 Here in general 
 $\Psi_{g;n_\Ra}(\varXi_{\alpha_1},\dots,\varXi_{\alpha_{n_\Ra}})$ is a holomorphic superstring measure for a genus $g$ super Riemann surface
 with $n_\Ra$ Ramond divisors and with the superconformal primary operators $\varXi_{\alpha_1},\dots,\varXi_{\alpha_{n_\Ra}}$ inserted at those
 divisors.  What appears in (\ref{unicorn}) is a special case with $n_\Ra=2$.
  $\Psi_{g;n_\Ra}(\varXi_{\alpha_1},\dots,\varXi_{\alpha_{n_\Ra}})$ can be understood as a generalized super Mumford form.  We give a short sketch of this in appendix \ref{superram}, but we will not need the details for our limited purposes here.
 Eqn. (\ref{unicorn}) is the Ramond sector analog of eqn. (\ref{supernon}) for a nonseparating degeneration of NS type, with a few differences
 that reflect the unusual properties of Ramond punctures.  The position parameters $a|\alpha$ and $b|\beta$ in (\ref{supernon}) have no
 Ramond sector analogs, because the Ramond sector formula is written in terms of unintegrated vertex operators $\varXi_\alpha$, but instead
 the Ramond formula has the fermionic gluing parameter $\zeta$.
 
 In (\ref{unicorn}), the expression $\sum_{\alpha=1}^{32} \varXi_\alpha\cdot \varXi^\alpha$ is constructed using the $SO(10)$-invariant
 inner product on the spinor representation of $SO(10)$.  This inner product is a pairing between spinors of opposite chirality, so if one of the two
 operators is GSO-even, the other is GSO-odd.  However, the expression on the right hand side of (\ref{unicorn}) is GSO-even because the fermionic
 gluing parameter $\zeta$ is also GSO-odd.  
 
 Before explaining the implications of these facts for $\pi_*(\Psi_{2,+})$,  let us first recall what happens in a superficially similar problem of
 computing the pole in a scattering amplitude due to an almost on-shell Ramond sector state.  In that case, the pole comes from $q_\Ra=0$
 and we represent its  residue 
 as an integral over the divisor $\fD_{\nonsep,\Ra}$.  We perform the integration by integrating first over the fibers of the fibration (\ref{zonely})
 or in other words by integrating over $\zeta$ while keeping fixed the moduli of the normalization $\Sigma^*$.  This has the effect of acting with $G_0$ on
 one or the other of the vertex operators on the two sides, after which the two operators
 are both GSO-even or both GSO-odd (and in effect the propagator is converted from $1/L_0$ to $G_0/L_0$).  The sum over the sign in the gluing law of eqn. (\ref{gluing}) or (\ref{thog}) projects onto the case that the
 two operators are both even.  
 
 A procedure like this will predict that integration over $\zeta$ will annihilate the right hand side of (\ref{unicorn}), because the operator
 $\varXi_\alpha$ is annihilated by $G_0$.  Indeed, for massless states at low energies, $G_0$ reduces to the Dirac operator $\gamma\cdot p$, 
 where $p$ is the momentum of the string state.  But the holomorphic spin fields $\varXi_\alpha$ have $p=0$.  The reason that nonetheless
 $\pi_*(\Psi_{2,+})$
 does have a pole at a nonseparating Ramond degeneration is that the $\pi_*$ operation is not defined by integrating over $\zeta$ keeping fixed the
 moduli of the normalization $\Sigma^*$.  Instead, we are supposed to integrate over $\zeta$ keeping fixed the super period matrix, which turns out
 to be a different procedure.   The last
 statement is analogous to the fact that, at a nonseparating NS degeneration, the super period matrix has a nontrivial dependence on $\alpha$
 and $\beta$ (for given moduli of $\Sigma^*$), as found in eqn. (\ref{zubz}).  
 
At a nonseparating Ramond degeneration in genus 2,
the answers to the two questions of section \ref{overview} 
 are both ``yes'': the super period
matrix is holomorphic along the divisor at infinity $\fD_{\nonsep,\Ra}$, and the projection $\pi$ defined by the super period matrix restricts
to a holomorphic projection from $\fD_{\nonsep,\Ra}$ to its reduced space $\D_{\nonsep,\Ra}$.  However, the answer to a refined version of
question (2) is ``no.''  To formulate this refined version, first observe that there is a natural projection $\pi_0:\fD_{\nonsep,\Ra}
\to \D_{\nonsep,\Ra}$ defined as the composition $\fD_{\nonsep,\Ra}\to \MM^*\to \D_{\nonsep,\Ra}$,
where the first map is the fibration of eqn. (\ref{zonely}), and the second is defined using the fact that  any supermanifold of odd dimension 1 
(such as $\Sigma^*$ for the case that $\Sigma$ has genus 2 and so $\Sigma^*$ has genus 1) has a unique projection to its reduced space.
The refined question is (2$'$): Does the projection $\pi:\fD_{\nonsep,\Ra}\to \D_{\nonsep,\Ra}$ coincide
with  $\pi_0$?

A ``no'' answer to question (1) or (2) 
would lead to $\pi_*(\Psi_{2,+})$ being more singular along $\D_{\nonsep,\Ra}$  than $\d q_\Ra/q_\Ra$,
similarly to what we explained in section \ref{elim} for separating degenerations.   A yes answer to those two questions and
also to question (2$'$) would lead to $\pi_*(\Psi_{2,+})$ being nonsingular along $\D_{\nonsep,\Ra}$, since the integral over $\zeta$ would
annihilate the right hand side of (\ref{unicorn}).  The actual behavior
$\pi_*(\Psi_{2,+})\sim \d q_\Ra/q_\Ra$, which has been assumed in writing (\ref{beffor}) 
(and demonstrated in \cite{DPHgold}), follows from ``yes'' answers to questions (1) and (2)
and a ``no'' answer to question (2$'$).   The reasoning here is the same as it was at the end of section \ref{nsnon}.

The calculations needed to answer questions (1), (2), and (2$'$) are also 
similar to what was explained in section
\ref{nsnon}.  We have to start with a split super Riemann surface of genus 2.  Turning on its odd moduli, the difference between the
super period matrix $\hat\Omega_{ij}$ and the ordinary period matrix $\Omega_{ij}$ is given in eqn. (\ref{zenilo}).  ``Yes'' answers to questions
(1) and (2) amount to the statement that $\hat\Omega_{ij}-\Omega_{ij}$ is holomorphic along $\D_{\nonsep,\Ra}$, in other words
that the right hand side of eqn. (\ref{zenilo}) has a limit at a nonseparating Ramond degeneration, along with general facts about nonseparating
degenerations that were explained in section \ref{generalities}.  A ``no'' answer to question (2$'$) amounts
to the statement that the limit  of eqn. (\ref{zenilo}) is nonzero and depends nontrivially on $\zeta$.  

The restriction of (\ref{zenilo}) to $\D_{\nonsep,\Ra}$ can be described as follows.  The objects $f_{s\,\t z}^\theta$ and
$f_{t\,\t z}^\theta$ become $(0,1)$-forms on $\Sigma^*$ with values in the sheaf of odd superconformal vector fields.  
Near a Ramond divisor $\F$ at which the superconformal structure is described in local coordinates $x|\theta$ by $D^*_\theta=\partial_\theta
+\theta\partial_x$,  an odd superconformal vector field takes the form $g(x)(\partial_\theta-\theta x\partial_x)$ for some function $g(x)$. 
Near a second Ramond divisor $\F'$ at which the local structure is similarly defined in local coordinates $y|\psi$ by $D_\psi^*=\partial_\psi
+\psi y\partial_y$, an odd superconformal vector field similarly takes the form $h(y)(\partial_\psi-\psi y\partial_y)$.  If $\F$ and $\F'$
are glued together to make a singular genus 2 surface $\Sigma$ by $\theta=\pm \sqrt{-1}\psi$ at $x=y=0$, then we must require 
\begin{equation}\label{zart}g(0)=\pm \sqrt{-1}h(0).\end{equation}
This defines the sheaf of odd superconformal vector fields over a split super Riemann surface $\Sigma$ that undergoes a nonseparating
Ramond degeneration, and hence explains what sort of objects are $f_{s\,\t z}^\theta$ and $f_{t\,\t z}^\theta$ of eqn. (\ref{zenilo}) in this
situation.   The limit of the Dirac propagator $S(z,z')$ in this situation can be described somewhat similarly.  The  fact that all the ingredients
in (\ref{zenilo}) have limits along $\D_{\nonsep,\Ra}$ is the essential reason for the ``yes'' answers to questions (1) and (2).  The ``no'' answer
to question (2$'$) comes from a calculation similar to that in eqn. (\ref{zubz}).  One uses the fact that the gluing parameter $\zeta$ corresponds
to a $(0,1)$-form $k_{\t z}^\theta$ valued in the sheaf of odd superconformal vector fields that is exact, $k_{\t z}^\theta=\partial_{\t z}k^\theta$
for some $k^\theta$,
but where $k^\theta$ does not obey the constraint (\ref{zart}), in the sense that $k^{\theta}|_{x=0}\not=\pm \sqrt{-1} k^\theta|_{y=0}$. (Thus $k^\theta$ makes sense as a smooth section of the sheaf of superconformal vector fields on $\SIgma^*$, but not after gluing
to make $\SIgma$. This condition on $k^\theta$
is analogous to the requirement in the discussion leading to eqn. (\ref{zubz}) that $f^\theta$ and $f^{'\theta}$ are nonzero at $p$ and $p'$,
respectively.)  Given this, (\ref{zenilo}) can be evaluated by an integration by parts analogous to that in (\ref{zubz}), and because of the potential
poles of $\omega_i^*$ and $\omega_j^*$, it does have a nontrivial
dependence on $\zeta$.

\vskip1cm

\appendix
\section{Mumford Isomorphism With A Marked Point}\label{telmo}

In this appendix, we will  discuss the analog of the Mumford isomorphism that arises if we replace $\M_g$ with $\M_{g,1}$, the moduli space of ordinary Riemann surfaces $\Sigma$
with 1 puncture $p\in\Sigma$. One could similarly include any number of punctures and one could likewise extend the following observations to NS punctures
on super
Riemann surfaces, though for brevity, we will consider only bosonic strings.  (For Ramond punctures on super Riemann surfaces, see appendix
\ref{superram}.)

We denote as $\pi:\M_{g,1}\to \M_g$ the natural projection
that forgets the marked point:
\begin{equation}\label{zombo}\begin{matrix} \Sigma& \to & \M_{g,1}\cr && \downarrow\pi \cr && \M_g . \end{matrix}\end{equation}
In a certain sense, the Mumford isomorphism says
nothing essentially new in this situation; the same information is simply packaged differently.  
The repackaging will give us a new perspective on the
 shift in exponent from 13 to 14 which occurs in genus 1, and which has a close analog for super Riemann surfaces.

The cotangent bundle $T^*\M_{g,1}$ to $\M_{g,1}$
consists of quadratic differentials on $\Sigma$ that may have a simple pole at $p$, so it is $H^0(\Sigma,K^2(p))$.  For $g\geq 1$, one has $H^1(\Sigma,K^2(p))=0$
so $\det H^*(K^2(p))\cong \det H^0(\Sigma,K^2(p))=\det T^*\M_{g,1}$.

We do not try to include the marked point $p$ in the definition of a holomorphic
differential (since in bosonic string theory, choosing a marked point does not give a pole to the matter fields), so in defining a  Mumford isomorphism for 
$\M_{g,1}$, we make use of $\det H^*(K)$, just as before.  However, another line bundle  over $\M_{g,1}$ is available, namely the line bundle
$\L_p\to \M_{g,1}$
whose fiber at a point corresponding to a given pair $\Sigma, p$ is $K|_p$, the fiber at $p$ of the canonical bundle $K=T^*\Sigma$.  
The Mumford isomorphism for $\M_{g,1}$ asserts the triviality of $\det T^*\M_{g,1}  \otimes \dets^{-13} H^0(\Sigma,K)\otimes \L_p^{-1}$.
We denote a trivialization as $\sPi_{g,1}$:
\begin{equation}\label{weldo} \sPi_{g,1}\in H^0(\M_{g,1},\det T^*\M_{g,1}\otimes \dets^{-13} H^0(\SIgma,K)\otimes \L_p^{-1}).\end{equation}
Because the line bundle $\L_p^{-1}$ does {\it not} have a natural hermitian metric, it is not possible to integrate the product $\bar\sPi_{g,1}\sPi_{g,1}$, as we did  in section \ref{what for} in the
case of the vacuum amplitude $\sPi_g$.
In string theory, if we were computing a 1-point function (or a more general scattering amplitude in the presence of several punctures), we would select a conformal primary field $\V$ of dimension 1.  Its expectation value $\langle \V\rangle$  would be a section of $\L_p$, so the product $\sPi_{g,1,\V}=\sPi_{g,1}\cdot \langle \V\rangle$
(which we can think of the path integral with an insertion of $\V$), would be a section  of $\det T^*\M_{g,1}\otimes \dets^{-13}H^0(\Sigma,K)$, just as in the absence
of the marked point.  So, just in section \ref{what for}, we can map $\bar\sPi_{g,1,\V}\sPi_{g,1,\V}$ to a top degree form on $\M_{g,1}$
that can be integrated, at least locally. (In bosonic string theory, we would face the usual infrared divergences in such an integral.) In eqn. (\ref{weldo}), as we have not introduced a conformal field $\V$, the operator inserted at $p$ is, in string theory
terms, the identity operator; that is why this formula gives no essentially new information.

To understand how the statement (\ref{weldo})  is related to the usual Mumford isomorphism, we start with the exact sequence of sheaves on $\SIgma$ (we 
identify a line bundle with its sheaf of sections):
\begin{equation}\label{eldo} 0 \to K^2\to K^2(p) \to{\mathcal K}|_p \to 0.\end{equation}
Here ${\mathcal K}|_p$ is the sheaf on $\Sigma$ associated to $K|_p$; it is defined by saying that its sections over an open set not containing $p$ vanish,
while its space of sections over an open set containing $p$ is  $K|_p$.
The map from $K^2(p)$ to ${\mathcal K}|_p$ is the residue map (a section of $K^2(p)$ is a quadratic differential
that may have a simple pole at $p$; its image in ${ K}|_p$ is the residue of the pole).  This leads to a long exact sequence in cohomology
\begin{equation}\label{keldo}0\to H^0(\SIgma,K^2)\to H^0(\SIgma,K^2(p))\to K|_p\to H^1(\Sigma,K^2)\to 0. \end{equation}
Here we identify $H^0(\Sigma,{\mathcal K}|_p)$ with $K_p$ and use the fact that $H^1(\Sigma,\mathcal K|_p)=0$ since $\mathcal K|_p$ is supported at a point, and the fact
that
$H^1(\Sigma,K^2(p))=0$ for all $g\geq 1$.

For  $g> 1$,  $H^1(\Sigma,K^2)=0$, and eqn. (\ref{keldo})
actually reduces to a short exact sequence:
\begin{equation}\label{okeldo}0\to H^0(\SIgma,K^2)\to H^0(\SIgma,K^2(p))\to K|_p\to 0. \end{equation}
Taking determinants, it follows that $\det H^*(\Sigma,K^2(p))\cong \det H^*(\Sigma,K^2)\otimes K|_p$.  In terms of line bundles over $\M_{g,1}$,
it follows that $\det H^*(K^2(p))\cong \det H^*(K^2)\otimes \L_p=\pi^*(\det T^*\M_g)\otimes \L_p$.  When this is incorporated in (\ref{weldo}),
$\L_p$ conveniently disappears, and we learn that  $\sPi_{g,1}$ is a trivialization of $\pi^*(\det T^*\M_g\otimes \dets^{-13}H^0(\Sigma,K))$.
  Comparing to (\ref{tobbo}),
we see that $\pi^*\sPi_g$ is a trivialization of the same line bundle, so we can fix the normalization of $\sPi_{g,1}$
such that $\sPi_{g,1}=\pi^*\sPi_g$.  In this sense, adding a marked point in the Mumford isomorphism does not give anything essentially new.

For $g=1$, $\M_{1,1}$ and $\M_1$ are the same space, since choosing a single marked point in a genus 1 surface does not add a modulus,
but instead removes the continuous automorphism group.  So $T^*\M_{1,1}=T^*\M_1=H^0(\Sigma,K^2)$.  The exact sequence (\ref{keldo}) for $g=1$
splits as a pair of isomorphisms, one between $H^0(\Sigma,K^2)$ and $H^0(\Sigma,K^2(p))$, and one between $K|_p$ and $H^1(\Sigma,K^2)$.
However, instead of using this directly, we can just go back to (\ref{weldo}) and observe that for genus 1, since $K$ is trivial,  a vector in the fiber $K|_p$ of $K$ at $p$ can
be extended in a unique way to a global holomorphic 1-form on $\Sigma$; this gives a natural isomorphism $K|_p\cong H^0(\SIgma,K)$, or in
 terms of line bundles over $\M_{1,1}$,  an isomorphism $\L_p\cong H^0(\SIgma,K)$.  Accordingly, for $g=1$, (\ref{weldo}) becomes
\begin{equation}\label{urto}\sPi_{1,1}\in H^0(\M_{1,1},T^*\M_{1,1}\otimes H^0(\Sigma,K)^{-14}),\end{equation}
again showing the shift in exponent.
Since $\M_{1,1}$ is the same as $\M_1$ and $T^*\M_{1,1}$ therefore coincides with $T^*\M_1$, we see on comparing to (\ref{obbo}) that
$\sPi_{1,1}$ and $\sPi_1$ are trivializations of the same line bundle.  We can choose the normalizations so that $\sPi_{1,1}=\sPi_1$.

 \section{Comparison To The Result Of D'Hoker and Phong}\label{comp}
 
 The genus 2 superstring measure $\pi_*(\Psi_{2,+})$ was first computed by D'Hoker and Phong in work that is surveyed in \cite{DPHgold}.
 Our aim here is to verify that our result agrees with their formula.
 
 One way that D'Hoker and Phong express their result is in terms of the ratio $\pi_*(\Psi_{2,+})/\Phi_2$ of the genus 2 superstring measure to the genus
 2 bosonic string measure.   Comparing the formulas  (\ref{relmit}) and   (\ref{beffor}), we see that in our notation this
 ratio is
 \begin{equation}\label{udof}\frac{\pi_*(\Psi_{2,+})}{\Phi_2}=\left(\d x/y\wedge x \d x/y\right)^8W(u,v)\end{equation}
 with
 \begin{equation}\label{nudof} W(u,v)=Q(u,v)\prod_{i<j}((u_i-u_j)^2(v_{i}-v_{j})^2)\prod_{k,l=1}^3 (u_k-v_l), \end{equation}
 where $Q(u,v)$ was given in eqn. (\ref{morzono}).
 Because of the factor   $\left(\d x/y\wedge x \d x/y\right)^8$ in (\ref{udof}), $W(u,v)$ is known as a modular form of weight 8.
 
 The formula given by D'Hoker and Phong for this object is
 \begin{equation}\label{wudof} W=\vartheta[\delta]^4\Xi_6[\delta], \end{equation}
 where $\delta$ is the spin structure associated with the division of the branch points into $u$'s and $v$'s, $\vartheta$ is the associated theta function, which
 transforms as a modular form of weight 1/2 
 so that $\vartheta[\delta]^4$ is a form of weight 2, and $\Xi_6[\delta]$ is a form of weight 6 that will be described momentarily.  
 In terms of the branch points $u_1,u_2,u_3$ and $v_1,v_2,v_3$, one has
 \begin{equation}\label{sudof}\vartheta[\delta]^4=(u_1-u_2)(u_2-u_3)(u_3-u_1)(v_1-v_2)(v_2-v_3)(v_3-v_1). \end{equation}
 Notice that $\vartheta[\delta]^4$ depends not just on the even spin structure $\delta$, but also on the cyclic orderings of the three $u$'s and of the three $v$'s.
 $\Xi_6$ has a similar property, so that $W(u,v)$ does not require any cyclic ordering.  
Like $W$ and $\vartheta[\delta]^4$, $\Xi_6$ will be invariant under exchange of all $u$'s and $v$'s.  

To define the function $\Xi_6[\delta]$, we need theta functions for the other even spin structures on $\Sigma$.  Any even spin structure other than $\delta$
is defined by removing one of the three $u$'s and adding one of the three $v$'s in its place.  For example, we could remove $u_1$ from the set
$\{u_1,u_2,u_3\}$ and replace it with $v_2$, giving a subset $\{u_1',u_2',u_3'\}=\{v_2,u_2,u_3\}$ (and a complementary set $\{v_1',v_2',v_3'\}=\{v_1,u_1,v_3\}$).
Let us write $\delta(i;j)$, $i,j=1,\dots,3$ for the new spin structure obtained by removing $u_i$ from the set $\{u_1,u_2,u_3\}$ and substituting $v_j$ in its place (and
likewise removing $v_j$ from the set $\{v_1,v_2,v_3\}$ and inserting $u_i$ instead).  When we remove a variable  from the set $\{u_1,u_2,u_3\}$ or
$\{v_1,v_2,v_3\}$ and insert a new variable instead, we insert the new variable at the place in the cyclic order formerly occupied by the variable that has been removed.
So each triple always has a cyclic order. Given this, we can use the same formula (\ref{sudof}) to define a function $\vartheta[\delta(i;j)]^4$ for each spin structure
$\delta(i;j)$.  

With this understood, we can define $\Xi_6$ by
\begin{equation}\label{zelbo}\Xi_6(u,v)=\sum_{i=1}^3\prod_{k=1}^3\vartheta[\delta(i;k)]^4, \end{equation}
which is equivalent to eqn. (8.8) of \cite{DPHgold}, though expressed slightly differently. $\Xi_6(u,v)$  has manifest symmetry under permutations of the $u$'s or $v$'s,
and is also invariant under exchange of the $u$'s and $v$'s (this is
more obvious in an alternative formula given in eqn. (8.9) of \cite{DPHgold}).
An elementary manipulation leads to
\begin{equation}\label{elbon}\Xi_6(u,v)= \sum_{i=1}^3(u_{i+1}-u_{i-1})^3\prod_{k=1}^3(u_i-v_k) \prod_{l,m=1}^3(u_l-v_m) \cdot (v_1-v_2)(v_2-v_3)(v_3-v_1).    \end{equation}
Now with the help of (\ref{sudof}) and  (\ref{elbon}), we see that to justify the claim $W=\vartheta[\delta]^4\Xi_6$, we need to show that
\begin{equation}\label{mexo}(u_1-u_2)(u_2-u_3)(u_3-u_1)Q(u,v)=\sum_{i=1}^3(u_{i+1}-u_{i-1})^3\prod_{k=1}^3(u_i-v_k). \end{equation}
The right hand side of (\ref{mexo}) vanishes if any two of the $u$'s are equal, so it is $(u_1-u_2)(u_2-u_3)(u_3-u_1)\hat Q(u,v)$, where $\hat Q(u,v)$
is a homogeneous cubic polynomial.  So to justify the desired formula, we just need to show that $\hat Q(u,v)=Q(u,v)$.  With computer
algebra, this is a very short exercise. 
To show it by hand, first observe that $\hat Q$ has the same symmetries as $Q$.  Symmetry under permutation of the $u$'s or the $v$'s is obvious, as is
translation invariance (that is, the condition (\ref{guelf})).  That  $\hat Q(u,v)$ is odd under the exchange of all $u$'s and $v$'s is less obvious from the formula
(\ref{mexo}), but follows from the fact that $\Xi_6$ is even under this exchange.   From our discussion following eqn. (\ref{donkeys}), we know that a homogeneous
cubic polynomial $\hat Q(u,v)$ with these symmetries is completely determined by its restriction to $v=0$, so we only need to show that $\hat Q(u,0)=Q(u,0)$, which is
a short exercise.
\vskip 1cm

\section{Super Mumford Form With Ramond Punctures}\label{superram}

\def\1{{\bf 1}}
\def\5{{\bf 5}}
\def\zone{{\bf {10}}}
Here we will give just a brief indication  of the generalization of the 
definition of the super Mumford form in the presence of Ramond punctures.

First of all, let $\Sigma$ be any complex supermanifold of dimension $1|1$, not necessarily a super Riemann surface.
Over any such $\Sigma$, we have the line bundle $\BBer=\BBer(\Sigma)$, the Berezinian of the tangent bundle $T\Sigma$,
and hence for any integer $k$ we can define the cohomology groups $H^i(\Sigma,\BBer^k(\Sigma))$, $i=0,1$, and the Berezinian line
$\BBer_k=\BBer(H^*(\Sigma,\BBer^k(\Sigma)))$.   The construction in \cite{RSV}
exhibits a natural vector $\Psi\in \BBer_3\otimes \BBer_1^{-5}$.  If $\Sigma$ varies over some parameter space $B$, then
the $\BBer_k$ become holomorphic line bundles over $B$, and the construction in \cite{RSV} gives a natural trivialization $\Psi$
of $\BBer_3/\BBer_1^5$, which we call the super Mumford form.  No global information is needed
to define $\Psi$.

All this is for any complex supermanifold of dimension $1|1$.  What is special to super Riemann surfaces is that if $\Sigma$ is a super Riemann
surface of genus $g$, then 
$H^0(\Sigma,\BBer^3(\Sigma))$
is the fiber at the point corresponding to $\Sigma$ of $T^*\MM_g$, where $\MM_g$ is the moduli space of super Riemann surfaces
of genus $g$.  As explained in section \ref{superwhatfor}, this fact is important in the usefulness of the super Mumford form.

Now suppose that $\Sigma$ is a super Riemann surface of genus $g$ with $n_\Ra$ Ramond punctures, and let $\MM_{g;n_\Ra}$ be the moduli
space of such objects. It is not true that $H^0(\Sigma,\BBer^3)$
is the cotangent space to $\MM_{g;n_\Ra}$ at the point corresponding to $\Sigma$, 
but it turns out that nonetheless  $\BBer \,H^*(\Sigma,\BBer^3)$ can be naturally identified with
$\BBer \,T^*\MM_{g;n_\Ra}$.  To see this, let $\F=\sum_{i=1}^{n_\Ra}\F_i$ be the Ramond divisor in $\Sigma$, with irreducible components $\F_i$.
The fiber of the cotangent bundle to $\MM_{g;n_\Ra}$ at the point corresponding to $\Sigma$ is\footnote{This is essentially shown
in section 4.2 of \cite{wittentwo}.  In the notation used there, the sheaf of superconformal vector fields is $\S\cong \D^2\cong \BBer^{-2}(-2\F)$.
The tangent bundle to $\MM_{g;n_\Ra}$ is $H^1(\Sigma,\S)$ and by Serre duality, the corresponding cotangent bundle is $H^0(\SIgma,\BBer\otimes \S^{-1})
=H^0(\SIgma,\BBer^3(2\F))$. The notation $\BBer^3(2\F)$ denotes a line bundle whose sections are sections of $\BBer^3$ that are allowed
to have a double pole along $\F$.} $H^0(\Sigma,\BBer^3(2\F))$.  So the claim that we need is that there is a natural isomorphism
$\BBer(H^*(\BBer^3(2\F)))\cong \BBer(H^*(\BBer^3))$.  For this we look at the exact sequence 
\begin{equation}\label{zonk}0 \to \BBer^3\to \BBer^3(2\F)\to \J\to 0, \end{equation}
where $\J$ is a sheaf supported on $\F$; a section of $\J$ is the polar part along $\F$ of a section of $\BBer^3(2\F)$ (viewed as a section of 
$\BBer^3$ that may have a double pole along $\F$).  The exact sequence (\ref{zonk}) leads to an isomorphism $\BBer( H^*(\SIgma,\BBer^3(2\F)))
\cong \BBer(H^*(\Sigma,\BBer^3))\otimes \BBer(H^*(\Sigma,\J))$, so to get the isomorphism we want, we need a natural trivialization of 
$\BBer(H^*(\Sigma,\J))$.  Since $\J$ has its support on a subvariety of $\Sigma $ of bosonic dimension 0, we have $H^1(\SIgma,\J)=0$, and
$\BBer(H^*(\Sigma,\J))=\BBer(H^0(\SIgma,\J))=\otimes_{\sigma=1}^{n_\Ra}\BBer(H^0(\SIgma,\J_\sigma))$, where $\J_\sigma$ 
is the subsheaf of $\J$ supported
on $\F_\sigma$.  Near each $\F_\sigma$, we can pick local coordinates $z|\theta$ such that the superconformal structure of $\Sigma$ is generated
by $D_\theta^*=\partial_\theta+\theta z\partial_z$.  Such coordinates are not unique, but along the Ramond divisor at $z=0$, $\theta$ is unique up to
$\theta\to \pm\theta+\zeta$ (with $\zeta$ an odd constant), 
and accordingly $D_\theta^*$ is uniquely defined up to multiplication by a function $f(z|\theta)$ that at $z=0$
is equal to $\pm 1$.  If we use $D_\theta^*$ to map even sections of $\J_\sigma$ to odd ones, this gives a trivialization of 
$\BBer(H^0(\Sigma,\J_\sigma))$
that does not depend on the choice of coordinates, establishing the desired result.

So finally, we have shown that $\Psi$ can be regarded as a trivialization of  $\BBer(T^*\MM_{g;n_\Ra})\otimes \BBer_1^{-5}$.
Accordingly, it is reasonable to call it a super Mumford form $\Psi_{g;n_\Ra}$.

However, this appears to give one distinguished procedure by which to treat the Ramond punctures, while from superconformal
field theory, we know that the simplest operators that can be inserted at a Ramond puncture transform in the spinor representation of $SO(10)$.
To gain insight, one must bear in mind that the 
 super Mumford form of superstring theory is natural in a framework in which 
$\R^{10}$ is identified with $\C^5$.    This obscures the $SO(10)$ symmetry;
 the natural symmetry group of $\C^5$ is only $U(5)$, or $GL(5)$ from a holomorphic point of view, 
 or more precisely (since spinors are present) the double
cover of $GL(5)$ that embeds in the complex form of $\mathrm{Spin}(10)$.   Therefore, instead of looking for
 spin fields that transform under $SO(10)$ as a sum of spinor representations of positive or negative chirality, we should expect them to transform
under $GL(5)$ as a sum of six pieces corresponding to the exterior powers of the fundamental five-dimensional representation $V$ of $GL(5)$,
in fact as $\det(V)^{-1/2}\oplus_{j=0}^5\wedge^j V$, or in a different language as 
\begin{equation}\label{doofus}\1^{-5/2}\oplus \5^{-3/2}\oplus \zone^{-1/2}\oplus \bar{\zone}^{1/2}\oplus \bar\5^{3/2}\oplus \bar \1^{5/2},\end{equation}
where for $j=0,\dots,5$, $\wedge^j V$ is denoted in boldface by its dimension, the dual of a representation is indicated by a bar,
and 
the exponent indicates the action of the center of $GL(5)$. 

Furthermore, by Serre duality, 
$\BBer_1=\BBer\, H^*(\Sigma,\BBer)$
is naturally isomorphic\footnote{The indicated Berezinian lines 
 are isomorphic, rather than dual, because we consider $\BBer$ to be a line bundle with odd
fibers, while $\O$ has even fibers.  This compensates for the minus sign (in the exponents of the cohomology groups) that
comes from using Serre duality.} to  $\BBer \,H^*(\Sigma,\O)$, 
where $\O$ is the sheaf of holomorphic functions
over $\Sigma$.    We can characterize $\BBer \,H^*(\Sigma,\O)$ as the
Berezinian line of the $\t\partial $ operator (the super Riemann surface analog of the usual $\bar\partial$ operator) 
acting on sections of $\O$.  So $\BBer_1^5$ is the
Berezinian line of the $\tilde\partial$ operator acting on the direct sum of 5 copies of $\O$, or equivalently on $\O\otimes V$, where $V$ is a fixed
five-dimensional vector space.

In the presence of  Ramond
divisors, we can generalize the sheaf $\O\otimes V$ in the following way.  To each Ramond divisor $\F_\sigma$, $\sigma=1,\dots,n_\Ra$, 
attach an integer  $j_\sigma$
in the range $0\leq j_\sigma\leq 5$, and a vector space $V_\sigma\subset V$ of codimension $j_\sigma$ (and so dimension $5-j_\sigma$).  
Then  setting $\h V=\{V_1,\dots,V_{n_\Ra}\}$, define $\W_{\hat V}$ to be the sheaf
of sections $s$ of $\O\otimes V$ with the property that, for each $\sigma$, 
when $s$ is restricted to $\F_\sigma$, its derivative along the $\F_\sigma$ direction
takes values in $V_\sigma$.   In more detail, if $\F_\sigma$ is defined in local coordinates $z|\theta$
by the condition $z=0$, then we require that $\partial_\theta s\in V_\sigma$ at $z=0$.  To understand this condition intuitively, we can think of $s$
as a collection of 5 chiral superfields $X^i(z|\theta)=x^i(z)+\theta\psi^i(z)$.  The condition on $\partial_\theta s$ places no constraint on the bosonic
fields $x^i(z)$, but constrains the fermions so that $\psi^i(z)\in V_\sigma$ at $z=0$.  That is what one expects in the presence of a spin field:
the bosonic fields are unaffected, and some linear combinations of the fermi fields are constrained to 
vanish.\footnote{The fermi fields $\psi^i$ can be expressed as $z^{1/2}\h\psi^i$, where 
$\h\psi$ is a conventionally normalized fermi field with a square root branch
point near a spin field.  While each component of $\h\psi^i$ is of order $z^{\pm 1/2}$ near the spin field, each component of
$\psi$ is of order 1 or $z$.}

Now we can define $\BBer(H^*(\W_{\h V}))$
to be the Berezinian line of the cohomology of the $\t\partial$ operator acting on 
$\W_{\h V}$.  $\BBer(H^*(\W_{\hat V}))$ coincides with $\BBer^5 H^*(\O)\cong \BBer_1^{5}$
 if $V_\sigma=V$ for all $\sigma$, so that $\W_{\h V}=\O\otimes V$.
The appropriate generalization of the super Mumford isomorphism  is the statement that $\BBer_3\otimes \BBer^{-1}(H^*(\W_{\hat V}))$
is naturally isomorphic to $\otimes_{\sigma=1}^{n_\Ra}\det^{-1}\,(V/V_\sigma)$ (and thus in particular is trivial if we keep the $V_\sigma$ fixed while
$\Sigma$ varies). 
This can be proved with an exact sequence analogous to (\ref{zonk}).   So we can define a super Mumford form $\Psi_{\h V}$ that is a trivialization of
$\BBer_3\otimes \BBer^{-1}(H^*(\W_{\h V})).$  $\Psi_{\h V}$ is a super Mumford form appropriate for a certain product of spin fields
inserted at the Ramond divisors.

To understand more concretely what this construction means, let us return to the case 
that all $V_\sigma$ are equal to $V$, so that $\W_{\h V}$ is just the direct
sum of 5 copies of $\O$.  On a super Riemann surface $\Sigma$
with even spin structure and without Ramond punctures, generically $H^0(\Sigma,\O)$ has dimension $1|0$, generated by the constant function
1.  However, in the presence of $n_\Ra$ Ramond punctures, generically $H^0(\Sigma,\O)$ has dimension $1|n_\Ra/2$.  (The follows
from the way eqn. (\ref{zork}) is modified in the presence of Ramond punctures; $K^{1/2}$ is replaced by a line bundle of degree $g-1+n_\Ra/2$,
with generically $n_\Ra/2$ global holomorphic sections.
See for example section 4.2.2 of \cite{wittentwo}.)  Physically, this means that if $\W_{\h V}=\O\otimes V$, then generically the matter
fermions have  $5n_\Ra/2$ zero-modes.  We can define a super Mumford form in this situation, but because of the fermion zero-modes, 
it is somewhat analogous to the super Mumford
forms with odd spin structure (and no Ramond punctures) described in section \ref{oddone}: it does not contribute directly to a correlation function
of the spin fields, but it can be an ingredient in a larger computation of a scattering amplitude (in which, for example, one  adds NS sector
vertex operators that can absorb the fermion zero-modes).

If we want to use $\Psi_{\h V}$ to directly compute an amplitude for a product of spin fields, along the lines of section \ref{superwhatfor}, we need
constraints to reduce the dimension of   $H^0(\Sigma,\W_{\h V})$ to $1|0$.  Taking the $j_\sigma$ to be positive precisely gives 
$\sum_\sigma j_\sigma$ constraints
on an odd section of $\W_{\h V}$.  Since there are generically $5n_\Ra/2$ odd zero-modes if the $j_\sigma$ all vanish, we need 
$\sum_\sigma j_\sigma\geq 5n_\Ra/2$ to eliminate fermion zero-modes coming from $H^0(\Sigma,\W_{\h V})$.  However, if 
$\sum_\sigma j_\sigma> 5n_\Ra/2$, we will have the opposite
problem of fermion zero-modes coming from $H^1(\Sigma,\W_{\h V})$.  So the super Mumford form can be used to directly compute the
expectation value of a product of spin fields only if $\sum_\sigma j_\sigma= 5n_\Ra/2$, or
\begin{equation}\label{toldoxx} \sum_{\sigma=1}^{n_\Ra}(-5/2+j_\sigma) =0.   \end{equation}
We interpret this physically to mean that there is a $U(1)$ symmetry -- or, from a holomorphic point of view, a 
 $\C^*$ symmetry  -- such that the spin field inserted at $\F_\sigma$ has $\C^*$ charge
$-5/2+j_\sigma$.  Comparing to (\ref{doofus}), we see an obvious interpretation: the $\C^*$ in question is simply the center of the $GL(5)$ symmetry
of $V$.  As support for this, we observe that if we let $V_\sigma$ vary in the Grassmanian $\mathrm{Gr}(5,5-j_\sigma)$ of subspaces of $V$
of codimension $j_\sigma$, then $\det V/V_\sigma$ is the fiber of the fundamental line bundle $\O(1)\to \mathrm{Gr}(5,5-j_\sigma)$,
whose space of sections is $\wedge^{j_\sigma} V$.  We take this to mean that the spin field inserted at $\F_\sigma$ transforms under $GL(5)$
as $\det V^{-1/2}\otimes \wedge^{j_\sigma}V$.

\vskip1cm
 \noindent {\it {Acknowledgements}}  Research supported in part by NSF Grant PHY-0969448.  I thank R. Donagi, S. Grushevsky, A. S. Schwarz, and D. Skinner for helpful comments, and P. Deligne for much advice over the years concerning supergeometry and super Riemann surfaces.

\bibliographystyle{unsrt}

\end{document}